\pdfoutput=1
\documentclass[a4paper,lastpage]{article}
\usepackage[utf8x]{inputenc}
\usepackage[english,russian]{babel}
\usepackage{indentfirst}
\usepackage{graphicx}
\usepackage{fullpage}
\usepackage{amsmath,amssymb,mathrsfs}
\usepackage{stmaryrd}
\usepackage{bm}
\usepackage[colorlinks]{hyperref}
\usepackage{pdfpages} %14.11.2019
\begin{document}

\selectlanguage{english}

\sloppy

\title{Modern Physics of the Condensed State:
Strong Correlations and Quantum Topology}
\author{V. Yu. Irkhin and Yu. N. Skryabin}
\date{\today}
\maketitle

In the space of solid material forms,

under the rule of their laws

and vortex fluxes,

in doubts of frustration

an electron secretly stores

in the wilds of the dual lattice,

even becoming bound,

its own triple nature

including boson, fermion and photon -

gauge field mediator.

Let confinement be obstacle,

monopoles are rampant,

it will not lose at dusk

the melody of hidden strings,

the message of another world,

memory of quantum topology.

\begin{abstract}
The topic of the review is the application of new ideas of unconventional quantum states to the physics of condensed matter, in particular of solid state, in the context of modern field theory. A comparison is made with classical papers on many-electron theory, including the formalism of many-electron operators. The essentially many-particle nature of the ground state, individual and collective excitations, and quantum fluctuations in the systems under consideration, as well as quantum phase transitions are discussed with an emphasis on topological aspects and with allowance for the effects of frustration. Variational approaches and representations of auxiliary particles, corresponding mean-field approximation and gauge field theory, confinement-deconfinement problem, violation of the Fermi-liquid picture, and exotic non-Fermi-liquid states are considered. An overview is given of the modern theory of entangled topological states, formation of spin liquid, strings and string networks.
\end{abstract}

\tableofcontents

\section{Introduction}

The 2016 Nobel Prize in Physics was awarded for
the theoretical discoveries of topological phase transitions and topological phases of matter, for the works carried out in 1970--1980, which ``opened the secrets of
exotic matter''. In the time that passed since the pioneering works by D. Haldane, M. Kosterlitz, and
D. Thouless, considerable progress was achieved in
the physics of the condensed state connected with the
application of new, substantially quantum topological
concepts, such as topological phases of substance,
including those protected by symmetry, spin liquids,
exotic excitations, strings, tensor networks, etc. These
achievements were connected with the activities of several
groups of researchers: Anderson \cite{633a,6331a}, Wen \cite{Wen,Wen3,wen11,Levin-RMP},
Sachdev \cite{Sachdev-r}, Coleman \cite{end}. A large contribution was
introduced here by Russian scientists A.M. Polyakov and A.Yu. Kitaev.

\subsection{Quantum Phases and the Concept of Quantum Topology}

Let us consider briefly the formal mathematical
definitions and approaches. As is known from contemporary mathematics \cite{Turaev}, the topological field theory and the theory of knots are closely connected.
Indeed, it is intuitively comprehensible that the topological entanglement is associated with the entanglement of quantum states (in particular, this gives a
beautiful analogy with the Feynman path integral).
Thus, from a practical point of view the quantum
topology proves to be closely related to the problem of
quantum computing \cite{Kauffman}.

Correspondingly, a mathematical basis was proposed for the classification of topological phases: each
of them is connected with a mathematical object
known as a ``tensor category'' \cite{Kassel,Levin-Wen2005}, which satisfies
some algebraic equations. This object characterizes
different topological phases and determines the universal properties of quasiparticle excitations, just as
the formalism of usual symmetry groups in the Landau
theory. Thus, the mathematical structure of the tensor
categories and the physical picture of string-net condensation ensures the general theory of topological phases.

This approach also provides exactly solvable
models and wave functions of the ground state for each
of the topological phases. These are the local boson
(or spin) models, which realize all discrete gauge theories in any dimensionality and the
Chern--Simons
theories (in the dimensionality 2 + 1). The topological
order is characterized by the stable degeneracy of the
ground state (quantum state is stabilized by topological invariants) and by nontrivial statistics of particles.
In contrast to the situations of broken symmetry, the
particles emergent in the topologically ordered states
include gauge bosons, and also fermions or anyons
(particles with a fractional statistics), which can appear
as collective excitations of purely boson models.

From the viewpoint of physics, the quantum
phases of a substance are the phases of matter at zero
temperatures and correspond to the ground states of
the quantum Hamiltonian of the system \cite{Sachdev-r}. In this
sense, the insulators, magnets, and superconductors
exist at a zero temperature $T$ and are examples of
quantum phases.

After the discovery of fractional quantum Hall
effect \cite{Tsui} and high-temperature superconductivity
(HTSC) \cite{Bednorz}, it was realized that there is a new type of
order---topological order (elementary excitations with
an energy gap) or quantum order (in the more general
case of gapless excitations). Since that time, the study
and classification of new types of order in the condensed media became a field of research that was
intensely studied.

According to \cite{Wen3}, the following classification is
possible. If the quantum state has a spectrum with a
gap, then the corresponding quantum order is called a
topological order. The low-energy effective theory of
the topologically ordered state is called the topological
quantum field theory (TQFT \cite{Witten}). The second class
of quantum orders corresponds to Fermi liquids or free
fermionic systems; in this case, the different quantum
orders are classified according to the topology of the
Fermi surface \cite{Volovik}. The third, most interesting class of
quantum order is described in terms of the string-net
condensation \cite{Levin-Wen2005}. Such states have a certain similarity with the usual situation of the broken symmetry
upon the condensation of single particles: they are
similar to Bose-condensed states, but the condensate
is formed from extended objects. The collective excitations in the string-net condensate are not usual scalar bosons; the vibrations of closed strings generate
gauge bosons, and upon the breaking of strings, their
ends can give fermions.

Thus, the quantum topology that describes the
internal degrees of freedom can be reformulated in
terms of the field-theory language of closed strings,
which are constructed from local spins or from
pseudo-spins (qubits---a concept that is widely used in
quantum calculations). In turn, from these strings it is
possible to pass to the concept of ``electric'' and ``magnetic'' gauge fields \cite{Kogut}.

The very name ``topological order'' was historically
borrowed from the low-energy effective theory of chiral spin states in TQFT \cite{Witten}. The term ``topological''
here means the long-range entanglement and therefore it relates to quantum topology \cite{wen11}; it is necessary
to distinguish it from the usual classical topology,
which operates with the vortices in the superfluid liquid, with the differences between the sphere and the torus, etc.

The term chiral spin state as a special type of quantum state was proposed in 1987 for the explanation of
high-temperature superconductivity \cite{Kalmeyer,Wen-Wilczek-Zee}. In contrast to quantum states known at that time, it contains
new exotic particles---spinons and holons---in the state
of deconfinement and corresponds to a stable phase at
a zero temperature. First, under the assumption of the
validity of the Landau theory of broken symmetry, the
chiral state was considered a state that violates the
symmetries of the time inversion and parity but preserves the spin-rotational and translational symmetries. However, it was rapidly discovered that there is a
set of various chiral states that have the same symmetry. Thus, it is insufficient to consider only the symmetry in order to characterize such states. This means
that they represent a new type of order, which was
called topological order.

From the macroscopic viewpoint, the topological
order is characterized by the strong ground-state
degeneracy and by non-Abelian geometric phases
\cite{Wen1990}. However, from a microscopic viewpoint the
topological order is determined by the state of quantum spin liquid (QSL) with gapped energy spectrum
\cite{1406.5090,1407.8203}, which cannot be represented by the product
of single-particle eigenstates if we do not take into
account phase transitions with a closing of the energy
gap. Such QSLs show a long-range entanglement,
which is precisely the essence of the topological order
\cite{0510092,0510613,1004.3835}.

The emergence of correlations in topological systems is very uncommon. There are no usual ``force''
correlations (forerunners of long-range order) here,
but specific correlations appear, caused by the topology of the sample. In this case, the ground state is
degenerate, not because of symmetry, but due to the
topological characteristics. Thus, the system ``feels''
the folding into a torus, the influence of a single
(pricked) point can prove to be radical \cite{Mermin}, etc.

\subsection{Quantum Phase Transitions}

It was traditionally assumed that the phenomenological Landau theory of phase transitions \cite{Landau,Ginzburg} can
describe all possible phases of matter and all phase
transitions. It seemed that all continuous phase transitions are associated with the broken symmetry within
the framework of the so-called Landau--Ginzburg--Wilson paradigm. For a long time there was confidence that the Landau theory of broken symmetry also
described all possible quantum phases and continuous
quantum phase transitions (QPTs) between them, i.e.,
the phase transitions at $T = 0$.

In contrast to classical phase transitions, the QPTs
take into account the fluctuations of the order parameter in the imaginary time. A distinctive feature of
QPTs is the presence of the zero-temperature quantum critical point, which separates the quantumordered and quantum-disordered phases in the temperature--parameter (coupling constant, pressure,
concentration, etc.) phase diagram that describes the
proximity to the quantum critical point. At a finite $T$,
these phases are separated by a quantum critical
region, limited by the lines of crossovers, which are
converged into the critical point. The presence of such
a region in the phase diagram leads to many uncommon phenomena, in particular, to the formation of
non-Fermi liquid in metallic systems.

In particular, the so-called projective groups were
proposed. As is known, the symmetry of the usual
ordered phases makes it possible to classify them in
terms of 230 space groups (in three dimensions),
and also it leads to the appearance of collective gapless
excitations---Goldstone bosons. The projective symmetry groups (PSGs) for the quantum orders have
similar applications, making it possible to classify
more than 100 different two-dimensional Z$_2$ spin liquids, moreover, they all have identical usual symmetry
\cite{wen11}. The projective symmetry group can also lead to
the appearance of gapless gauge bosons and fermions.
In the PSG method, the wave functions are not
defined concretely, but are classified according to the
symmetry by an analysis with the aid of local unitary
transformations. The transformations of the corresponding renormalization group make it possible to
grasp the simple essence of the arising complicated phenomena. The PSG method makes it possible to determine the phase diagrams that include phases with both
long-range and short-range topological order and
entanglement. In addition, in the theory of topological
systems, terms and analogies can be used from the
information theory and theory of electrical chains and
also of neuron networks \cite{Orus,Deng}.

After the discovery of HTSCs, the hope appeared
that the exotic highly correlated states and quantum
spin liquids could play the decisive role in its understanding \cite{633a}. In such states the separation of spin and
charge is possible: electron is decomposed into two
quasiparticles---a spinon (spin 1/2, charge 0) and a
holon (spin 0, charge $e$). The Bose-condensation of
holons leads to the superconductivity, and its new
mechanism explains the great interest in the study of
different spin liquids.

However, despite concrete achievements, the theory of phases with broken symmetry cannot explain
the existence of QSLs in the ground state because of
the mutual retention (confinement) of spinons and
holons caused by the arising gauge field \cite{Baskaran}. Indeed,
it was found (see, e.g., \cite{Wen1}) that in the U(1) gauge
theory the ground state is either antiferromagnetic or
superconductive, or is a Fermi liquid---all these phases
are in the state of confinement.

Nevertheless, in the phase diagram of high-temperature superconductors at finite temperatures
regions (``pseudogap'' phases) can exist where the fermions and bosons are in the deconfined state. It
should be noted that here the fermions and bosons
cannot be considered to be free because of their strong
interaction with the gauge field. In the ground state,
the region of deconfinement is as a rule reduced into a
``deconfined'' quantum critical point \cite{0311326,0312617}. A number of important physical phenomena are connected
with this point, including the separation of the spin
and charge degrees of freedom of the electron.

Both phenomenological approaches in quantum
topology and microscopic models for the systems with
strong correlations will be considered in this survey.
We will try to trace the connections between the contemporary concepts, which use the latest achievements of the quantum field theory, and the classical
ideas of the many-electron theory, which started in the
works of Shubin and Vonsovskii, Gutzwiller, Hubbard, and other authors.

\section{Development of the Ideas and Methods of the Many-Electron Theory}

The first successes of the theory of metals in 1920--1930 were connected with the development of quantum mechanics and with the discovery of the Fermi
statistics. Within the framework of the approximation
of free electrons and then of the one-electron band
theory---in the works of Pauli, Bloch, Wilson, Peierls,
Sommerfeld---an explanation was given of paramagnetism, of the behavior of heat capacity, and of kinetic
properties \cite{1}. However, for describing ferromagnetism and a number of other phenomena (for example, metal–insulator transition), these ideas proved to
be insufficient. On the other hand, the attempt to use
for the magnetic metals the Dirac--Heisenberg model
based on the atomistic picture of localized spins also
did not give good results (in particular, it was unable to
explain the fractional values of magnetic moments).
Thus, a definite synthesis was required of the many-electron Heisenberg model and one-electron band
model.

The polar model of Shubin and Vonsovskii was
proposed \cite{662,Shubin} in 1934, and in 1946, the $s-d$
exchange model. Both these models played an exclusively important role in the theoretical description of $s$
and $f$ metals and their compounds. The further development of many-electron approaches went mainly in
the directions of diagram methods and of the phenomenological Landau theory of Fermi liquids.

The new impulse to the many-electron theory of
crystals was given by the ideas of Hubbard \cite{28,31},
who separated in his model the most essential part of
the Coulomb interaction---strong repulsion $U$ of electrons located at the same site. In the case of the nondegenerate band, its Hamiltonian is written down as
follows:
\begin{equation}
\mathcal{H}=\sum_{ij\sigma
}t_{ij}c_{i\sigma }^{\dagger }c_{j\sigma
}+ U\sum_ic_{i\uparrow }^{\dagger
}c_{i\uparrow }c_{i\downarrow }^{\dagger }c_{i\downarrow }, \label{eq:G.1}
\end{equation}
where $t_{ij}$ are the transfer integrals. Beginning from the
Hubbard works, this model was widely used for examining the ferromagnetism of itinerant electrons, the
metal--insulator transition, and other physical phenomena, which include strong correlations and phenomena that are not described by the methods of perturbation theory.

In the lecture ``Many Body Physics: Unfinished
Revolution'' \cite{end}, P. Coleman speaks about three eras
of the physics of the condensed state: the first successes after the discovery of quantum mechanics (free
fermions); many-body physics of the middle of the
20th century (use of the perturbation theory and
Feynman diagram technique; the beginning of the
study of collective phenomena); and the contemporary era of physics of the highly correlated
matter---the
study of ``exotic'' systems, in which the collective phenomena come to the foreground.

At present, the role of the quantum correlations is
actively discussed, where the topological degeneracy
comes to the foreground, which leads to a radical complication of the ground state of the system.

\subsection{Second Quantization: Fermions and Many-Electron Operators}

Upon the transition to the standard representation
of second quantization \cite{651}, the wave functions of a
crystal $\Psi (x_1\ldots x_N)$ ($x=\{\mathbf{r}_is_i\}$, $s_i$
are the spin coordinates) are selected in the form of linear combinations
of Slater determinants. They are comprised from the
one-electron wave functions $\psi _\lambda (x)$ ($\lambda =\{{\nu \gamma \}}$, $\nu $
are
the indices of cells in the lattice, and $\gamma$ are the one-electron sets of quantum numbers):
\begin{equation}
\Psi (x_1\ldots x_N)=\sum_{\lambda _1\ldots \lambda _N}f(\lambda
_1\ldots \lambda _N)(N!)^{-1/2}\sum_P(-1)^P P\prod _i\psi _{\lambda _i}(x_i),
 \label{eq:A.1}
\end{equation}
where $P$ runs over all possible transpositions of $x_i$. The
idea of second quantization is introduced by using
one-electron occupation numbers $n_x$ as new variables:
\begin{equation}
\Psi (x_1\ldots x_N)=\sum_{\{n_\lambda \}}f(\ldots n_\lambda
\ldots )\Psi _{\{n_\lambda \}}(x_1\ldots x_N).
 \label{eq:A.3}
\end{equation}
Then, $f(\ldots n_\lambda \ldots )$ plays the role of a new wave function.
The one-electron Fermi operators of creation and
annihilation are determined as follows:
\[
c_\lambda f(\ldots n_\lambda \ldots )=(-1)^{\eta _\lambda
}n_\lambda f(\ldots n_\lambda -1\ldots ),
\]
\begin{equation}
c_\lambda ^{\dagger }f(\ldots n_\lambda \ldots )=(-1)^{\eta
_\lambda }(1-n_\lambda )f(\ldots n_\lambda +1\ldots ),
 \label{eq:A.4}
\end{equation}
where
\[
\eta _\lambda =\sum_{\lambda ^{\prime }>\lambda }n_{\lambda
^{\prime }},\qquad c_\lambda ^{\dagger }c_\lambda =\hat{n}_\lambda .
\]

It should be noted that the very introduction of
one-electron operators, which became customary for
us, is a very nontrivial step. Alternatively, the operators
of fermions in the lattice (pseudo-) spin models can be
constructed from the operators of the pseudo-spin \cite{Wen}
\begin{equation}
c(i_{a})=\sigma ^{+}(i_{a})\prod\limits_{b<a}\sigma ^{z}(i_{b})
\end{equation}
(the product is taken along the lattice sites ordered in
a specific manner on the contour), which generalizes
the Jordan--Wigner transformation for the one-dimensional lattice to the lattices of higher dimensions.
Thus, a fermion appears as a substantially many-particle and nonlocal formation. This gives another trend
in the development of the theory of many-particle systems, connected with the string picture, which uses
the language of qubits (see Section 5).

Now let us generalize the method of second quantization, introducing the quantum numbers of electron groups $\Lambda =\{{\nu \Gamma }\}$, where $\Gamma$ are the many-electron (ME) levels. In particular, it is possible to unite electrons at each site in the lattice
($\Lambda =\{{\nu \Gamma }\}$, $\Gamma_i$ are the ME levels), to obtain
\begin{equation}
\Psi (x_1\ldots x_N)=\sum_{\{N_\lambda \}}f(\ldots N_\lambda
\ldots )\Psi _{\{N_\lambda \}}(x_1\ldots x_N).
 \label{eq:A.5}
\end{equation}

In the case of an atomic configuration of equivalent electrons $l^n$, the ME wave function of the electron group is determined by the recursion relation (see \cite{20}):
\begin{equation}
\Psi _{\Gamma _n}(x_1\ldots x_N)=\sum_{\Gamma _{n-1},\gamma
}G_{\Gamma _{n-1}}^{\Gamma _n}C_{\Gamma _{n-1},\gamma }^{\Gamma
_n}\Psi _{\Gamma _{n-1}}(x_1\ldots x_{n-1})\psi _\gamma (x_n),
 \label{eq:A.6}
\end{equation}
where ${C}$ are the Clebsch--Gordan coefficients. In the case of the $LS$ coupling
\begin{equation}
C_{\Gamma _{n-1},\gamma }^{\Gamma _n}\equiv
C_{L_{n-1}M_{n-1},lm}^{L_nM_n}C_{S_{n-1}\mu _{n-1},\frac 12\sigma
}^{S_n\mu _n},
 \label{eq:A.7}
\end{equation}
the summation over $\gamma =\{lm{\sigma }\}$ occurs instead of the summation over one-electron orbital projections $m$ and spin projections $\sigma $, rather than over $l$;
$G_{\Gamma _{n-1}}^{\Gamma _n}\equiv G_{S_{n-1}L_{n-1}}^{S_nL_n}$ are the fractional parentage coefficients. If the added electron belongs to another shell, it is possible to simply write
\begin{multline}
\Psi _{\Gamma _n}(x_1\ldots x_n)=n^{-1/2}\sum_{i,\Gamma
_{n-1},\gamma }(-1)^{n-i}C_{\Gamma _{n-1},\gamma }^{\Gamma
_n}\times
\\
\times \Psi _{\Gamma _{n-1}}(x_1\ldots x_{i-1},x_{i-1}\ldots
x_{n-1})\psi _\gamma (x_i)
 \label{eq:A.10}
\end{multline}
(in contrast to the case of equivalent electrons, the antisymmetrization is required here). The wave function of the entire crystal (\ref{eq:A.5}) can now be obtained as the antisymmetrized product of ME functions for the electron groups, which in principle makes it possible to introduce ME states for the entire crystal.

By analogy with (\ref{eq:A.6}), (\ref{eq:A.10}), it is possible to introduce ME operators of creation for the electron groups \cite{652}. For the equivalent electrons, we have
\begin{equation}
A_{\Gamma _n}^{\dagger }=n^{-1/2}\sum_{\Gamma _{n-1},\gamma
}G_{\Gamma _{n-1}}^{\Gamma _n}C_{\Gamma _{n-1},\gamma }^{\Gamma
_n}a_\gamma ^{\dagger }A_{\Gamma _{n-1}}^{\dagger }.
 \label{eq:A.11}
\end{equation}

Upon the addition of an electron from another shell, we obtain
\begin{equation}
A_{\Gamma _n}^{\dagger }=\sum_{\Gamma _{n-1},\gamma }C_{\Gamma
_{n-1},\gamma }^{\Gamma _n}a_\gamma ^{\dagger}A_{\Gamma
_{n-1}}^{\dagger}.
 \label{eq:A.12}
\end{equation}

The operators (\ref{eq:A.11}), (\ref{eq:A.12}) are convenient only for the operations with the configurations with fixed numbers of electrons. In order to avoid the problems of ``nonorthogonality'' in the case of different $n$, projection coefficients should be introduced \cite{653,654}:
\begin{equation}
\tilde{A}_\Gamma ^{\dagger }=A_\Gamma ^{\dagger }\prod_\gamma
(1-\hat{n}_\gamma ).
 \label{eq:A.16}
\end{equation}

Formally, the product in (\ref{eq:A.16}) is taken over all permissible one-electron states $\gamma $. However, because of the identity $c_\gamma ^{\dagger }\hat{n}_\gamma =0$, it suffices to preserve only those $\gamma $ that do not enter into the corresponding products of operators in $A_\Gamma $. Introducing ME operators that depend on all one-electron quantum numbers (of both occupied and free states), we make the next step in the quantum-field description after usual second quantization. Thus, it is possible to come to the concept of ME occupation numbers $N_\Gamma = \tilde{A}_\Gamma ^{\dagger }\tilde{A}_\Gamma $ at a given site.

Now, we can introduce the generalized projection operators---the Hubbard $X$ operators \cite{31}:
\begin{equation}
X(\Gamma ,\Gamma ^{\prime })=\tilde{A}_\Gamma ^{\dagger
}\tilde{A}_{\Gamma ^{\prime }},
 \label{eq:A.21}
\end{equation}
which convert state $\Gamma ^{\prime }$ into state $\Gamma $:
\begin{equation}
X(\Gamma ,\Gamma ^{\prime })=|\Gamma \rangle \langle \Gamma
^{\prime }|, \, X(\Gamma ,\Gamma ^{\prime })X(\Gamma ^{\prime \prime },\Gamma
^{\prime \prime \prime })=\delta _{\Gamma ^{\prime }\Gamma
^{\prime \prime }}X(\Gamma ,\Gamma ^{\prime \prime \prime }),
 \label{eq:A.22}
\end{equation}
where $|\Gamma \rangle $ are the exact eigenstates of the Hamiltonian. In the simplest case of $s$ electrons,  $\gamma  =\sigma =\pm (\uparrow ,\downarrow )$, $\Gamma =0,\
\sigma ,\ 2$ and $|0\rangle $,  is the free state (hole), and $|2\rangle $ is the %twice
occupied singlet state at the site (a double).

An arbitrary operator $\hat{O}$ acting on the electrons at a given site $i$, is expressed through the $X$ operators as follows:
\begin{equation}
\hat{O}=\sum_{\Gamma \Gamma ^{\prime }}\langle \Gamma |\hat{O}|\Gamma
^{\prime }\rangle X(\Gamma ,\Gamma ^{\prime }).
 \label{eq:A.29}
\end{equation}
Then, finding the matrix elements of one-electron Fermi operators from (\ref{eq:A.11}), we obtain the following representation \cite{32}:
\begin{equation}
c_\gamma ^{\dagger }=\sum_nn^{1/2}\sum_{\Gamma _n\Gamma
_{n-1}}G_{\Gamma _{n-1}}^{\Gamma _n}C_{\Gamma _{n-1},\gamma
}^{\Gamma _n}X(\Gamma _n,\Gamma _{n-1}).
 \label{eq:A.31}
\end{equation}

In particular, for $s$ electrons
\begin{equation}
c_\sigma ^{\dagger }=X(\sigma ,0)+\sigma X(2,-\sigma ).
 \label{eq:A.32}
\end{equation}

In fact, the ME states in general can be treated as the primary states that are obtained when solving the complete ME problem, rather than constructed from the one-electron states (as it was done above based on the example of atomic approach). In this case, an important role is played by the nonlocality, and the nature of collective states can be very complicated and include long-range interaction.

The existence of nontrivial many-particle states is confirmed by the generalized Lieb--Schultz--Mattis theorem \cite{Lieb}, according to which in the system with the half-integer spin per cell and global U(1) symmetry the excitation spectrum in the thermodynamic limit cannot obey simultaneously to two requirements: (a) that the ground state is unique; and (b) that there is a finite gap for all excitations. This means that the gapped state with the unbroken symmetry must have a degenerate ground state, which is topological by its nature.

With an increase of the strength of correlations in ME systems, there occurs a change of the statistics of elementary excitations from the band-type to the atomic type, which is manifested as a correlational Hubbard splitting and metal--insulator transition (for the half-filled band). This transformation is accompanied by the formation of local magnetic moments \cite{353} and by a change in the number of states located below the Fermi surface if the latter generally exist (see Section 3.6).

As was indicated by Anderson \cite{Zou1}, the formation of Hubbard subbands is connected with the orthogonality catastrophe, as a result of which the quasiparticle residue disappears and the Landau Fermi-liquid picture is violated. The disappearance of the residue means that the addition of one particle to the system leads to a global change in its state, which is characteristic of ME systems. In this case, in particular, the formation of exotic particles and of a state of the spin-liquid type can be expected. All these problems will be discussed below.

\subsection{Variational Approaches}

Different approximate methods to work with ME
states were developed for many years. In the works of
Shubin and Vonsovskii, a simple quasi-classical
approximation \cite{662,81} was used. It consists in the
replacement of $X$ operators by $c$-number functions,
which determine the amplitude of the probability of
the existence of a site in the states of a singly occupied
site, doubly occupied site, or a hole:
\begin{equation}
X_{i}(+,0)\rightarrow \varphi _{i}^{\ast }\Psi _{i},~X_{i}(2,-)\rightarrow
\Phi _{i}^{\ast }\psi _{i},~X_{i}(2,0)\rightarrow \Phi _{i}^{\ast }\Psi _{i}
\label{quasicl}
\end{equation}
with an additional condition
$
|\varphi _{i}|^{2}+|\psi _{i}|^{2}+|\Phi _{i}|^{2}+|\Psi _{i}|^{2}=1.
$
This corresponds to the definition of the total energy of the system based on the variational principle with the wave function
\begin{equation}
\phi =\prod_{i}[\varphi _{i}^{\ast }X_{i}(+,0)+\psi _{i}^{\ast
}X_{i}(-,0)+\Phi _{i}^{\ast }X_{i}(2,0~)+\Psi _{i}^{\ast }])|0\rangle .
\label{dops}
\end{equation}

It mixes the excitations of the Bose and Fermi types and, therefore, does not satisfy the Pauli principle. Nevertheless, the quasi-classical approximation makes it possible to roughly describe the metal--insulator transition, which indeed was subsequently made by a number of authors (see, e.g., the survey \cite{81}). Thus, Caron and Pratt \cite{caron} even considered the mean field for the fermions, so that in fact it is the usual Fermi operators that were replaced by the $c$-number functions rather than the $X$ operators.

Anticipating the Hubbard works, the equations of quasi-classical approximation give a change in the width of an energy band and also in the relative location of different bands depending on the number of doubles and magnetic moments.

The variational Gutzwiller method proved to be more rigorous and more consistent (see \cite{Vollhardt}). The corresponding wave function can be written as
\begin{equation}
\Psi = \prod_{i}[1-(1-\hat{n}_{i\uparrow}\hat{n}_{i\downarrow})]|\psi_0\rangle ,
\label{Eq.111}
\end{equation}
where the variation parameter $g$ ($0<g<1$) takes into account the decrease of the probability of states with a large number of doubles, and $|\psi_0\rangle$ is the wave function of the uncorrelated state.

A different version of the variational approach, which takes into account long-range interaction, is the exciton approach, in which the condensation of electron--hole or electron--electron pairs is considered \cite{exciton,exciton1}. Most simply, it can be illustrated using the description of superconductivity. The standard Bardeen--Cooper--Schrieffer (BCS) wave function is written as
\begin{equation}
| \Psi_0\rangle =\prod_{\bf k}(u_{\bf k}+v_{\bf k}c_{\bf k \uparrow}^\dagger c_{-\bf k \downarrow}^\dagger) | 0\rangle
\end{equation}
and is a superposition of states with different numbers of particles $N$, i.e., in this case there are introduced Bogolyubov anomalous averages. In \cite{BCS}, the calculations with the wave function with a fixed $N$ were carried out,
\begin{equation}
| \Psi_0\rangle =
\biggl[ \sum_{\bf k}  \phi_{\bf k} c_{\bf k \uparrow}^\dagger c_{-\bf k \downarrow}^\dagger
\biggr]^{N/2} | 0\rangle ,
\end{equation}
which explicitly introduces the condensate of Cooper pairs with the use of the theory of the enumeration of graphs. In \cite{692}, the ground state of a Hubbard antiferromagnet was constructed analogously without the introduction of anomalous averages (magnetic sublattices):
\begin{equation}
| \Psi_0\rangle =
\biggl[ \sum_{\bf k}  \phi_{\bf k} c_{\bf k+Q \downarrow}^\dagger c_{\bf k \uparrow} ,
 \biggr]^{N/2}
| F\rangle ,
\end{equation}
where $| F\rangle$ is the ferromagnetic state, and $\bf Q $ is the wave vector of the magnetic structure. This description, according to the observed results, corresponds to Slater's antiferromagnetism. Thus, the state with a long-range order can easily be described; however, the description of a correlated paramagnetic state requires a solution of a very complicated problem---the taking into account of excitons with different momenta.

An analogous description of a long-range order in an antiferromagnet can be given within the framework of the Heisenberg model with the aid of a function $| \Psi_0\rangle = (S_{\bf Q}^+)^{N/2} | F\rangle$ \cite{Heis}.

A variational description of the state of the resonating valence bonds (RVBs), which combines the approaches mentioned above, was developed by Anderson \cite{633a}. In the case of a half-filled band, the following trial function can be used here:
\begin{equation}
\Psi = P_G|\psi_0\rangle,
\label{Eq.12}
\end{equation}
where $P_G = \prod_{i}(1-\hat{n}_{i\uparrow}\hat{n}_{i\downarrow})$ is the Gutzwiller projection operator in the limit of strong correlations, which completely removes the doubly occupied states at the site. Since $\phi({\bf k}) = v_{\bf k}/u_{\bf k} \simeq {\rm sign} (t_{\bf k}-E_F)$ ($t_{\bf k}$ is the band energy), the function (\ref{Eq.12}) can be written, also, in the form of
\begin{equation}
| \Psi_0\rangle =\prod_{k<k_F}c_{\bf k \uparrow}^\dagger c_{-\bf k \downarrow}^\dagger| 0\rangle . \label{Eq.121}
\end{equation}

It is precisely under the action of the projector that the BCS function, which describes the superconductive state, is converted into the RVB state---i.e., there appears a Fermi surface, since the Bogolyubov transformation factors take values of 0 and 1. Thus, the function of the ground state must be constructed just based on ME excitations. However, as it will be shown in Section 3.1, in practice there is frequently used a special function of the mean field, and the sum rule at the site can be taken into account through the topological vortex excitations. The particular selection of the function $\phi({\bf k})$ (for example, in the approximation of nearest neighbors) makes it possible to take into account a special short-range pairing (formation of singlets).

The N\'eel state possesses a long-range order and an infinite degeneracy of the ground state, which leads to the appearance of Goldstone modes---magnons. On the other hand, the RVB state is a nondegenerate singlet state with a short-range antiferromagnetic order. The appearance of sublattices can be described as decoherence (violation of quantum coherence in the singlet state) \cite{decoh}.

According to the approach described in \cite{Wen1}, the trial function for the half filling in the case of an antiferromagnet can be written as follows:
\begin{equation}
| \Psi_0\rangle = P_G
\biggl[ \sum_{\bf k} ( A_{\bf k} a_{\bf k \uparrow}^\dagger a_{-\bf k \downarrow}^\dagger
+ B_{\bf k} b_{\bf k \uparrow}^\dagger b_{-\bf k \downarrow}^\dagger )\biggr]^{N/2}
| 0\rangle ,
\end{equation}
where the operators $a_{\bf k \sigma}^\dagger$ and $b_{\bf k \sigma}^\dagger$ correspond to upper and lower Slater subbands with the energies $\pm \xi_{\bf k}$,  $A_{\bf k} = (E_{\bf k} + \xi_{\bf k})/\Delta_{\bf k}$, $B_{\bf k} =  (-E_{\bf k} + \xi_{\bf k})/\Delta_{\bf k}$, where $E_{\bf k} = \sqrt{ \xi_{\bf k}^2+ \Delta_{\bf k}^2}$ and $\Delta_{\bf k} = (3/8)J \Delta( \cos k_x - \cos k_y)$. Due to the introduction of the singlet RVB pairing, which is determined by the magnitude of $\Delta$, this wave function gives a substantially better value of the energy in comparison with the usual Slater antiferromagnetism.

The wave function (\ref{Eq.12}) can also be written through many-electron $X$ operators \cite{633a}:
\begin{equation}
| \Psi_0\rangle =\prod_{\bf k}(u_{\bf k}+v_{\bf k}X^{+0}_{\bf k } X_{-\bf k} ^{-0} | 0\rangle).
\end{equation}

The ideas of the exciton condensate were also applied to systems with the intermediate valence of the SmB$_6$ and SmS type within the framework of the two-band Falikov–Kimball models with hybridization or of the Anderson lattice \cite{590}. Depending on the values of the parameters in these problems, excitons of a large and a small radius can appear.

Such approaches to the problem of intermediate
valence were recently also developed in the modern
language of exotic particles---spinons and holons \cite{SmB6,Assaad}. According to \cite{SmB6}, the neutral fermionic composite exciton is formed as the bound state of a conduction electron and holon.

\subsection{Ideas of the Auxiliary Particles}

The direct work with the many-electron $X$ operators is difficult (in particular, the usual expansions of the perturbation theory and the diagram technique are absent, although the formal expansions in terms of the reciprocal coordination number $1/z$ \cite{654} may be used). Therefore, their representations through the auxiliary (``slave'') Bose and Fermi operators are frequently used.

Depending on the specific physical problem and situation, different variants of such representations are used. In connection with the theory of two-dimensional high-temperature superconductors, Anderson \cite{633a} suggested the following representation:
\begin{equation}
c_{i\sigma}^{\dagger}=f_{i\sigma}^{\dagger}e_i+\sigma d_i^{\dagger}f_{i-\sigma}.
 \label{eq:6.131}
\end{equation}

Expression (\ref{eq:6.131}) corresponds to the partition of the one-electron operator into two many-electron $X$ operators (\ref{eq:A.32}), i.e., to the formation of Hubbard subbands. Here, $f_{i\sigma }^{\dagger}$ are the creation operators for the neutral fermions (spinons), and $e_i^{\dagger}$ and $d_i^{\dagger}$ are the creation operators for the charged spinless bosons. The physical meaning of such excitations can be explained as follows. Let us consider a lattice with one electron at the site with a strong Hubbard repulsion, so that each site is neutral. In the ground RVB state, each site participates in one bond. When the bond is broken, two unpaired sites appear, which possess spins equal to 1/2. The corresponding excitations (spinons) are not charged. At the same time, the vacancy site (hole) in the system bears a charge, but not a spin. This picture describes the state of a spin liquid, where the separation of the spin and charge degrees of freedom of an electron occurs and gauge fields play an essential role.

The introduction of four bosons $p_{i \sigma}$, $e_i$,  $d_i$, which
correspond to the projection onto the singly occupied
states, holes, and doubles, respectively, was suggested
as an extension of the physical space in \cite{Kotliar}. As a
result, the Hubbard Hamiltonian takes on the following form:
\begin{equation}
\mathcal{H}=\sum_{ij \sigma}t _{ij}f_{i \sigma}^{\dagger }f_{j \sigma}z_{i \sigma}^{\dagger }z_{j \sigma} + U\sum_{i}d _{i}^{\dagger }d _{i},
 \label{eq:133}
\end{equation}
where $f_{i\sigma },f_{i\sigma }^{\dag }$ are the auxiliary Fermi operators,
\begin{equation}
z_{i\sigma }^{\dag}=(1-e_{i}^{\dag }e_{i}-p_{i-\sigma}^{\dag }p_{i-\sigma})^{-1/2}(p^{\dag }_{i\sigma }e^{}_{i} + d^{\dag}_{i}p^{}_{i-\sigma})
(1-d_{i}^{\dag }d_{i}-p_{i\sigma }^{\dag }p_{i\sigma
})^{-1/2},
\label{zzz}
\end{equation}
and additional constraints are superimposed:
\[
\sum_{\sigma}p_{i \sigma}^{\dagger }p_{i \sigma}
+d _{i}^{\dagger }d _{i}+d _{i}^{\dagger }d _{i}=1, \,
f_{i \sigma}^{\dagger }f_{i \sigma}=p_{i \sigma}^{\dagger }p_{i \sigma}+d _{i}^{\dagger }d _{i}.
\]
Here, there is a definite arbitrariness: the additional operator coefficients have their own eigenvalues equal to 1 in the physical subspace, but upon the decoupling make it possible to pass to the limit of free electrons, which is important for constructing the mean-field theory. This representation made it possible to qualitatively reproduce a number of previous results (for example, just as in the case of the quasi-classical approximation (\ref{quasicl}), to obtain the description of the metal--insulator transition \cite{Lavagna1}). The mean-field theory for the representation of the Kotliar--Ruckenstein bosons (\ref{eq:133}) gives results identical to the Gutzwiller theory in the limit of an infinite coordination number. Within the framework of this representation, the magnetic ordering can also be considered \cite{Igoshev:2015}.

A rotationally invariant version \cite{Fresard:1992a} also exists
\begin{equation}
c_{i\sigma }^{\dag }=\sum_{\sigma ^{\prime }}f_{i\sigma }^{\dag }z_{i\sigma
\sigma ^{\prime }}^{\dag },~\hat{z}_{i}=e_{i}^{\dag}\hat{L}_{i}M_{i}\hat{R}_{i}%
\hat{p}_{i}+\widehat{\tilde{p}}_{i}^{\dag}\hat{L}_{i}M_{i}\hat{R}_{\hat{i}%
}d_{i},
\label{zz}
\end{equation}
where the coefficients $L$, $M$, and $R$ are analogous to (\ref{zzz}), the scalar and vector auxiliary boson fields and  are introduced as $\hat{p}_{i}=\frac{1%
}{2}(p_{i0}\sigma _{0}+\mathbf{p}_{i}\mbox{\boldmath$\sigma $})$, and $%
\widehat{\tilde{p}}_{i}$ is the time-inverted operator $\hat{p}_{i}$. This variant is convenient in the case of magnetically ordered phases, making it possible to take into account corrections due to spin fluctuations. In particular, it makes it possible to rather simply describe the non-quasiparticle states caused by electron--magnon scattering, which previously were considered in the many-electron representation of Hubbard operators (see \cite{Irkhin:1990}) (in this case, the $p^-$ operators describe the magnons, and $f$, describe the spinons.

Let us note also the representation of slave rotors \cite{Florens}
\begin{equation}
c_{i\sigma }^{\dag }=f_{i\sigma }^{\dag }b_i=f_{i\sigma }^{\dag }\exp (i\theta _{i}),
\label{rotor}
\end{equation}
in which the initial Hubbard interaction is replaced by a simple kinetic term for the phase field $ (U / 2) \hat {L} _ {i} ^ {2} $ with the angular momentum $\hat {L} = - i \partial / \partial \theta $. This representation is convenient for describing the metal--dielectric transition in the paramagnetic phase, and it was also generalized onto the case of doping \cite{rotor,rotor1,rotor2}.

A representation was proposed in \cite{Wang} that contains two types of Fermi operators---holons $ e_{i}$ and doublons $d_{i}$ which correspond to holes and doubles:
\begin{eqnarray}
X_{i}(+,0) &=&e_{i}(1-d_{i}^{\dagger
}d_{i}^{{}})(1/2+s_{i}^{z}),~X_{i}(-,0)=e_{i}(1-d_{i}^{\dagger
}d_{i}^{{}})s_{i}^{-},  \nonumber \\
X(2,-) &=&d_{i}^{\dagger }(1-d_{i}^{\dagger
}d_{i}^{{}})s_{i}^{+},~X(2,+)=d_{i}^{\dagger }(1-d_{i}^{\dagger
}d_{i}^{{}})(1/2+s_{i}^{z}).  \label{ed}
\end{eqnarray}

In this case, the operators of physical spins are connected with the pseudo-spin operators $s_{i}^{\alpha}$ by the relationship $\mathbf{S}_{i}=\mathbf{s}_{i}(1-d_{i}^{\dagger
}d_{i}^{{}}-e_{i}^{\dagger}e_{i}^{{}})$.  Different special cases of this representation were considered previously, which correspond to the limit of large $U$ ($t-J$ models, see survey \cite{Izyumov1}).

Supersymmetrical representations were also proposed \cite{lavagna} at a later date.

\subsection{$t-J$ Model and $s-d$ Exchange Model with Strong Correlations}

Upon the examination of strongly correlated compounds, for example, of copper–oxygen high-temperature superconductors, the $t-J$ model (Hubbard's model with $U\rightarrow \infty $ and with the allowance for the Heisenberg exchange) is widely used. The Hamiltonian of this model in the ME representation takes on the following form:
\begin{equation}
\mathcal{H}=-\sum_{ij\sigma }t_{ij}X_i(0,\sigma )X_j(\sigma,0)
+\sum_{ij}J_{ij}{\bf S}_i {\bf S}_j.
 \label{eq:I.7}
\end{equation}

Upon the derivation of the $t-J$ model from the Hubbard model with large $U$, the elimination of doubles via the canonical transformation \cite{Zou} leads to the Anderson antiferromagnetic super-exchange $J=-2t^2/U$; however, sometimes it is convenient to consider that $J$ is an independent variable. In the simple model (\ref{eq:I.7}) a rich phase diagram appears, which includes spiral magnetic structures and inhomogeneous states (see survey \cite{Izyumov1}).

According to Anderson \cite{Casey}, the $t-J$ Hamiltonian is not only a convenient alternative of the Hubbard model; it reflects the physical fact that the low-energy states are located in the subspace, which is overfilled and is described by one band of electronic states, since the anti-bonding states of doubles from the upper part of the band are neglected. There is no such convergent perturbation method, which could connect low-energy states with the initial band of the Hubbard model, since they exist in the Hilbert spaces of different dimensionalities.

The first-principles model for describing cuprates is a three-band $p$---$d$ model
\begin{eqnarray}
\mathcal{H}&=&\sum_{\mathbf{k}a\sigma }\left[ \varepsilon
p_{\mathbf{k}a\sigma }^{\dagger }p_{\mathbf{k}a\sigma }+\Delta
d_{\mathbf{k}\sigma }^{\dagger }d_{\mathbf{k}\sigma
}+V_{\mathbf{k}}(p_{\mathbf{k}\sigma }^{\dagger
}d_{\mathbf{k}\sigma }+h.c.)\right]
\nonumber\\
&+&U\sum_id_{i\uparrow }^{\dagger }d_{i\uparrow }d_{i\downarrow
}^{\dagger }d_{i\downarrow },
 \label{eq:6.108}
\end{eqnarray}
where $\varepsilon $ and $\Delta $ are the positions of the $p$ and $d$ levels for O and Cu ions, and  are the matrix elements of the $p$---$d$ hybridization (cf. \cite{Raimondi}). In the limit of large $U$, we can use a slave-boson representation $d^\dag_{i\sigma} \rightarrow X{_i(\sigma 0)}=f^{\dag}_{i\sigma}e_i$.

At $|V_{pd}|\ll \varepsilon -\Delta $ (large charge transfer gap), the Hamiltonian (\ref{eq:6.108}) is again reduced by the canonical transformation to the $t-J$ model with $t_{\text{eff}}=V_{pd}^2/(\varepsilon -\Delta )$. It is interesting that the $t-J$ model obtained from the one-band Hubbard model is also formally reduced to an analogous structure in the representation (\ref{zzz}) with the auxiliary rather than physical particles $ p_i $. Thus, the Hubbard model and the model (\ref{eq:6.108}) can be examined in a parallel manner \cite{Raimondi}.

The model (\ref{eq:I.7}) describes the interaction of charge carriers with the local moments. In order to explicitly demonstrate the separation of these degrees of freedom, let us indicate its equivalence to the $s-d$ exchange model with the exchange parameter $|I| \rightarrow \infty $. Indeed, after the passage to the ME representation the Hamiltonian of this model is written as \cite{699,6991}
\begin{equation}
\mathcal{H}=\sum_{ij\sigma }t_{ij} g_{i\sigma \alpha }^{\dagger
}g_{j\sigma \alpha }+\mathscr{H}_d,\quad \alpha ={\rm sign}{}
I,
 \label{eq:I.5}
\end{equation}
where
\begin{equation}
g_{i\sigma \alpha}^{\dagger }=\sum_M\{[S+\sigma \alpha M+(1+\alpha)/2]/(2S+1)\}^{1/2}
X_i\left( M+\frac \sigma 2,\alpha;M\right).
\end{equation}

The model (\ref{eq:I.5}) corresponds to the case where the charge carriers do not belong to the energy band in which the magnetic moments are formed. This situation takes place in some magnetic semiconductors and insulators. However, it can easily be shown that at $S=1/2$, $\alpha =-$, the Hamiltonian (\ref{eq:I.5}) coincides with (\ref{eq:I.7}), with $t_{ij}$ in (\ref{eq:I.5}) being replaced by $2t_{ij}$ (factor of 2 appears as a result of the equivalence of electron jumps with both opposite spins in the Hubbard model).

The Hamiltonian (\ref{eq:I.5}) can be expressed through the Fermi operators and operators of localized spins. Using (\ref{eq:A.21}) and (\ref{eq:A.12}), it is possible to obtain the operators of conduction electrons from the $X$ operators:
\begin{equation}
g_{i\sigma \alpha }^{\dagger }=\sum_{\sigma ^{\prime }}c_{i\sigma
^{\prime }}^{\dagger }(1-n_{i,-\sigma ^{\prime }})\left[ P_\alpha
\delta _{\sigma \sigma ^{\prime }}+\frac \alpha
{2S+1}(\mathbf{S}_i\mbox{\boldmath$\sigma $}_{\sigma ^{\prime }
\sigma })\right],
 \label{eq:I.8}
\end{equation}
\begin{equation}
P_{+}=\frac{S+1}{2S+1},\qquad P_{-}=\frac S{2S+1}.
 \label{eq:I.9}
\end{equation}
Then, the use of the properties of the Pauli matrices gives
\begin{multline}
\mathscr{H} =\sum_{ij\sigma \sigma ^{\prime }}t_{ij}\Biggl\{
\left[ P_\alpha
^2+\frac{\mathbf{S}_i\mathbf{S}_j}{(2S+1)^2}\right] \delta
_{\sigma \sigma ^{\prime }}+\frac \alpha {(2S+1)^2}P_\alpha
(\mathbf{S}_i+\mathbf{S}_j)\mbox{\boldmath$\sigma $}_{\sigma
\sigma ^{\prime }}+
\\
+\frac{i}{(2S+1)^2}\mbox{\boldmath$\sigma $}_{\sigma \sigma
^{\prime }}[\mathbf{S}_i\times \mathbf{S}_j]\Biggr\} c_{i\sigma
}^{\dagger }(1-\hat{n}_{i,-\sigma })(1-\hat{n}_{j,-\sigma ^{\prime
}})c_{j\sigma ^{\prime }}+\mathscr{H}_d.
 \label{eq:I.10}
\end{multline}
For the spin operators, we have
\begin{equation}
\mathbf{S}_i=\frac 12\sum_{\sigma \sigma ^{\prime }}a_{i\sigma
}^{\dagger }\mbox{\boldmath$\sigma $}_{\sigma \sigma ^{\prime
}}a_{i\sigma ^{\prime }}, \,
 \label{eq:O.1}
\end{equation}
where, depending on the problem, it is possible to use the representation of Schwinger bosons ($a_{i\sigma }^{\dagger } = b_{i\sigma }^{\dagger }$) or Fermi spinons ($a_{i\sigma }^{\dagger } = f_{i\sigma}^{\dagger }$). The terms in (\ref{eq:I.10}) that are linear in spin operators provide the opportunity of the effective hybridization of electrons with spinons.

The representation of the Hamiltonian in the form of (\ref{eq:I.10}) (of course, it is correct in the $t-J$ model as well) was obtained in \cite{653,654}. Subsequently, it was used for constructing the special mean-field theory in application to the phase diagram of HTSC cuprates as a ``new formulation of the $t-J$ model'' \cite{Ribeiro}. In the mean-field theory for the $t-J$ model of these authors, the electronic spectral weight is comprised from two bands: the low-energy spinon band and high-energy electron band. The spectral weight from the spinon band is a sharp coherent peak. The wide spectral weight from the electron band corresponds to the incoherent background.

Upon the examination of the spectrum of cuprates, a nodal--antinodal dichotomy \cite{Wen1} arises: the nature of the spectrum is different in the different regions of the Fermi surface. The spectrum is gapless near the nodal points $(\pm\pi/2,\pm\pi/2)$ (where the excitations are described as Dirac fermions, see Section 3.3) and is gapped near the anti-nodal point $(0,\pi)$. Near a nodal point, there appears a strong hybridization mixing between spinons and electrons (dopons, in the terminology of \cite{Ribeiro}), while near the anti-nodal point, the mixing is absent. In essence, the described picture is close to the hybridization two-band model of Kondo lattices, where the localized and itinerant states are initially separated, but are mixed up in the mean-field approximation for the auxiliary $f$ pseudo-fermions, so that the latter participate in the formation of the Fermi surface \cite{I17} (see also Section 3.5).

Thus, the use of the concept of a two-band model helps to obtain the physical picture of the exotic states of the strongly correlated systems.

The terms that contain vector products describe the anisotropic electron scattering and they can be important upon the examination of kinetic phenomena in the case of narrow bands, for example, of the anomalous Hall effect. The Hamiltonian (\ref{eq:I.10}) can be useful upon a study of states with the anomalous ``chiral'' order parameters, which were investigated within the framework of the two-dimensional Heisenberg model and of the $t-J$ model (see, e.g., \cite{703}).

\section{Fermi Liquid and non-Fermi-Liquid Phases}

The concept of a Fermi liquid assumes the one-to-one correspondence of the low-energy states between the interacting system in question and the system of the noninteracting electrons. This implies, in particular, the existence of only quasiparticle excitations with a charge $ \pm e $ and spin 1/2. This means the existence of a Fermi surface determined by the poles of the single-particle Green's function at the energy $E = 0$. This Fermi surface obeys the Luttinger theorem, i.e., it surrounds the volume of momentum space determined by the total electron density.

The violation of the Fermi-liquid behavior at low temperatures can be caused by different factors. In clean systems, effects of interaction can give stable non-Fermi-liquid phases. One of the scenarios consists in the fact that the low-energy excitations acquire quantum numbers different from the usual electrons or holes, which leads to the uncommon low-temperature properties. In the case of the dimensionality $d = 1$, such a behavior is usual and is well understood (Luttinger liquids with the separation of spin and charge). Similar separation is expected in the spin liquids in the case of higher dimensions.

In the case of the half-filled band, the decrease of Coulomb repulsion leads to the transition of the Mott insulator into the metal. Near this transition, the metallic state is strongly correlated, including Hubbard subbands and local moments, which are subjected to frustration. The combined action of these factors can lead to a significantly reduced scale of coherence and, consequently, to a heavy-fermion behavior. For examining the metal--dielectric transition, it is convenient to use the rotor representation (\ref{rotor}) of the electron annihilation operator in the form of a product of the charged boson $b_i$ and the neutral fermion with a spin (spinon) $f_{i\sigma}$ \cite{Senthil,Sachdev4}. For the odd filling of the band, the system of spinless bosons passes with an increase of the Hubbard parameter $U$ into the superfluid Mott insulator. In the mean-field approximation, the spinons are free (noninteracting), in spite of the strong correlations in the electronic system. If the bosons $b_i$ are condensed, i.e., $\langle b \rangle \neq 0$ there is obtained a Fermi-liquid phase of physical electrons: upon the replacement of $b$ by its $c$-number average $\langle b \rangle $, the fermions $f_ \sigma$ acquire the same quantum numbers as the initial electrons, so that their Fermi surface describes a usual metal. If a boson is gapped and, therefore, is not condensed, there is formed a spin-liquid Mott insulator, in which the Fermi surface of neutral fermionic excitations (spinons) survives. The Mott insulator for the bosons is also a dielectric state for the electrons with a gap for all charged excitations, and there occurs a continuous transition to the insulator with a ``ghost'' Fermi surface of spinons.

A similar scenario of the metal--insulator transition can be considered in the representation of the Kotliar--Ruckenstein bosons (\ref{zzz}) \cite{Lavagna1,Castellani}. In this case, when $\langle b \rangle \neq 0$ or when it is small (the quasiparticle residue is small), the lacking spectral weight is determined by fluctuations---by the incoherent contribution of Hubbard subbands \cite{Castellani}.

Besides the decrease of interaction, one more additional method to convert the frustrated Mott insulator into metal is doping. Upon small levels of doping, local moments can exist and, therefore, effects of frustration can be important. Therefore, a number of ideas were suggested on the appearance of uncommon metallic phases with a non-Fermi-liquid behavior; in particular, such a metal can represent a fractionalized state with spinon and holon excitations (Section 3.6).

It can be supposed that far from the half-filling of the band, especially at low electron densities, Fermi-liquid state appears, since the acts of particle scattering become rare, so that the effects of even strong Coulomb interaction $U$ are weak. However, the situation is by no means so simple. Indeed, according to the spectral representation of Green's functions, for the number of doubles that determines the local moment we have \cite{353}
\begin{equation}
N_2 =\langle c_{i\uparrow }^{\dagger }c_{i\uparrow }c_{i\downarrow
}^{\dagger }c_{i\downarrow }\rangle
=-\frac 1{\pi U}\int {\rm Im} \sum_{\mathbf{k}}\frac{\Sigma _{\mathbf{k}}(E)}{E-t_{\mathbf{k}}}f(E)\,dE .
 \label{eq:H.21}
\end{equation}
Since the self-energy $\Sigma _{\mathbf{k}}(E)$ is finite in the limit of $U \rightarrow \infty$ (in particular, in the Kanamori $T$-matrix approximation), $N_2$ behaves as $1/U$. Then, the Hellmann--Feynman theorem $N_2 = \partial \mathcal{E}/\partial U$ gives the divergence of the energy of the ground state $\mathcal{E}$:
\begin{equation}
\mathcal{E}(U)-\mathcal{E}(0)=\int\limits_0^\infty N_2(U)\,dU\sim \ln{U}.
 \label{eq:H.22}
\end{equation}

This divergence indicates the inadequacy of the one-electron picture of the Fermi liquid at large $U$ and the formation of Hubbard subbands. On the other hand, the calculation in the many-electron representation of $X$ operators \cite{353} gives a correct asymptotic behavior $N_2\sim 1/U^2$. An accurate result is also obtained in the representation of the Kotliar--Ruckenstein bosons (\ref{eq:133}).

The authors of \cite{YRZ} suggested a phenomenological representation for the one-electron Green's function, which satisfies the Luttinger's theorem in a different way: the electron density is connected with the number of singularities of the logarithm of the Green's function at the zero energy rather than with the volume of the Fermi surface. Thus, both the usual poles and the zeros of the Green's function (i.e., the quasiparticle poles of the electron self-energy) are taken into account; the last surface is called the Luttinger surface.

From the phenomenological form of the Green's function, the dispersion of quasiparticles and the spectral weight are obtained, which characterize the coherent part of the one-electron Green's function. In this case, a critical concentration of holes, above which the spin-liquid anomalous self-energy disappears. As a consequence, there is a quantum critical point, which separates two topologically different phases. At a density lower than the critical density of holes exists, the $G(\bf{k}, 0)$ is characterized by the coexistence of the Luttinger surface and of $p$-type pockets of the Fermi-surface; at higher densities, there is only a closed Fermi surface, as in the Landau's theory.

\begin{figure}[tbp]
\centering
\includegraphics[width=9.0cm,height=7.0cm]{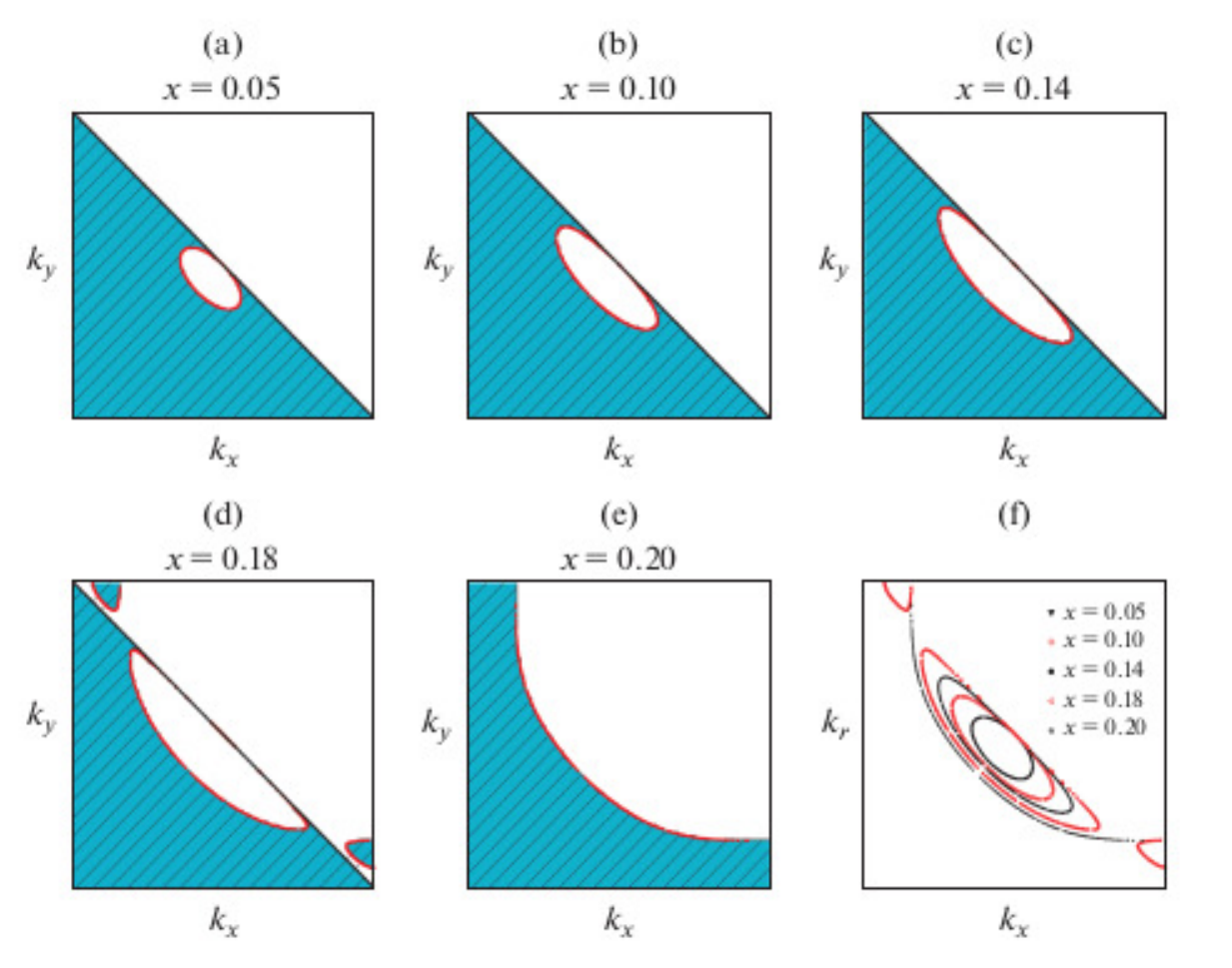}
\caption{
Evolution of the Fermi surface with a hole concentration $x$ in the model of \cite{YRZ}. In the shaded regions, $G({\bf k}, 0)> 0$.  The thick lines in (a)--(d) are the lines of Luttingers's zeroes; near the critical doping $x_c = 0.18$ in (d), there is an additional electron pocket.
}
\label{fig1_fig}
\end{figure}

As is shown in Fig. \ref{fig1_fig}, in a narrow range of $x\lesssim x_{c}$ a new set of electron Fermi-pockets appears near the saddle points, outside the Luttinger surface of zeroes. These pockets merge with the hole pockets inside the Luttinger's surface at $x = x_{c}$ forming a united Fermi surface. Thus, there are two closely spaced topological changes: a transition of the Lifshitz type, upon which the form of the Fermi surface of the poles undergoes topological changes; and a quantum critical point $x = x_{c}$ connected with opening of a single-particle gap with the appearance of an RVB spin liquid at  which leads to the formation of a Luttinger surface.

The form of the Green's function \cite{YRZ} is close to the result of \cite{Ng} obtained on the basis of the slave-boson mean-field theory in the $t-J$ model. In this approach, the spinons and holons interact with each other through the gauge field, and for describing the gauge fluctuations there is introduced a phenomenological attraction between them, which is constant in a certain range of distances. This attraction leads to the binding of the spinon and holon and, consequently, to the quasiparticle pole of self-energy.

\subsection{Mean-Field Theory and Gauge Fields}

Let us discuss the picture of a spin liquid in the framework of the representation (\ref{eq:6.131}). In the case of a half-filled band, there are present only spinon excitations with a kinetic energy of order $|J|$. Upon the doping of the system by holes, there appear charge carriers that are described by the holon operators. Thus, a spin-liquid state appears with a suppressed long-range magnetic order. In this case, in the simplest gapless variant of the RVB theory in the purely spin (undoped) systems, a low-energy scale $J$ and a large linear term in the specific heat capacity with $\gamma \sim 1/|J|$ appear, connected with the existence of the Fermi surface of spinons \cite{633a}.

Upon the further development of the theory, another variant of the RVB theory was proposed, in which a gap on the Fermi surface appears, so that a structure of the $d$-wave type is formed with the maximum of the gap near $(0, \pi)$; this is in agreement with photoemissive experiments (see \cite{6331a}). More complicated variants of the RVB theory were also developed, in which topological approaches and analogies with the fractional quantum Hall effect are used (see, e.g., \cite{633}). These ideas led to the uncommon and beautiful results. In particular, the spinons can obey fractional statistics, i.e., the wave function of the system acquires a complex coefficient upon the transposition of two quasiparticles.

Let us examine a mean-field approximation in the $t-J$ model in the representation of auxiliary bosons   $c^\dagger_{ i\sigma} \rightarrow X_i(\sigma 0)= f_{ i\sigma}^\dagger b_{ i}$ which corresponds to the U(1) theory. Here, it is possible to introduce the pairings
\begin{equation}
\chi_{ij} = \sum_\sigma \langle
f^\dagger_{i\sigma} f_{j\sigma} \rangle, \,\,
\Delta_{ij} = \langle  f_{i\uparrow}f_{j\downarrow} - f_{i\downarrow}f_{i\uparrow}  \rangle .
\label{Eq.40}
\end{equation}

\begin{figure}[t]
\centering
\includegraphics[width=2.5in]{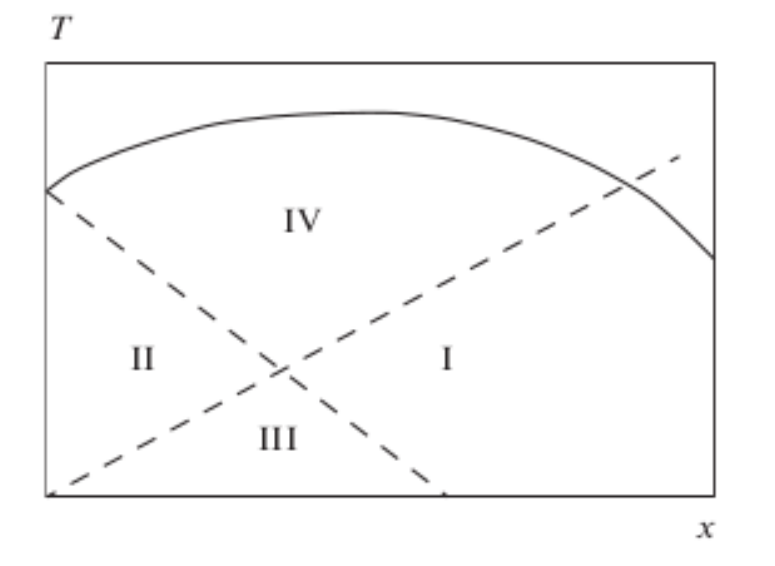}
\caption{
Schematic phase diagram of the mean-field theory U(1). The solid line designates the beginning of the uniform RVB state $ (\chi \neq 0)$; the dashed line corresponds to the appearance of a fermionic pairing $(\Delta \neq 0)$; and %short
dashed (startng from zero), Bose condensation in the mean-field approach $(b \neq 0)$. Four regions appear: (I) Fermi liquid with $\chi \neq 0$, $b \neq 0$; (II) the spin-gap state with $\chi \neq 0$, $\Delta \neq 0$; (III) $d$-wave superconductor with $\chi \neq 0$, $\Delta \neq 0$, $b \neq 0$; and (IV) strange metal with $\chi \neq 0$ \cite{Lee}.
}
\label{U1}
\end{figure}

In the uniform RVB phase (uRVB), $\chi_{ij} = \chi$ for all bonds and is real, and the gap is $\Delta_{ij} = 0$ so that the spectrum of $f$ fermions has the form $E_{\bf k} = - 2 {J}\chi (\cos k_x + \cos k_y)$. However, there are phases with a lower energy, including a $d$-wave superconductor. The phase diagram with taking into account the last state is shown in Fig. \ref{U1}.

It should be again be noted that the description of the Fermi-liquid phase in the $t-J$ model is self-contradictory in some sense, since some part of states is already excluded because of the projection onto the singly occupied states. In this sense, a more consistent is the approach that takes into account the formation of doubles and both auxiliary bosons ($e$ and $d$) within the framework of the representations (\ref{eq:6.131}) or (\ref{eq:133}). The latter representation makes it possible also to consider the formation of an upper and a lower Hubbard subbands as an incoherent contribution---a bound state of a spinon and a holon \cite{Castellani}.

More complicated spin-liquid phases are obtained upon the allowance for the SU(2) invariance of the $t-J$ Hamiltonian, which makes it possible to remove a number of difficulties of the U(1) theory \cite{Wen1}. The SU(2) slave-boson theory, which describes the doped spin liquids, includes two SU(2) doublets $\psi_{ i}$ and $h_{ i}$ with two boson fields $b_{1 i}$ and $b_{2 i}$ with spin 0, which form the SU(2) singlets. Then, we have the following representation of the electronic operator:
\begin{align}
\label{cpsib}
c_{\uparrow i}
\rightarrow& {1\over \sqrt 2} h^\dagger_{ i}\psi_{ i}
= {1\over \sqrt 2} \left (b^\dagger_{1 i} f_{\uparrow i}
+ b^\dagger_{2 i}f^\dagger_{\downarrow i}\right ),
\nonumber\\
c_{\downarrow i}
\rightarrow& {1\over \sqrt 2} h^\dagger_{ i} \bar \psi_{ i}
= {1\over \sqrt 2} \left ( b^\dagger_{1 i}f_{\downarrow i}
- b^\dagger_{2 i}f^\dagger_{\uparrow i}\right ).
\end{align}

In the sfL state (staggered flux liquid), we obtain
\begin{equation*}
 E_{\bf k}=\pm \frac34 J\sqrt{\chi^2 (\cos k_x+\cos k_y)^2
 + \Delta^2 (\cos k_x - \cos k_y)^2 }.
\end{equation*}

This phase has a Dirac line spectrum at the points $(\pm \pi/2,\pm \pi/2)$. In the $\pi$fL phase,
\begin{equation*}
 E_{\bf k}=\pm \frac34 J|\chi| \sqrt{\sin^2 k_x + \sin^2 k_y}.
\end{equation*}

Besides these two gapless phases, there can also appear a gapped Z$_2$ state.

\begin{figure}
\centering
\includegraphics[width=3in]{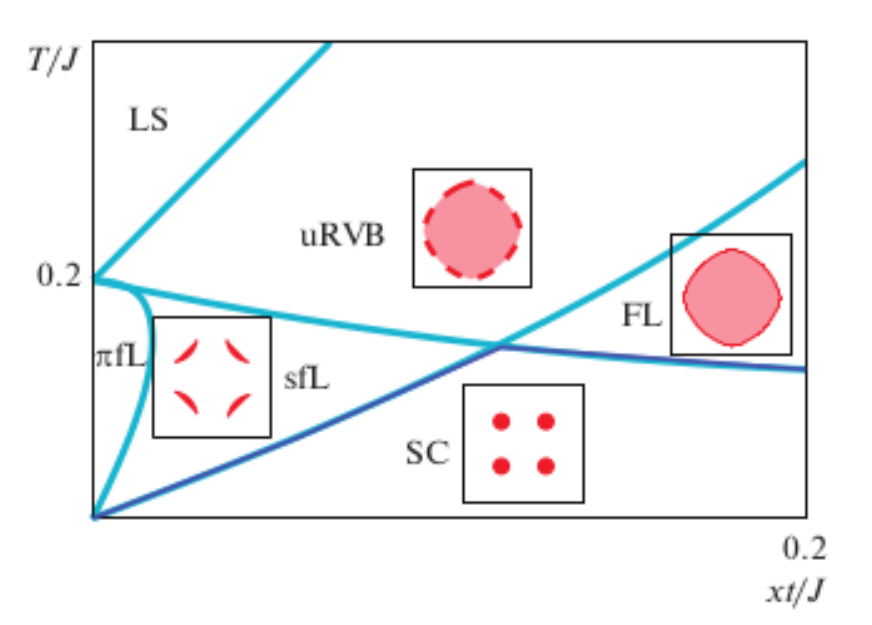}
\caption{
SU(2) mean-field phase diagram for $t/J = 1$ \cite{Lee}. There are shown a Fermi surface, Fermi arcs, and Fermi points in different phases.
}
\label{fig3}
\end{figure}

The phase diagram in the SU(2) slave-boson theory of the doped Mott insulator is shown in Fig. \ref{fig3}. This theory also is not free from difficulties. In particular, the area of the Fermi surface approaches to $1 - x$ very slowly with an increase in the doping $x$ and a decrease in temperature.

The intuitive picture of a spin liquid with the symmetry SU(2) is given exactly by the Anderson RVB state, which includes the singlet pairing of spins and the formation of a quantum superposition of many different pairs, i.e., different dimeric covering of the lattice, with the retention of the symmetry of the Hamiltonian. This picture includes fractional excitations, since the destruction of a dimer leads to two excitations with the independent dynamics: the monomers obtained are the objects, which carry charge 0 and spin 1/2, i.e., are the spinons. In the spin liquid Z$_2$ they are connected with the appearing Z$_2$ excitations, which are the Z$_2$ vortices (or fluxes) called ``visons'' (see also Section 5.1).

The wave function $|\Psi_\text{MF}{(\chi_{i j})}\rangle$ of the mean-field Hamiltonian
\begin{align}
\label{HmeanU1}
\mathcal{H}_\text{MF}=J \sum_{ij\sigma} (\chi_{ij}f^\dag_{i\sigma}f_{j\sigma}+h.c.)
+\sum_{i\sigma} a_0(f^\dag_{i\sigma}f_{i\sigma}-1),
\end{align}
is not a completely satisfactory function of a spin system, since it satisfies the condition of single filling of sites only on average. In order to take this condition into account more precisely, it is necessary to introduce U(1) gauge fluctuations of the phase of the quantity $a_0$, which correspond to the fluctuations of chirality, $\chi_{i j}\rightarrow \bar \chi_{i j}\exp(-i a_{ij})$. Performing projecting onto the singly occupied states, we obtain the following physical wave function \cite{Wen1}:
\begin{equation}
\Psi_\text{spin}{(\chi_{i j},\{\sigma_{i}\}})
= \langle0_f|\prod_{i} f_{i\sigma_{i}} |
\Psi_\text{MF}{(\chi_{i j})}\rangle ,
\label{PsiPsichi}
\end{equation}
where $|0_f\rangle$ is the state without $f$ fermions: $f_{\sigma_i}|0_f\rangle=0$. This construction makes it possible to explicitly obtain the wave function corresponding to the gauge fluctuations $a_{ij}$
\begin{equation}
\Psi_\text{spin}{(a_{ij})}
= \langle0_f|\prod_{i} f_{i\sigma_{i}}
|\Psi_\text{MF}{(\bar \chi_{i j}\exp{(i a_{ij})})}\rangle,
\end{equation}
so that there arise a gauge field in a natural way.

Let us discuss the topological nature of quasiparticle excitations and the gauge field in a chiral spin state \cite{Wen,Wen1}. A simplest excitation in the mean-field approximation can be obtained by adding a spinon into the conduction band. However, this excitation is not yet physical, since a spinon in the conduction band violates the limitation $\sum_{\sigma} \langle f^\dag_{i\sigma}f_{i\sigma}\rangle=1$. The additional density of spinons can be removed by introducing a vortex flux of the gauge field
$$\Phi= -\pi \sum_{i\sigma} (\langle f^\dag_{i\sigma}f_{i\sigma}\rangle-1/2).$$

Therefore, the physical quasiparticles in the chiral spin state are spinons dressed by a $\pi $ vortex. The dressed spinons carry a spin 1/2, since a vortex cannot induce arbitrary spin quantum numbers. However, a spinon that bears a unit charge of the gauge field has a fractional statistics, being a bound state of a charge and a vortex.

On the whole, the mean-field theory proves to be extremely rich. Within the framework of the gauge theory, the appearance of a large number of spin liquids is possible, both gapped and gapless. Besides of the spin-boson representation, which describes the deconfinement phase with gapless spinons, there can be used a spin-fermionic representation, which however describes only gapped phases \cite{Wen}.

\subsection{Confinement and Deconfinement}

Formally, the representation (\ref{eq:6.131}) is always valid; however, in reality the picture of spinons and holons can be realized only under specific conditions (for example, in the case of exactly solvable models of the Kitaev type \cite{Kitaev1,Kitaev2}). Here, the fundamental question is the problem of confinement and deconfinement \cite{Wen,Wen1}. In the confinement phase, the entire picture of spin liquid is destroyed upon the consideration of fluctuation effects as a result of the interaction of a spinon with a gauge boson because of the proliferation of topological defects described by the corresponding homotopic groups \cite{Mermin}. As a result, the spinons interact with each other through the potential linearly increasing with distance, while for the bosons there is a large energy gap, so that all these exotic particles in no way are manifested in the low-energy spectrum. As a result, the spinons again become ``glued'' with the bosons into the usual electrons and we return to the trivial state of a usual Fermi liquid---effective single-particle theory. Thus, the utilized complicated formalism agrees with the customary physical concepts of the one-electron theory \cite{Wen}.

The interaction of the gauge fields on the classical level does not lead to confinement. However, if we go beyond the limits of the classical approximation and examine the interaction between the gauge fluctuations, the picture will completely change. In the dimensionality 2 + 1, such an interaction transforms the potential $g \ln R$ into a linear one, irrespective of the coupling constant $g$. Thus, the mean-field SU(2) states with the gapped fermions are unstable, and the mean-field theory cannot be trusted.

A different picture appears in the deconfinement phase (gapless topological phase). Here, the topological defects are absent, the gauge field is noncompact, and the flux of the electric field and the topological charge are conserved. As a result, the spinons and the gauge bosons are well defined particles, and strings of the electric flux and topological order (string networks) appear, which indicates the appearance of a many-particle quantum entanglement.

The interactions and mechanisms that lead to the deconfinement can be different: the frustrations of the exchange parameters, Kondo effect (suppression of magnetic moments by conduction electrons), tunneling, pseudo-spins, etc. Frequently, their description can be reduced to the effective isotropic or anisotropic Heisenberg model \cite{PL}.

As the order parameter for the confinement–deconfinement transition, the concept of the Wilson's loop can be used (see \cite{Wen1}). It is connected with the gauge potential $V(R)$ between two static gauge charges $\pm q$ with the opposite signs placed at a distance $R$ as $W(C) = \exp[-V(R)T]$ where the loop $C$ consists of the paths of length $T$ along the time direction and of length $R$ along the spatial direction. There are two types of the behavior of $W(C)$ the area law with $ W (C) \sim \exp{[- \alpha RT]} $; and the law of perimeter with $W(C) \sim \exp {[-\beta (R + T)]} $ where $\alpha$ and $\beta$ are constants. In the first case, the potential $V(R)$ grows linearly with $R$ and, therefore, the two gauge charges cannot be free. Therefore, this case corresponds to the confinement, while the second case corresponds to the deconfinement.

The compact quantum electrodynamics (purely gauge model) in the dimensionality 2 + 1 is in confinement no matter how low is the coupling constant because of the appearance of instantons \cite{Polyakov1987}. In this case, the instanton behaves as a positive magnetic charge, and the anti-instanton, as a negative magnetic charge. A Coulomb gas in the 2 + 1 dimensionalities is always in the screening phase: the Coulomb interaction becomes short-range because of the screening of the charge by a cloud of surrounding charges of the opposite sign. Therefore, the creation energy of an instanton is finite and there are freed magnetic charges, which disorder the gauge field, leading to the area law and to the confinement. However, the confinement can disappear at some temperature higher than a certain transition temperature.

In the presence of material fields, the problem of the confinement becomes even more complicated. The particle–antiparticle pairs can be induced from the vacuum, screening static charges. At finite temperatures, the transition to the deconfinement is substituted by a gradual crossover into a plasma phase. Therefore, it can be expected that the slave-boson mean-field theory without confinement describes the physics on the intermediate energy scale, although the ground state is of confinement type (antiferromagnetic, superconducting, or Fermi-liquid). As an example, there can serve a ``deconfined'' psevdo-gap state discussed in the case of the HTSC systems, which exists only at finite temperatures and is described by fermions and bosons connected with noncompact gauge fields.

One additional method of obtaining the phase of deconfinement is the transition into higher dimensions. In the dimensionality 3 + 1, the gapless U(1) fluctuations do not create confining interactions. In dimensionality 4 + 1 and above, even a non-Abelian gauge theory can be in the deconfinement phase. Thus, it is possible to construct boson models on the cubic lattice, which have gapless photons---U(1) gauge bosons \cite{Wen1}.

The phenomenon of the deconfinement recently is studied actively in the case of calculations by the quantum Monte Carlo method for the specific models of the Kitaev type, which include frustration \cite{Sandvik}; these calculations are sufficiently reliable, since, in contrast to the standard fermionic tasks, the problem of sign is absent here. The picture of deconfinement can be rather complicated: not all particles prove to be involved in it. The size of spinons is obtained to be sufficiently large---comparable with the size of their bound state, which differs from the ``classical'' RVB with the short bonds.

The deconfinement in two-dimensional systems is also investigated experimentally in the case of organometallic compounds with a square lattice. It is manifested as the appearance of spinons in the high-energy excitation spectrum even in the absence of frustrations \cite{exper}. With moving from the corresponding anomalous wave vector, these fractional particles can become bound, forming usual magnons.

\subsection{Dirac Fermions and the Algebraic Spin Liquid}

There is one additional way to the deconfinement state---it is the interaction of the compact gauge field U(1) with gapless fermions, in particular, with Dirac fermions appearing in the sfL phase. In this case, the correlated metallic state is semi-metallic, which contains gapless electron excitations only at separate Fermi points of the Brillouin zone; this determines the special topology of the problem \cite{Volovik}. These points are the Fermi surface of the electrons. This model can be applicable to HTSC cuprates \cite{Wen1} or to bilayer graphene \cite{IScr}.

Since the states near the Fermi points have a Dirac spectrum, the problem can be analyzed within the framework of the appropriate relativistic formalism. The low-energy action for neutral Dirac spinons $\Psi$ in the insulator phase has the following structure:
\begin{eqnarray}
S&=&\int d^{3}x \sum_\mu \sum_{\sigma=1}^N\bar{\Psi}_{\sigma}
 (\partial_{\mu}-ia_{\mu})\gamma_{\mu}\Psi_{\sigma},
\label{Q}
\end{eqnarray}
where the integration is carried out in 2 + 1 dimensions, $\gamma_\mu$ are the Dirac matrices, and
$\bar{\Psi}_{\sigma} \equiv \Psi^{\dag}_{\sigma} \gamma_{0}$, and $a_\mu$ is the emergent gauge field.

Depending on the details of the lattice, $a_\mu$ can be a gauge field of the U(1) or SU(2) type. For a large number of ``colors'' $N$ (determined by the number of Dirac points in the Brillouin zone), the action $S_D$ describes the conformal field theory (CFT). Thus, we have a scale-invariant quantum state with a strong interaction and a power-law spectrum for all excitations, and the absence of well defined quasiparticles. This state is called an algebraic spin liquid (see \cite{Sachdev4,Wen1}).

Thus, the gapless excitations, which carry gauge charges, can screen the gauge interaction, making it less confining \cite{Wen1}. In contrast to the true deconfinement phases, where the noninteracting quasiparticles become free at low energies, here the deconfinement means only that the gapless charged particles remain gapless, but are not completely free. The corresponding gapless spin liquids are obtained from the phase of the exotic liquid sfL and from the uRVB.

At the level of the mean field, the electron Green's function of the uRVB state in the coordinate space is a product of the fermion and boson Green's functions, and in the momentum--frequency representation is their convolution. The spectral function consists of a broadened quasiparticle peak with a weight of the order of doping and an incoherent background. Upon the doping of the uRVB phase, a strange-metal state with a large Fermi surface appears, so that the picture of the isolated spinon Dirac points proves to be somewhat changed \cite{Wen1}.

The mean-field sfL phase leads to a gapless U(1) spin liquid. It gives an example of an algebraic spin liquid, in which the correlation functions have branch cuts instead of poles, and the spin correlations decrease according to an algebraic power law. For other widely known gapless states, such as solids, superfluid liquids, Fermi liquids, etc., the gapless excitations are always described by free bosons or free fermions. The only exception is the one-dimensional Luttinger liquid; the algebraic spin liquid can be considered as its generalization beyond the limits of one dimension. It is important that the algebraic spin liquid is a phase of the matter, rather than simply appears at the critical point between two phases.

The Fermi-liquid phase includes boson condensation, which restores the quasiparticle picture. Therefore, the low-energy excitations in this phase are described by electron-like quasiparticles, and this phase corresponds to the Fermi liquid of electrons.

The dynamics for the gauge field U(1) appears as a result of the screening by bosons and fermions; moreover, they both bear a gauge charge. In the case of low doping, it is possible to take into account screening only by fermions. After integration with respect to $\Psi$ in (\ref{Q}), the effective partition function for the gauge field U(1) has the following form \cite{Kim}:
\begin{eqnarray}
\cal Z&=&\int Da_{\mu}\exp\Big( -\frac{1}{2}\int\frac{d^3q}{(2\pi)^3}a_{\mu}
({q})\Pi_{\mu\nu}a_{\nu}(-{q})\Big), \nonumber \\
\Pi_{\mu\nu}&=&\frac{N}{8}\sqrt{{q}^2}\Big(\delta_{\mu\nu}
- \frac{q_{\mu}q_{\nu}}{{q}^{2}}\Big).
\label{Pi}
\end{eqnarray}
The polarizability $\Pi$ converts the gauge coupling $a_\mu j^\mu$ ($j^\mu$ is the current) into a marginal disturbance at a fixed point of free fermions.

Let us now examine the electron Green's function. In the first order in $1/N$, we obtain
\begin{equation}
 G(x)=\langle b^\dagger( x) b(0)\rangle_0
\langle f( x) f^\dagger(0) \exp (i\int_0^{x} dx \cdot a) \rangle,
\end{equation}
where $ \int_0^{x}dx$ is the integral along the direct path of return and means the integral over gauge fluctuations \cite{Wen1}. Then, we have
\begin{equation}
 G(x)
\propto (x^2)^{-(2-\alpha)/2}
\label{b}
\end{equation}
with the exponent $\alpha \sim 1/N$ which has an anomalous dimensionality; for the antiferromagnet with a square lattice, $\alpha = 32/(3\pi^2N)$. These results describe a partial confinement of spinons and bosons coupled by a gauge field.

\subsection{Model of Phase Strings}

Since in the presence of Hubbard projectors the Fermi--Dirac statistics is not applicable, Weng proposed ``the statistics of phase strings,'' which describes the process of doping in the $t-J$ model \cite{Weng2007,Weng20071,Zaanen-Overbosch}. The entire fermionic sign structure on simple lattices for the half-filled band in the Mott insulator is completely reducible (removable), since the problem of interacting electrons is converted into the problem of interaction of localized spins. When the lattice is frustrated, the spin problem can, as before, include the problem of sign, which hampers calculations of the Monte Carlo type, but the nature of the spin in the Mott state implies localization, so that the electrons become distinguishable.

An injected hole remains stable because of the confinement of spinons and holons by the field of the phase shift, in spite of the fact that the background is a spinon--holon sea. The exact deconfinement appears only in the limit of the zero doping, when the hole loses its integrity and is decomposed into a holon and a spinon. Upon the doping of a Mott insulator by holes, there appears an irreducible sign structure, but upon the low doping this structure must be very rarefied in comparison with the equivalent system of free fermions with the same density.

A simple method to formulate the Weng statistics consists in systematically calculating the irreducible signs in the language of world lines \cite{Zaanen-Overbosch}. For the noninteracting Fermi gas, the empirical signs enter into the partition function as
\begin{equation}
Z_{FG} = \sum_c (-1)^{N_\text{ex} [c]} Z_0 [c],
\label{fermionsigns}
\end{equation}
where $Z_0[c]$ designates the partition function for the world line of one particle, the sum is taken over the configurations of lines $c$, $Z_0[c]> 0$ and $N_{\text ex}[c] = \sum_w w C_w [c] - \sum_w C_w [c]$ is an integer, which numerates the number of exchanges (number of revolutions $w$ around the time axis); $C$ are possible cyclic expansions. On the contrary, the irreducible signs entering into the partition function of the $t-J$ model can be calculated as
\begin{equation}
Z_{t-J} = \sum_c (-1)^{N^h_\text{ex} [c] + N^{\downarrow}_h [c]} {\cal Z} [c].
\label{tJsigns}
\end{equation}

The sum now is taken over the configurations of the world lines of spin particles (``spinons'') and holes. Relative to each other, the holes behave as fermions, and $N^h_\text{ex}[c]$ calculates the exchanges in the configuration $c$. The spinons here are bosons with a solid core (the upward spins are considered as the background). The new aspect of the Weng statistics consists in the fact that the sign structure is determined by the parity of the number of hole--spinon collisions, which are connected with the concrete configuration of world lines.

Formally, the projected operator of annihilation of electrons is written in the slave-fermionic representation; moreover, the sign coefficient $(-\sigma)^j$ is considered explicitly:
$ c_ {j \sigma} = (-\sigma)^jf^{\dagger}_j b_{j\sigma}$, where $f^{\dagger}_j$ is the fermionic holon, and Schwinger bosons $b^{\dagger}_{j\sigma}$ are used for the spin system. Because of the sign coefficient, the spin-spin superexchange acquires the general negative sign; therefore, the wave function of the ground state of the purely spin system upon the half filling does not have nodes.

Weng suggested one more formulation, which is strictly equivalent to the representation of world lines. It includes boson operators $h^{\dagger}_{i\sigma}$ and $b_{i\sigma}$ for ``holons'' and ``spinons,'' which satisfy the limitation $h^{\dagger}_i h_i + \sum_\sigma b^{\dagger}_{i\sigma} b_{i \sigma} = 1$ which can be interpreted in the pseudo-spin language. Using the formalism of gauge fields, it is possible to write down the projected electronic operator as
\begin{equation}
c_{i\sigma}  =  h^{\dagger}_i b_{i\sigma} \exp (\frac{1}{2} \left[ \Phi^s_i - \Phi^0_i - \sigma \Phi^h_i \right] ) (\sigma)^{N_h} (-\sigma)^i,
\label{elopweng}
\end{equation}
where $N_h$ is the total operator of the number of holons. This representation of the Jordan--Wigner type includes $\Phi^{s, h, 0}$ phases nonlocally connected with the position of all other particles. This nonlocal effect is taken into account by the introduction of a ``mutual statistics'' between the spins and holes.

\begin{figure}[ht]
\centering
\includegraphics[keepaspectratio,width=3.5in]{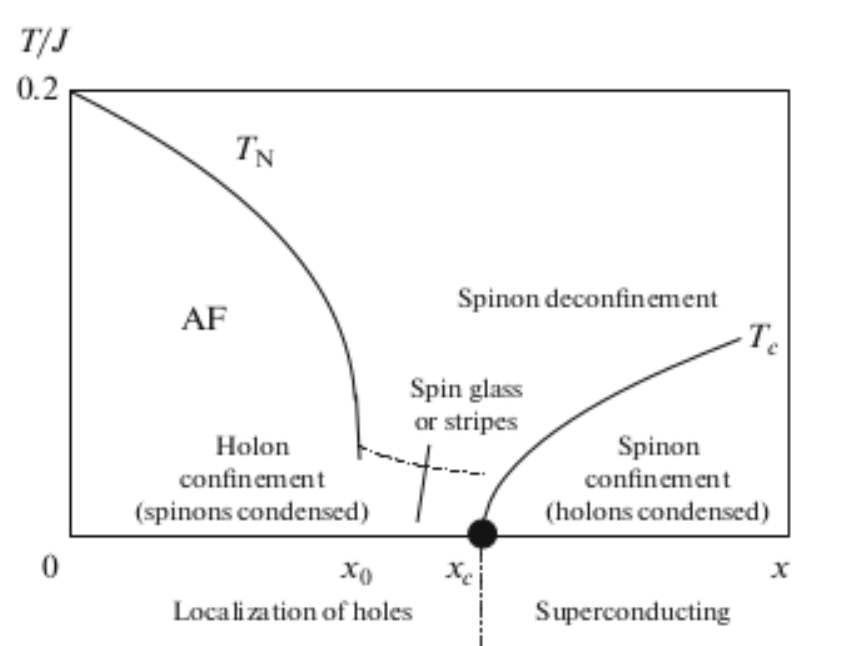}
\caption{
Phase diagram of the model of phase strings upon low doping \cite{Weng2003}. Dual topological gauge structure determines different phases by means of confinement–deconfinement.
}
\label{inter}
\end{figure}

According to \cite{Ye}, the antiferromagnetic and superconducting phases in the Weng picture are dual: in the first phase, the holons are in the confinement and the spinons are in the deconfinement; in the second phase, vice versa. These two phases are separated by a new phase---by the so-called Bose insulator, where both the holons and the spinons are in the deconfinement.

\subsection{Frustrations and Spin Liquid in Kondo Lattices}

Due to the suppression (screening) of the magnetic moment, the Kondo effect can lead to the appearance of exotic states---heavy Fermi liquid or spin liquid. The energy gain of the nonmagnetic state in comparison with the one-band Hubbard model or Heisenberg model is determined by the Kondo temperature $T_K$ \cite{711}. In turn, the tendency toward the state of spin liquid (formation of RVB singlets) gives an additional gain for the Kondo state in comparison with the magnetically ordered phases.

To describe the ground state of a Kondo lattice in the regime of strong coupling, a special mean-field approximation was developed, in which a pseudo-fermionic representation was used for the operators of localized spins $S=1/2$. Within the framework of this approximation, Coleman and Andrei \cite{711}, taking into account the SU(2) symmetry in the formalism of path integral, examined the formation of the state of spin liquid in the two-dimensional situation. Then, this approximation was applied to the case of ferro- and antiferromagnetic ordering \cite{608,I17,Sachdev}. In this case, the mixing between the local moments and conduction electrons can be represented via an auxiliary boson $b_i \sim \sum_{\sigma}c^{\dagger}_{i\sigma} f_{i\sigma}$. Its condensation means the formation of a heavy Fermi liquid: in the region, where the electron states have predominantly an $f$ nature, the band is almost flat.

\begin{figure}
\centering
\includegraphics[width=2.7in]{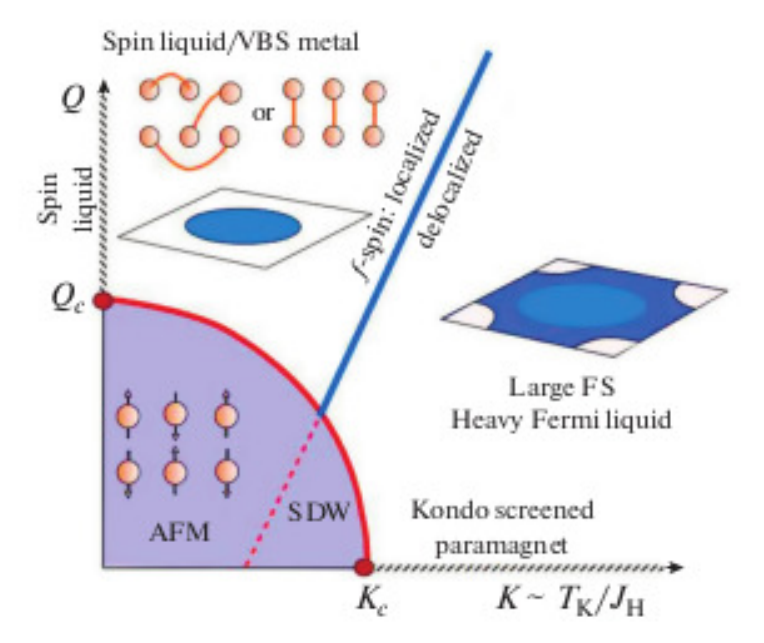}
\caption{
Schematic diagram of a Kondo lattice in the space of parameters of quantum frustration $Q = 1/S$ and
$K=T_K/J^H$ ($J^H$ is the Heisenberg interaction) \cite{7111}. The transition between the small and large Fermi surfaces can also occur through the intermediate phase of the ``strange metal.''
}
\label{coleman}
\end{figure}

The role of frustrations in Kondo lattices was investigated in detail in \cite{7111}. As can be seen from Fig. \ref{coleman}, the Kondo screening decreases the magnitude of the local moment, thus decreasing the critical value $Q_c$ necessary for the formation of the spin liquid. The boundary of the antiferromagnetic phase extends from $K = K_{c}$ on the Kondo axis to $Q = Q_{c}$ on the axis of frustrations. At the large $Q$ and small $K$, a spin-liquid metal is realized with the localized $f$ electrons and small Fermi surface, while at large $K$ a heavy Fermi liquid arises with a large Fermi surface and delocalized $f$ electrons. Since the volume of the Fermi surface is retained, it cannot smoothly evolve from small to large state; therefore, the spin-liquid metal and the heavy Fermi liquid must be separated by a quantum phase transition at the zero temperature.

In \cite{7111}, there was also examined a simple example of a frustrated spin model---a Shastry--Sutherland--Kondo lattice model, whose Hamiltonian takes on the form $\mathcal{H}_{SSK} =\mathcal{H}_{K}+ \mathcal{H}_{SS},$ where the terms
\begin{eqnarray}\label{SSK2}
\mathcal{H}_K &=& \sum_{{\bf k} \sigma} t_{\bf k} c^\dagger_{{\bf k} \sigma} c_{{\bf k} \sigma}
 + 2J_K \sum_{i}{\bf s}_{c} (i) \mathbf{S}_i , \cr
\mathcal{H}_{SS}&=&
J_{1}\sum_{\boxslash,\boxbslash}\mathbf{S}_k \mathbf{S}_l+
J_{2}\sum_{\langle i,j \rangle}\mathbf{S}_i\mathbf{S}_j .
\end{eqnarray}
contain the Kondo and Heisenberg interactions. Here, $\boxslash, \boxbslash$ means the sum taken over plaquettes with the alternating diagonal bonds. The ratio of the diagonal interaction to the interaction of the nearest neighbors $Q = J_{1}/J_{2}$ plays the role of the parameter of frustration.

The Shastry--Sutherland lattice has the advantage the ground state at the large frustration is a state of the valence bond with spin singlets located along the diagonal bonds and with a well defined wave function. In contrast to other types of spin liquid, it can be shown that this dimeric state is an \textit{exact} ground state of the model under the condition $J_1 \gtrsim 2J_2$.

\subsection{Fractionalized Fermi Liquid}

One more scenario of the formation of a non-Fermi liquid was called the fractionalized Fermi liquid FL*. In this phase, the charged excitations have the usual quantum numbers (charge $\pm e$ and spin 1/2), but they coexist with the additional fractional degrees of freedom. It is important to note that this phase has a Fermi surface (FS) with a volume that violates the Luttinger theorem.

In the works of Sachdev et al. \cite{Sachdev,Sachdev1} the concept of the fractionalized Fermi liquid was introduced initially for the $s$---$d(f)$ exchange model (Kondo lattice). In the FL* phase, the Kondo pairing, i.e., the condensation of Higgs boson $b$ ($\langle b_i \rangle=\langle f^\dagger_i c_i\rangle=0$); however, there is an anomalous average of the RVB type $\chi_{ij}=\langle f^\dagger_i f_j\rangle$. Thus, the localized moments do not take part in the formation of the FS (they form a separate spinon Fermi surface), but are adiabatically connected with the spin liquid described by the gauge theory and possessing corresponding exotic excitations in the phase of deconfinement. In the spatial dimensions $d \geq 2$, a spin liquid of the type Z$_2$ is stable, and at $d \geq 3$ there exists a spin liquid U(1). In this phase, the coefficient of electronic specific heat $C/T$ diverges logarithmically. Against the background of this state, there can appear a magnetic instability for the spinon FS---a metallic magnetic state of the spin-density wave SDW*, which can be characterized by a small moment.

Let us mention the difficulties of the theory: in \cite{Sachdev}, it was necessary to artificially introduce a field that stabilizes the unsaturated state with a small moment. In \cite{Isaev}, the introduction of such a field was avoided, and in the numerical calculations there was obtained a state, where the Kondo effect coexists with antiferromagnetism, but the spin-liquid order parameter disappears upon the transition to ``small'' FS. Note that in the simple renormalization-group consideration of quantum phase transitions in the magnetic Kondo lattices an intermediate region can appear with a non-Fermi-liquid behavior and partially suppressed magnetic moments \cite{I17}.

In subsequent works \cite{Sachdev1}, the FL* phase as the ground state has already been examined only at the quantum critical point, and beyond these limits it was considered as a phase unstable relative to phase with the local moments (see Section 4.1).

Subsequently, following the idea of Ribeiro and Wen \cite{Ribeiro} about the appearance of two hybridized subsystems (spinons and dopons) in the $t-J$ model (Section 2.5), the concept of the fractionalized Fermi liquid was generalized onto the one-band Hubbard model \cite{Punk,Punk1}. In this case, in the antiferromagnetic and spin-liquid (disordered) phases the spinons are described as Schwinger bosons, and in the Fermi-liquid phase, as fermions (see (\ref{eq:O.1})).

A change in the statistics of spinons occurs at the point of the quantum phase transition between two confinement phases---magnetic phase and a phase with large FS. The description of this transition requires further studies. Here, supersymmetrical concepts can prove to be useful (see, e.g., \cite{lavagna}). Let us mention also a completely fractionalized representation of the electron, where it is divided into quasiparticles that carry its spin and charge, and also the Majorana fermions, which carry Fermi statistics \cite{Majorana}.

\begin{figure}
\centering
\includegraphics[keepaspectratio,width=3.0 in]{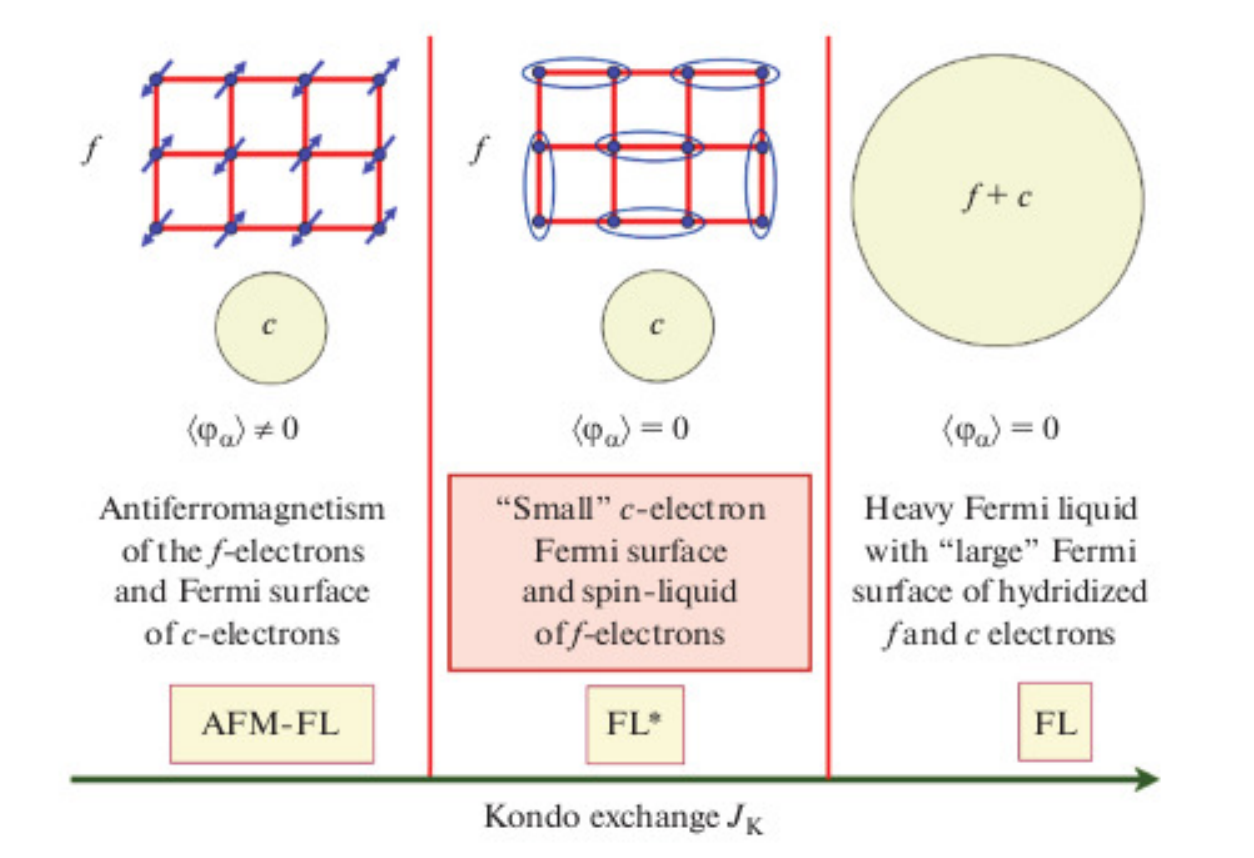}
\caption{
Fermi-liquid phases of the Kondo-lattice model \cite{Punk1}. In the FL* state, the spin liquid of local $f$ moments is separated from the small FS of $c$ conduction electrons, which does not obey the usual Luttinger theorem.
}
\label{fig:fig6}
\end{figure}

\begin{figure}[!hbp]
\centering
\includegraphics[keepaspectratio,width=3.0in]{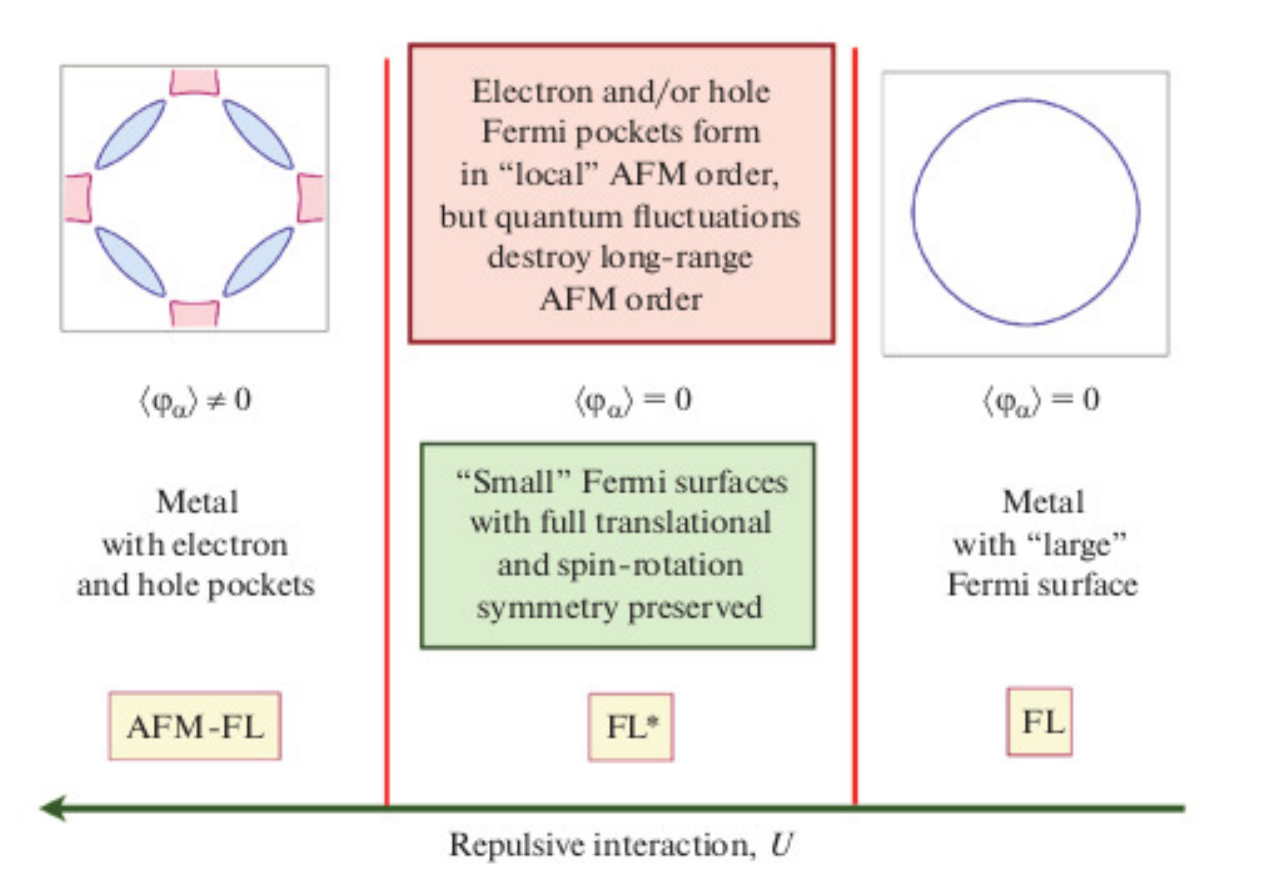}
\caption{
Phase diagram of the ground state of the Hubbard model according to \cite{Punk1}; $\varphi_\alpha $ is the parameter of the antiferromagnetic order. In the AFM-FL phase, the doubling of the unit cell due to the AFM order guarantees the fulfillment of the Luttinger theorem. The FL* phase does not have an antiferromagnetic order ($ \langle \varphi_\alpha \rangle = 0$), however, it inherits the small pockets of the FS from the phase with $ \langle \varphi_\alpha \rangle \neq 0$, and the Luttinger theorem it not fulfilled.
}
\label{fig:sdw2}
\end{figure}

Upon the comparison of Figs. \ref{fig:fig6} and \ref{fig:sdw2}, it is interesting to note that in contrast to Hubbard model, where Coulomb interaction leads to the destruction of the Fermi-liquid state (to the formation of Hubbard subbands with small FS), the effect of the $s-d$ exchange is opposite: with an increase in the $s-d$ exchange parameter, a heavy Fermi liquid and a large FS appear. The reason lies in the fact that these interactions lead to the different types of pairing of electrons and spinons: to diagonal and to nondiagonal (hybridization).

The concept of Schwinger bosons can be introduced not only in the $t-J$ model \cite{Punk}, but also in the antiferromagnetic phase of the spin-fermionic model---the modification of the one-band Hubbard model with finite interaction \cite{Punk1}, which makes it possible to speak about the realization in the latter of the state of unconventional Fermi liquid FL*.

The spin-fermionic model \cite{Abanov} used upon the interpolation description of the itinerant magnetism (its analogies with the $s-d$ exchange model are discussed in \cite{2008}), also makes it possible to separate spin and electronic degrees of freedom.

\begin{figure}
\centering
\includegraphics[width=3.1in]{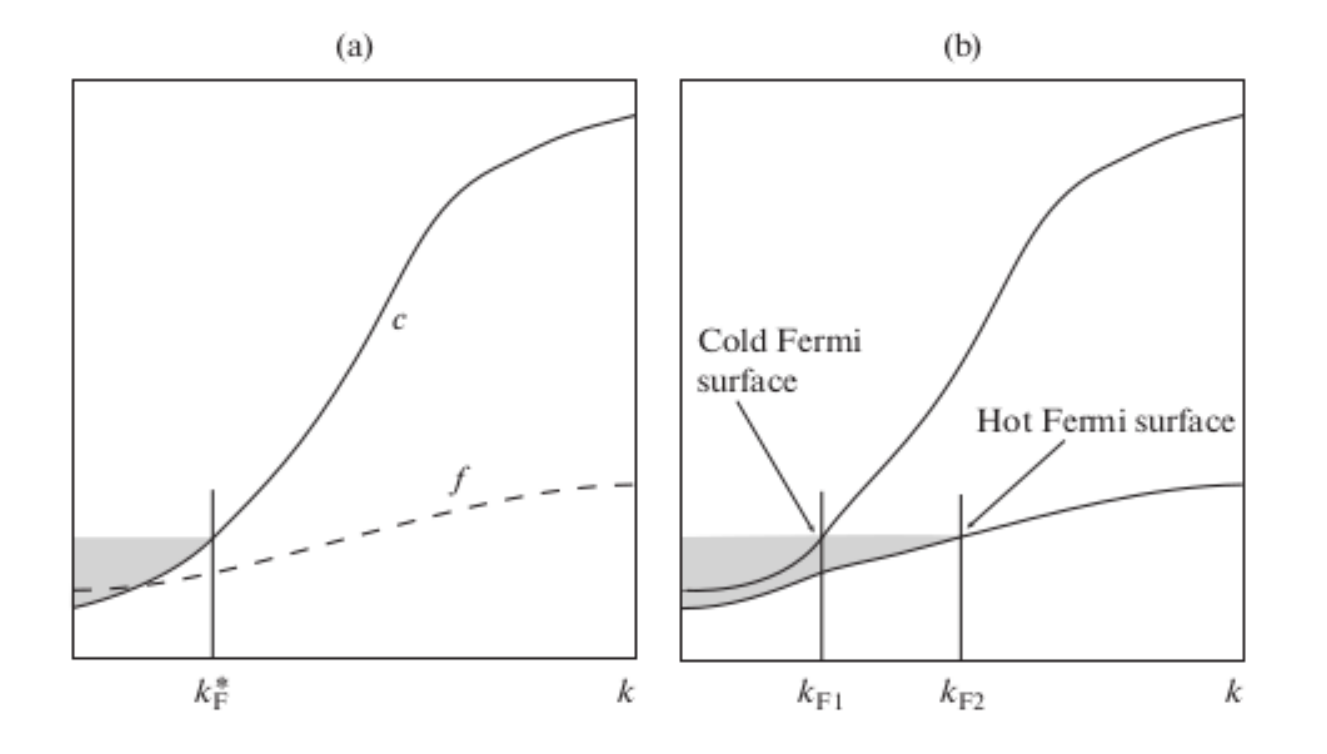}
\caption{\label{fig8}
(a) The FL* state with a cold FS and a Fermi momentum $k_\text{F}^{\ast}$. The fermions $f$, which form the spin liquid, are schematically represented by the dashed line. (b) The FL state, which obeys the usual Luttinger theorem, but has two FSs.
\label{fig:kondo3}
}
\end{figure}

The mechanism of transition and the description of the quantum critical point between the FL and FL* phases is discussed in \cite{Sachdev1} (Fig. \ref{fig:fig6}). When we move to the FL* phase in the limits of the FL phase, the two Fermi surfaces with $k_\text{F1}$ and $k_\text{F2}$ behave differently. The FS with $k_\text{F1}$ evolves into the FS with $k_\text{F*}$---into the FL* state, and remains ``cold'' near the quantum critical point: there is no strong scattering of these quasiparticles, and the quasiparticle residue of electrons upon the transition remains finite. On the contrary, the FS with $k_\text{F2}$ becomes ``hot'': the lifetime of quasiparticles is shortened; the residue decreases and finally disappears at the quantum critical point, so that the dynamics of the quasiparticles manifests a non-Fermi-liquid behavior. This anomalous behavior extends onto the entire hot FS, in contrast to the isolated ``hot points'' that appear in the theory of spin-density waves.

In the FL* state, there are additional low-energy excitations of local moments, which give an additional topological contribution to a change in the momentum of the crystal. Indeed, the action of a vortex flux is analogous to the Lieb--Schultz--Mattis transformation \cite{Lieb}, so that the spin-liquid state in the dimensionality $d = 2$ acquires this change of the momentum. Usually, the volume of the Fermi surface is determined by the total number of electrons in the system. However, there is a common topological analysis \cite{Oshikawa}, based on the threading the quantum of the flux (due to the cyclic boundary conditions, the system is considered as a torus, within which a crystal momentum appears; this is analogous to the appearance of a Faraday force with a change in the flux) and of the global gauge symmetry U(1) (charge conservation). Hence, it follows that the existence of a nonmagnetic FL* state with another volume of the FS is permitted if we permit the existence of global topological excitations. Such excitations naturally appear in the gauge theories for the FL* state. Thus, the violation of the Luttinger theorem must be accompanied by the appearance of a topological order \cite{Sachdev}. Consequently, the formation of a small FS, of the Hubbard splitting (note that the Oshikawa theorem \cite{Oshikawa} is valid for the insulator as well), and of a state with disordered local moments can be connected with the topological order in the state of spin liquid.

In a multiband system, there is an opportunity for a partial Mott transition between two metallic phases, where the Fermi surface undergoes a quantized change. In the simplest case, one subband (or orbital) changes its nature from the metallic to the Mott character, while the other bands remain metallic (orbital-selective Mott transition, see \cite{Vojta}).

Being stable in the low-temperature limit, the partial Mott phase violates the Luttinger theorem and, therefore, is a non-Fermi-liquid metal. According to the Sachdev concept, this is the fractionalized Fermi liquid (FL*), and the transition between FL and FL* is the orbital-selective Mott transition.

The partial Mott transition was obtained from the numerical results for the doped one-band Hubbard model of cuprate superconductors. This behavior is, apparently, momentum-selective, i.e., some regions in the momentum space behave similar to a Fermi liquid, and others, as Mott insulators.

\section{Theory of Quantum Phase Transitions. Spin Liquid and the Topological Order}

From the viewpoint of quantum phase transitions, of special interest are frustrated magnetic systems, in which the long-range order is suppressed (the simplest example here is a triangular lattice, where, according to numerical calculations, there is a finite region of deconfinement caused by frustration).
In these systems, with the allowance for quantum fluctuations, there can be realized states with local moments without breaking of the symmetry---up to the lowest temperatures, there exists only short-range order. In the case of half-filling of the band, these systems can demonstrate metal--insulator transition in the paramagnetic phase.

Such quantum spin liquids (QSLs) noticeably differ from their classical analogs in a number of positions \cite{Vojta}.
The quantum fluctuations usually remove the strong degeneracy of the ground state of the frustrated systems due to the quantum tunneling, which leads to the existence of a unique ground state (with an accuracy to the global transformations of symmetry or topological degeneracy).
The QSLs are the thermodynamically stable phases of matter, which are characterized by dynamic gauge fields and by topological order.
This implies the existence of fractional excitations coupled to the gauge field.
Despite of this coupling, the fractional excitations are asymptotically free (deconfined).
The wave functions in a QSL can be characterized by a long-range entanglement.
In this case, the QSLs should be differed from the ``trivial'' quantum paramagnets without a topological order and fractionalization (for example, singlet magnets materials of the TlCuCl$_3$ type, where spins 1/2 are connected into pairs).

\subsection{Deconfined Quantum Criticality}

\begin{figure}
\centering
\includegraphics[width=2in]{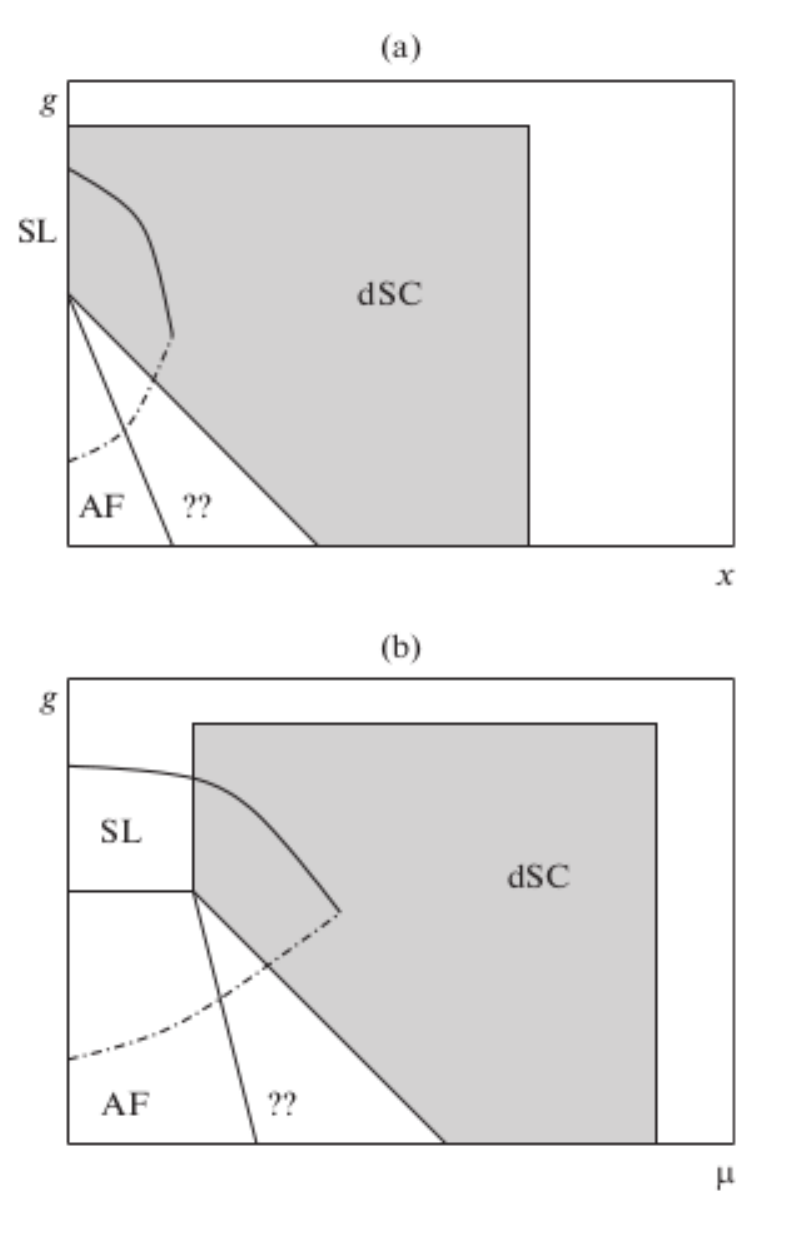}
\caption{
(a) Schematic zero-temperature phase diagram, which shows the path between the antiferromagnetic Mott insulator and $d$-wave superconductor as a function of doping $x$. The vertical axis is marked by the parameter $g$, which indicates the measure of the frustration of the spin system in the insulator phase. AF is the antiferromagnetic ordered state; SL, insulator spin liquid, which is obtained by an increase of the frustration. (b) Same as in (a) but as a function of the chemical potential.
}
\label{sldsc1}
\end{figure}

Let us turn now to a question of how the gauge-theory approach can be applied to the problem of high-temperature superconductivity.
In the ground state, the latter appears upon doping of the Mott insulator, when the antiferromagnetic order disappears, and at temperatures higher than the point of the superconducting transition the normal state is described by a pseudo-gap picture.
In order to connect the simultaneous disappearance of magnetic order with the appearance
of pseudo-gap behavior and of superconductivity, an approach was proposed which makes it possible to describe the appearance of, first, a nonmagnetic Mott insulator at a zero doping (spin liquid), which upon the doping is converted into a superconductor (see Fig. \ref{sldsc1}).

\begin{figure}
\centering
\includegraphics[width=2.4in]{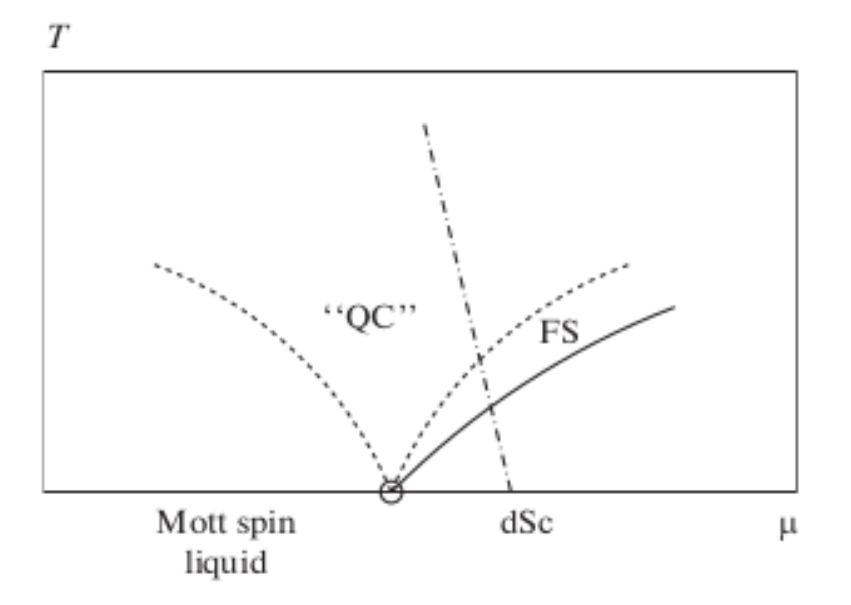}
\caption{
Schematic phase diagram for the doping-induced Mott transition between the spin-liquid insulator and the $d$-wave superconductor. The semi-solid line is the path of the system with the hole density $x$, which has the superconducting ground state. The region designated by FS corresponds to the fluctuation regime of the superconducting transition. QC is the quantum critical region connected with the Mott critical point. This region can be identified with the high-temperature pseudo-gap phase.
}
\label{fig10}
\end{figure}

Let us consider a certain spin-liquid Mott state. Upon a change of the chemical potential in it a zero-temperature Mott phase transition can occur, which is caused by doping rather than by a change in the interaction.
The quantum critical point will determine the behavior of the system in a finite interval of parameters. The lines of the crossovers, which appear near this transition, are shown in Fig. \ref{fig10}.

It is interesting to note that the rotor representation also makes it possible to describe an intermediate pseudo-gap phase, giving a transition to the metallic state only at the finite concentration of holes \cite{rotor2}.
A similar picture is given by the model of phase strings (Section 3.4), and also by calculations in the representation of ME $X$ operators (incoherent states upon small doping) \cite{Irkhin:1990}.

Among the possible candidates for the spin-liquid state in the literature there are mentioned a dimeric phase (valence-bound state, VBS), and Z$_2$ and U(1) spin liquids.
Since it is quite probable that the translational-invariant spin-liquid state upon doping can bring to $d$-wave superconductivity with nodal quasiparticles, the most suitable phase appears to be the U(1) phase \cite{Wen1}.
Moreover, even if a physical system does not possess a stable U(1) spin-liquid state, the latter can exist as a critical state, which separates the antiferromagnetic phase from the Z$_2$ spin liquid, which favors the appearance of pseudo-gap and superconducting phases \cite{0406066}.

The simplest situation that is beyond the framework of the Landau theory is the QPT at the zero temperature, when between different quantum phases with broken symmetry at the quantum critical point a transition occurs forbidden according to Landau.
One example is a frustrated antiferromagnet \cite{0311326,0312617}, where the deconfined quantum critical point separates the ordered phases with the violated rotational and translational symmetry.
For such points, the best starting point upon the description of the critical behavior is not the usual parameter of order, but rather a set of fractional excitations, which are specific for the critical point but are not present in any of the phases near it.
Moreover, there is an additional topological structure, which is present at the critical point and connected with the topological law of conservation.
It is convenient to interpret this additional conservable quantity as the total flux of the gauge field that appears at the critical point.
It is retained only asymptotically---at low energies at the quantum critical point.
This law of conservation ensures a sharp difference between the deconfinement and usual critical points.

In the continual limit, the topological flux is defined as
$$ Q = \frac{1}{4\pi} \int dxdy \epsilon_{\alpha \beta \gamma} \phi_{\alpha}\partial_x \phi_{\beta} \partial_y \phi_{\gamma} $$
and is connected with the number of skyrmions in the spin configuration.
The conservation of $Q$ means the appearance of an additional global U(1) symmetry at the critical point and the presence of an additional set of gapless gauge excitations.
It naturally leads to large values of the critical exponent of the correlation function $\eta$ for the deconfinement critical point.

\begin{figure}
\centering
\includegraphics[keepaspectratio,width=3.5in]{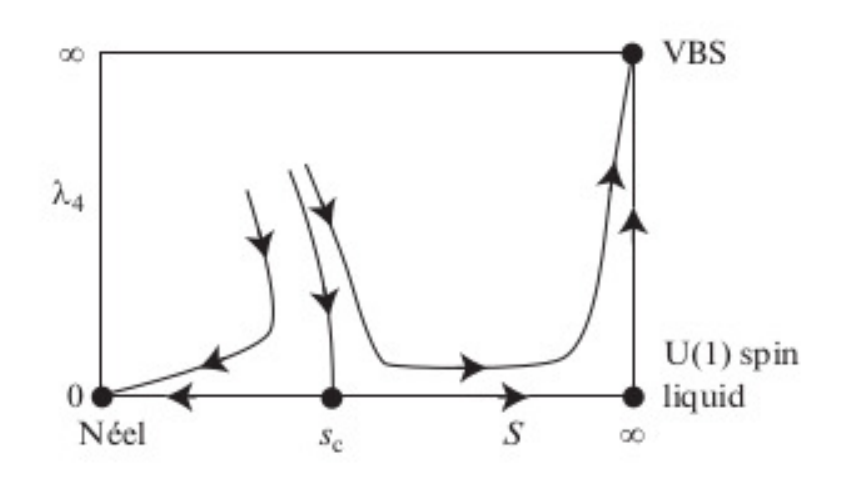}
\caption{
\label{fig:fig11} Trajectories of the renormalization group for the antiferromagnet with a spin S = 1/2 in square lattice.
}
\end{figure}

It is known that the compact U(1) gauge theory in 2 + 1-dimensional space-time allows the presence of point defects (monopoles), which change the topological charge (number of skyrmions).
Following \cite{0404718}, let us examine the trajectories of the two-dimensional renormalization group with the driving parameter $s$ and the parameter $\lambda_4$ describing the efficiency of monopoles (monopole fugacity); they are shown in Fig. \ref{fig:fig11}.

There is an unstable fixed point $s = s_c$ located on the line $\lambda_4 = 0$, separating the lines of renormalization-group trajectories in the direction of the magnetically ordered N\'eel state from the lines in the direction of paramagnetic spin liquid with a gapless U(1) photon and gapped spinon excitations.

Let us now turn to the case of $\lambda_4 \neq 0$.
At small $s$, the magnitude of $\lambda_4$ proves to be insignificant (irrelevant) for the case of the fixed point $\lambda_4 = 0, s_c=0$.
However, for $s > s_c$, the trajectories deviate and are attracted to the side of greater values of $\lambda_4$.
This is connected with the fact that the parameter $\lambda_4$ becomes a dangerously relevant disturbance for the fixed point at large $s$, which describes the spin-liquid phase.
Near $s_c$, the value of $\lambda_4$ is very small on the scales of the order of the correlation length $\xi_{\text{spin}}$, but becomes on the order of unity and even greater on the second length scale, which is designated as $\xi_{\text{RVB}}$.
Then, we have a relation $\xi_{\text{RVB}}\sim \xi_{\text{spin}}^\lambda$ where $\lambda >1$ therefore, $\xi_{\text{RVB}}\gg \xi_{\text{spin}}$.

\begin{figure}
\centering
\includegraphics[keepaspectratio,width=3.in]{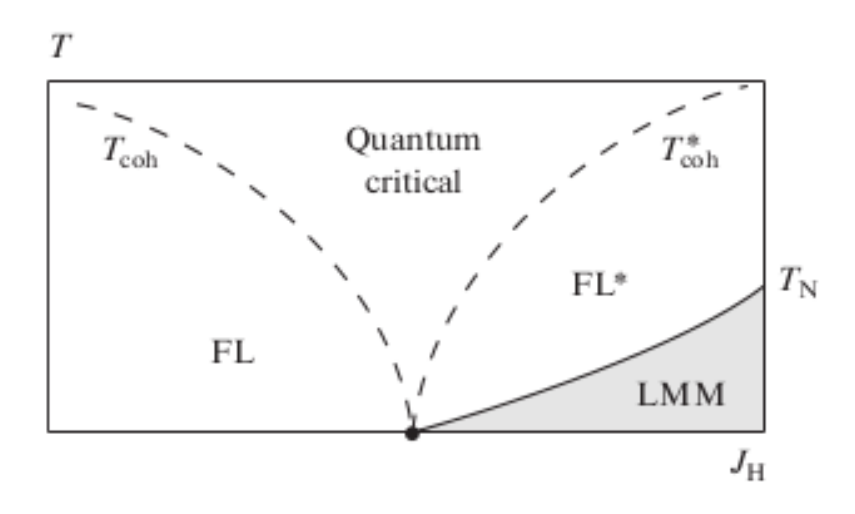}
\caption{
Phase diagram near a quantum transition between the Fermi liquid FL and the state with local moments LMM \cite{Sachdev1}. Two different energy scales lead to two different critical exponents, according to which the N\'eel point $T_N$ and coherence temperature $T_\text{coh}$ (or $T^*_\text{coh}$) approach to a quantum critical point.
}
\label{fig:fig12}
\end{figure}

A second example of a deconfined quantum critical point is the QPT in a heavy Fermi liquid; let us examine its instabilities, following \cite{Sachdev1} (Fig. \ref{fig:fig12}).
In the insulator magnets, there is a direct second-order transition between the N\'eel state and state with another order---the valence-bound solid state (VBS).
It can be considered as an analog of the direct transition between the heavy Fermi liquid and metal with local magnetic moments (LMM), where a loss occurs of the “Kondo” order simultaneously with the appearance of a magnetic order.
Indeed, the N\'eel state and the FL state (Kondo lattice) are stable Higgs phases of the compact U(1) gauge theory.
They are unstable relative to the ``deconfinement'' transition into the U(1) spin liquid and into the FL* phase, respectively.
However, such deconfinement phases are rather labile states of matter and can be unstable with respect to confinement phases with the usual order; for example, the instability of the U(1) spin liquid to the transition into the VBS.
It is natural to examine the instability of the FL* state with respect to the state with LMMs, but in such a way that the quantum criticality would remain deconfined.
In the insulator magnets, the transition between two competing orders (N\'eel and VBS) occurs not as a function of a certain parameter of tuning, but dynamically, as a function of the scale of the length--time.
Indeed, in the case of a paramagnet the loss of N\'eel correlations occurs on the scale of $\xi$, whereas the pinning of the VBS order appears on a much larger scale $\xi_{\rm VBS}$, which diverges as some degree of $\xi$.
Similarly, the direct second-order transition between the competing Kondo order in the FL and magnetic order in the metal with local moments is possible, but it requires at least two divergent scales of length--time at the quantum critical point, and the deconfinement is observed only on the shorter scales.

\subsection{Higgs Criticality}

As it was mentioned above, beginning from the work \cite{Sachdev}, the transitions between the Fermi liquid and non-Fermi-liquid states are discussed in terms of the concept of the Higgs boson $b$.
As in the theory of elementary particles, this formalism is convenient upon the description of broken symmetries.

In the Higgs phase, the gauge field in the Coulomb gas (Section 3.2) becomes gapless according to the Anderson--Higgs mechanism, so that in the case of the charge $q  = 1$ the Higgs phase and the confinement phases are smoothly connected with each other.
The situation differs sharply at $q = 2$, i.e., when the field of bosons corresponds to pairing, for example, in the superconductor.
In this case, there can exist a boundary between the confinement phase and Higgs phases, so that latter (phase of pairing) is deconfined and has a residual gauge symmetry Z$_2$, so that the parameter of order is invariant upon the change of the sign of charges $q = 1$ composing the pair \cite{Wen1}.

Following \cite{Chowdhury}, let us discuss the evolution of the phase U(1)--FL* with its small Fermi surface toward the usual ``large'' FS upon strong doping.
There is a usual way of transition, connected with the disappearance of the antiferromagnetic order.
This is the transition between two states of the Fermi liquid, which is described by the standard Hertz--Millis theory \cite{Hertz76, Millis93}.

\begin{figure}[ht!]
\centering
\includegraphics[keepaspectratio,width=3.5in]{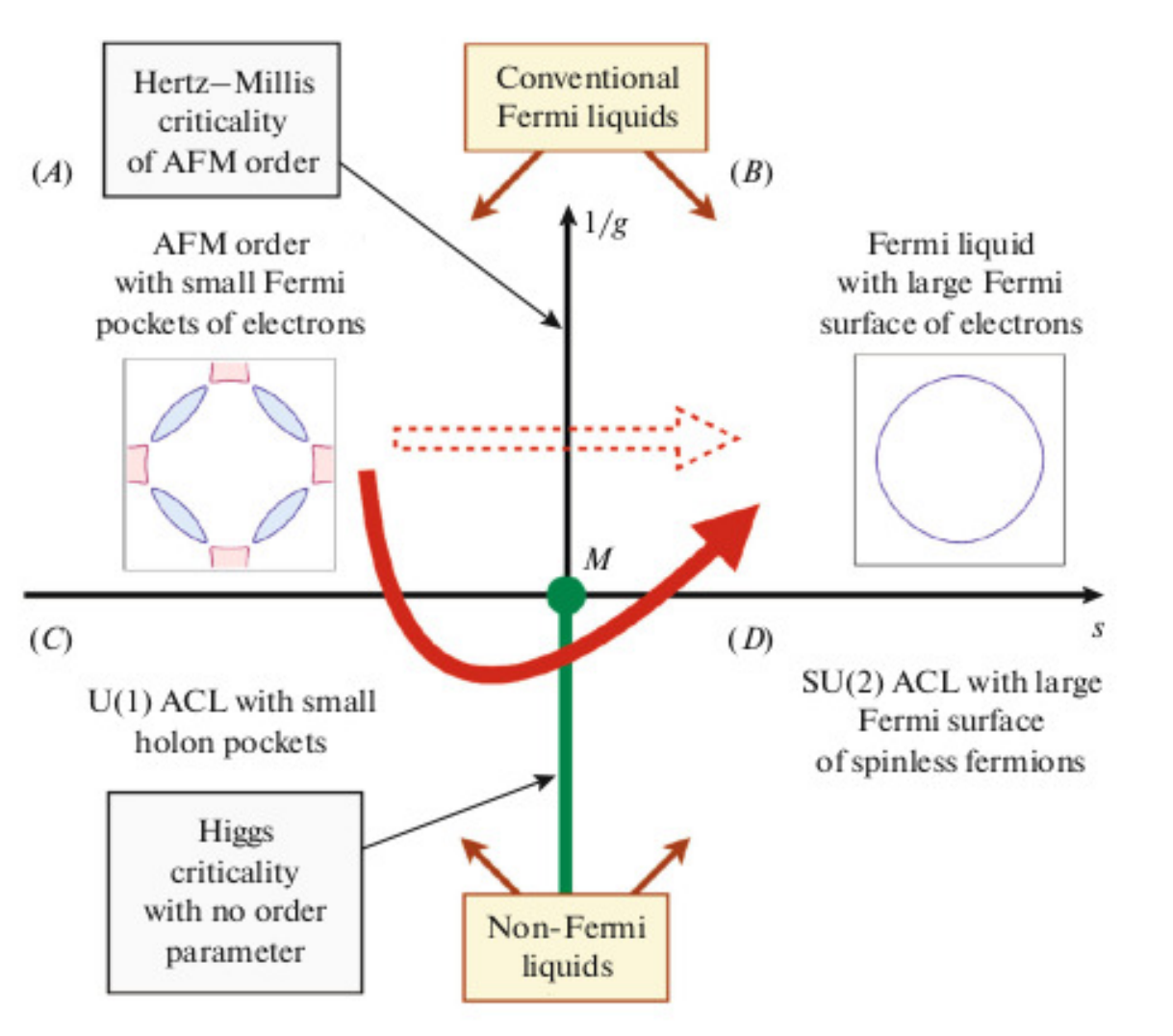}
\caption{
Diagram of metallic phases according to \cite{Chowdhury} in the reciprocal-coupling-constant--Higgs-criticality-parameter coordinates. Only the $A$ phase has the broken global symmetry connected with the presence of long-range antiferromagnetic (AFM) order. The usual Fermi-liquid phases on top undergo the transition from small to large Fermi surfaces accompanied by the loss of the AFM order. The dashed arrow shows the direct path between these phases, which can give the description of cuprates with electron doping. The arrow around the point $M$ is the path with an increase in the hole doping. The phase U(1)--FL* descends from the phase U(1) ACL and has a small Fermi surface of electrons because of the presence of a topological order, whereas phase $A$ above it has a ``small'' FS of electrons because of the breaking of translational symmetry.
}
\label{fig:phase2}
\end{figure}

The simple SU(2) field theory \cite{Chowdhury} allows the existence of condensates of two boson fields: of the gauge boson $R$ and of the Higgs boson $H^a$. The corresponding phases are obtained by changing the driving parameters $s$ and $g$ (Fig. \ref{fig:phase2}) and are designated according to their condensates.

(1) The Higgs phase designated as ($A$) in Fig. \ref{fig:phase2}, where both symmetries
(SU(2)$ _{\rm spin}$ and SU(2)$_{\rm gauge} $) are violated, which brings to
$\langle R \rangle \neq 0, \langle H^a \rangle \neq 0$.
Here, the gauge excitations--photons-- are gapped. This phase describes an AFM metal, where the large Fermi surface is reconstructed into hole and electron pockets because of the condensation of the boson $H^a $ along the vector of the N\'eel parameter of order $ {\bf{n}}$.

(2) The confinement phase SU(2) designated as ($B$) in Fig. \ref{fig:phase2}, where the symmetry SU(2)$_{\rm spin}$ is not violated.
Here, we have $\langle R \rangle \neq 0, ~ \langle H^a \rangle = 0$  which is necessary for retaining the spin rotational invariance, since ${\bf{n}} = 0$.
This is a usual Fermi liquid with a large FS.

(3) In the Higgs phase ($C$), the symmetry SU(2)$_{\rm gauge} $ is violated, but
SU(2)$_{\rm spin}$ is not broken, which leads to
$\langle R \rangle = 0, \langle H^a \rangle \neq 0$.
This means the existence of developed local AFM fluctuations without any long-range order.
If we select $H^a$ along the direction (0,0,1) by conducting a gauge transformation, the symmetry of the U(1) subgroup SU(2)$_{\rm gauge}$ is not violated, so that the gauge  photon remains gapless.
Thus, this phase describes a U(1) algebraic charge liquid (ACL), or a holon metal \cite{acl}. However, because of the existence of a local AFM order, the FS is reconstructed into pockets of holons $\psi_p$ which are minimally coupled to with the gauge field U(1).
Depending on the temperature, there can occur a continuous crossover from the U(1) ACL into the U(1) FL* (or ``holon--hole'' metal), where some of holons ($\psi_{\pm}$) begin forming bound states with gapped spinons.

(4) The last phase ($D$) has complete symmetry, so that no one of the fields is condensed:
$\langle R \rangle = \langle H ^ a \rangle = 0$.
In contrast to U(1) ACL, where only the $A^z$ photon was gapless, in this phase there is a triplet of gapless SU(2) photons, connected with a large FS.
This phase can be described as an SU(2) algebraic charge liquid.
At sufficiently low temperatures, it is unstable with respect to the superconductivity.

\begin{figure}[ht]
\centering
\includegraphics[keepaspectratio,width=3.5in]{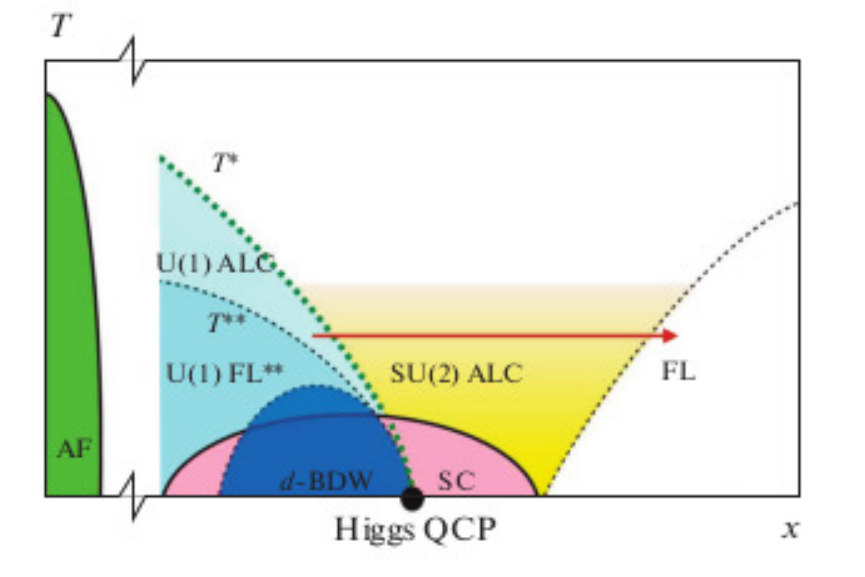}
\caption{
Phase diagram for cuprates, based on the Higgs theory of criticality. The algebraic charge liquids (ACL) possess Fermi surfaces of spinless $\psi$ fermions, which carry the electromagnetic charge: in the SU(2), the ACL Fermi surface is large and is connected with SU(2) gauge field, whereas in the U(1) the ACL Fermi surface is small and is connected with the U(1) gauge field. The Fermi liquid FL* descends from U(1) ACL upon the binding of $\psi$ fermions with neutral spinons. The $d$-BDW is the wave of the density of bonds; SC is the $d$-wave superconductor,
}
\label{fig:fig14}
\end{figure}

The comparison of the calculations of the electron Green's function in the topologically ordered Higgs phase SU(2) of the gauge theory of fluctuating antiferromagnetism on a square lattice with the results of first-principles calculations in the dynamic mean-field theory (DMFT) and quantum Monte Carlo method is carried out in \cite{Scheurer}.

\subsection{Phase Transitions and Incoherent States in Conducting Magnets}

In the conducting magnets, upon a change in the parameters of interaction there appear additional quantum phase transitions in comparison with the Heisenberg model---the appearance of magnetic ordering in the ground state, a qualitative change in the behavior of the magnetic moment, etc.
In the limit of strong correlations, such transitions are not described by simple approaches of the Stoner or Hertz--Millis type \cite{Hertz76,Millis93}.
In particular, in a ferromagnet there can occur a transition from a saturated into an unsaturated phase. In the saturated (half-metallic) state, the Fermi surface disappears for one of the projections of spin, so that this transition can be considered topological.
Note that a strict proof of the Nagaoka theorem about the saturated ferromagnetic ground state of the Hubbard model with an infinite $U$ has a topological nature; the Weng theory of phase strings in the $t-J$ model can be considered as a generalization of this approach to the case of finite $J$.

States with the spins down in the Hubbard model in a saturated ferromagnet give the simplest example of non-quasiparticle states. Formally, in the limit of $U=\infty$ it is convenient to factorize the Hubbard operator $X_i(-,0)=X_i(-,+)X_i(+,0)$ which is analogous to the separation of holon and spinon. Then, the calculation gives \cite{Irkhin:1990}
\begin{equation}
\langle \!\langle X_{\mathbf{k}}(-,
0)|X_{-\mathbf{k}}(0,- )\rangle \!\rangle _E=
\left\{ E-t
_{\mathbf{k}}+\left[ G_{\mathbf{k}\downarrow }^0(E)\right]
^{-1}\right\} ^{-1}, \,
G_{\mathbf{k}\downarrow
}^0(E)=\sum_{\mathbf{q}}\frac{n_{\mathbf{k}+\mathbf{q}}}{E-t
_{\mathbf{k}+\mathbf{q}}+\omega _{\mathbf{q}}},
 \label{eq:J.23}
\end{equation}
where $t_{\mathbf{k}}$ and $\omega _{\mathbf{q}}$ are the bare spectra of holes and magnons, respectively.
The Green's function in the zero approximation $G_{\mathbf{k}\downarrow}^0(E)$ is the convolution of the Green's function of the free carrier with the spin up and of a magnon, and describes the completely non-quasiparticle states.
With an increase in the concentration of holes, the expression (\ref{eq:J.23}) acquires a pole at the Fermi level, which means the destruction of the saturated ferromagnetism.

It is interesting that historically the concept of spinons was introduced by Anderson in connection with the problem of spin polarization in ferromagnetic metals \cite{Polarization} (see also \cite{Corrias}).
By spinons, there were called incoherent contributions to the electronic spectral density caused by the interaction with the magnons. The corresponding formalism of the description of strong band ferromagnetism based on the Ward identity was developed in the work of Edwards and Hertz \cite{Hertz1}.
It was also used for describing the metal--insulator transition in the paramagnetic phase and for the treatment of the non-Fermi-liquid state \cite{Hertz-para}.
The paramagnetic state with disordered magnetic moments obtained in \cite{Hertz-para} in a similar approximation can be identified with the Anderson state of the RVB.

A further development of the theory of the instability of ferromagnetism is connected with the analysis of three-particle contributions \cite{Igarashi:1985}.

In the $t-J$ model of antiferromagnets, a large part of states is incoherent, although at finite $J$ the states near the bottom of the band form a narrow coherent band with a small residue of order $|J/t| \ll 1$ and with a heavy mass $\sim |t/J|$ \cite{Kane}.

\begin{figure}[!ht]
\centering
\includegraphics[width=0.5\textwidth]{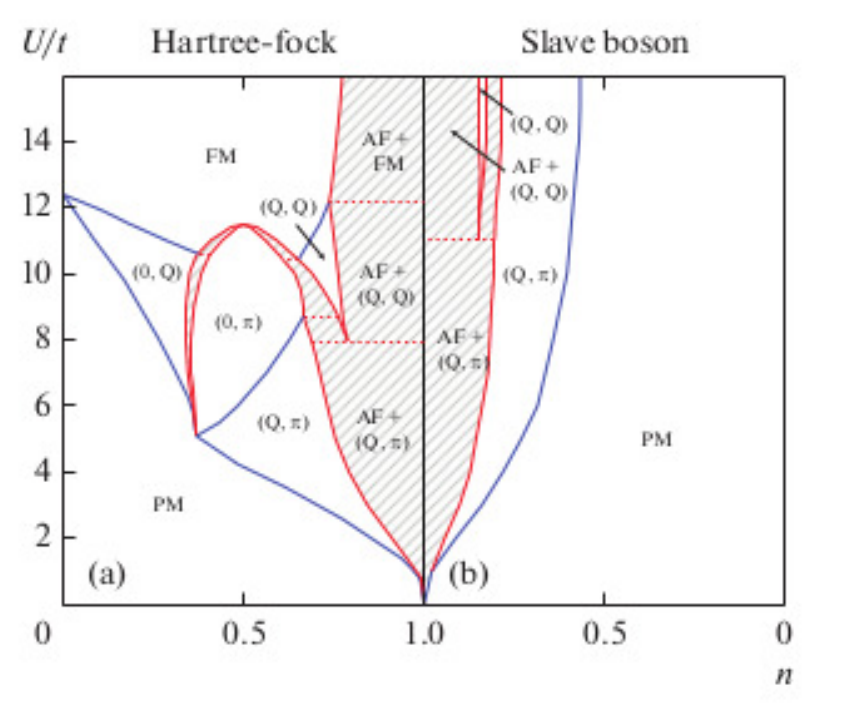}
\caption{
Magnetic phase diagram of the ground state of the two-dimensional Hubbard model within the framework of (a) the Hartree--Fock approximation and (b) with taking into account correlations for the square lattice with the concentration of electrons $n < 1$ \cite{Igoshev:2015}. The spiral phases are designated in accordance with the form of their wave vector. Heavy lines designate second-order phase transitions. The shading shows the region of phase separation; the dotted lines correspond to the boundaries between different pairs of phases. The solid lines correspond to the boundaries between the uniform phase and regions of phase separation; $\mathbf{Q}_{\rm AFM} = (\pi,\pi)$, PM and FM are the paramagnetic and ferromagnetic phases, respectively.
}
\label{fig:sq}
\end{figure}

The competition of ferro- and antiferromagnetism upon doping of a Mott insulator leads to first-order transitions, to the formation of inhomogeneous states and of spiral magnetic structures (see Fig. \ref{fig:sq}).
The corresponding calculations in the generalized Hartree--Fock approximation and with allowance for correlations in the Kotliar--Ruckenstein method of auxiliary bosons were carried out in \cite{Igoshev:2010,Igoshev:2015,Igoshev}.
Different antiferromagnetic phases appear in this case, including some phases that are characterized by a small residue \cite{Igoshev}.
The methods of the Hartree--Fock type overestimate the stability of magnetic phases, whereas the allowance for correlations makes it possible to successfully describe the paramagnetic phase at large $U$, i.e., a phase of the spin-liquid type \cite{Igoshev:2015}.

\subsection{Superconductivity and Topological Order}
\label{sect:supercon}

The experimental discovery of the superconductive order by Kamerlingh Onnes historically led to the theory of broken symmetry; however, a sequential quantum topological examination complicates this picture.
In fact, the superconductivity can be described by the Ginzburg--Landau theory with a dynamic U(1) gauge field \cite{wen11}.
The condensation of an electron pair of charges $2e$ violates the U(1) gauge theory, reduces it to the Z$_2$ gauge theory at low energies.
The latter is the effective theory of topological order Z$_2$, so that a real superconductor has precisely such topological order.
Frequently, the superconductivity is described by the Ginzburg--Landau theory without the dynamic gauge field U(1); instead, a violated symmetry U(1) is examined \cite{Wen3}.
However, the real superconductors in an electromagnetic gauge field are not states with violated symmetry, but are topologically ordered states \cite{Wen:1991,0404327}.
The decisive importance here belongs to the Elitzur theorem \cite{Elitzur} about the impossibility of the existence of the gauge-invariant local order parameter for the state in the electromagnetic gauge field.

By itself, a superconductive order \textit{in a field} is not an order that breaks symmetry, but it is rather a topological order, which is beyond the scope of the Landau theory of symmetry (although, of course, when without a field, this theory gives a complete description); ironically, it is precisely this order was discovered in 1911.
If the gauge U(1) electromagnetic field is dynamic, the superconductor has string excitations, which characterize a topological order--loops of the flux $hc/2e$.

It turns out that the destruction of a superconductive order occurs due to the usual Berezinski--Kosterlitz--Thouless mechanism, which includes the proliferation of topological vortices.
In a spin liquid, the instantons, which destroy topological order, are irrelevant. Then, the U(1) gauge flux is a conserved quantity, and we have a deconfinement state.

The theory of phase fluctuations in two dimensions is described well by the Berezinski--Kosterlitz--Thouless approach (BKT); therefore, the destruction of the superconducting order occurs by the usual BKT mechanism, due to the propagation of vortices and thermal ``uncoupling'' of the vortex--antivortex pairs.

One of the consequences of the phenomenological Ginzburg--Landau approach is a competition between the BKT transition for the fermionic pairing and condensation of bosons. Therefore, the lines of phase transition for the fermionic pairing and boson condensation in the phase diagram become the lines of crossover, and only the superconductive transition remains a real BKT transition.

In an external electromagnetic field, the vortices are pushed out. This field upon the condensation acquires a mass (hence, there appears a Meissner effect); the U(1) symmetry is reduced to Z$_2$ (i.e., there appears a spin liquid). Thus, in a field the Landau mechanism of breaking symmetry, which takes place in the case of superfluidity, does not work---there appears a topological transition.

Let us now turn to the vortex structures in the superconducting state. The most intriguing problem here is the quantization of the magnetic flux.
Since the boson has a charge $e$, whereas a fermionic pair has a charge $2e$, the problem consists in whether a vortex $hc/e$ can be more stable than the usual vortex $hc/2e $.
As it turns out, a vortex of type B with the quantization of the flux $hc/e$ will have a lower energy than the vortices $hc/2e$ especially upon a low doping.
This conclusion is the common specific feature of the gauge theory U(1).
However, the $hc/2e$ vortices are naturally explained in the two-boson SU(2) theory \cite{Wen1}.

The Weng statistics leads to a change of the hidden topological structure in comparison with the Fermi-liquid BCS superconductor: a superconductor is a condensate of bosons with a charge $e$ and, at the first glance, it seems that the vortices bear flux $hc/e$.
However, if we examine the influence of the mutual Chern--Simons statistics on the topology of the superconductor, we will see that a holon with a charge $e$ acquires a phase jump $\pi$ upon the bypass of an isolated spinon.
Thus, the binding of a spinon with a vortex leads to a twofold decrease of the quantum of the flux to the value $hc/2e$ as in the case of the usual condensate of Cooper pairs \cite{Weng2007,Zaanen-Overbosch}.

\section{Lattice Gauge Theories and Strings}

In the field theory, the transition to lattice models, which include the discrezation of space and time, is frequently used.
This transition is even more natural in the quantum theory of solids with the use of imaginary time $t=i \tau$; therefore, a number of the difficulties of the gauge theory are solved more easily.
In the limit of zero temperature, the time interval becomes infinite, so that an additional space dimension appears.

The physical space of states of the Abel lattice gauge theory consists of closed loops of the electric flux.
Let us discuss their origin, following the survey \cite{Kogut}.
In this case, it is convenient to use a Hamilton approach in the approximation of continuous time. Let us consider an action with anisotropic bonds
\begin{equation}
S=\beta _{\tau }\sum_{nk}[1-\cos \theta _{0k}(n)]-\beta \sum_{n,ik}\cos \theta _{ik}(n),
\end{equation}
where $\theta(n)$ is the angular continuous variable, the spatial bonds are designated by Latin indices, and the time direction $\tau$, by index $0$ (this theory is built analogously to the gauge theory of the Ising model).
In the classical continual limit, the electric field is expressed via the operator of angular momentum as $E_{k}^{{}}(\mathbf{n})=(g/a^{2})L_{k}^{{}}(\mathbf{n})$, where $a$ is the lattice parameter, $g$ is the gauge charge, which plays the role of the parameter of interaction (here, there is an analogy with the rotor representation (\ref{rotor})). Then, we find
\begin{equation}
\mathcal{H}=(a^{3}/2) \sum_{nk}E_{k}^{2}(\mathbf{n})-(1/g^{2}a)\sum_{n,ik}\cos \theta _{ik}(%
\mathbf{n}).
\label{9.1}
\end{equation}
After the decomposition in the continual limit, we have
\begin{equation}
\mathcal{H}=(a^{3}/2)\sum_{nk}[E_{k}^{2}(\mathbf{n})-B_{k}^{2}(\mathbf{n})],
\end{equation}
where $\theta_{jk}=a^2 g B_i$ and the indices $ijk$ are determined cyclically.
The first term in (\ref{9.1}) is the lattice form of the volume integral of the square of the electric field, and the second term, of the square of the magnetic field.

The physical space of states is locally gauge-invariant and satisfies the Gauss law.
Therefore, only closed loops of the electric flux are permitted: the theory does not have sources or sinks.
In the case of a strong coupling, only the electric term is important in Eq. (\ref{9.1}); the energy of the closed loop of the flux is proportional to its length. The magnetic term in (\ref{9.1}) makes possible for loops to fluctuate, but always leaves them closed.

The fluxes of electric charge correspond to the potential energy; the magnetic part, to the kinetic energy.
The conservation of the flux in the topological phase means the conservation of the charge, and consequently the presence of bosons.
The lines of the electric flux determine strings (or, in another terminology, loops).
The details of the string picture are determined by the value of the coupling constant $g$. For the electric field, the coupling constant is $g^2$; for the magnetic field, $1/g^2$.
Therefore, at large $g$ it is potential energy that dominates; the closed loops of the electric flux become constricted; their number is small.
On the contrary, at small $g$ (in the case of weak coupling), it is the kinetic energy that prevails, whose role is played by the magnetic field, so that the loops are easily excited, their number is large.

A. M. Polyakov revealed deep correspondences between the lattice gauge theories in three and four dimensions and the two-dimensional spin systems %\ref{9.1}
\cite{Polyakov79}. The two-dimensional O(n) Heisenberg spin systems allow exact solution in the continual limit for $n \geq 3$.
The reason is in that these theories have ``hidden symmetry,'' which leads to the infinite number of laws of conservation.
In two dimensions, these laws of conservation forbid the formation of particles in the processes of scattering.
Such laws of conservation cannot exist in the nontrivial four-dimensional theories.
However, Polyakov supposed that there is a nontrivial generalization of the laws of conservation on the non-Abelian gauge fields in four dimensions, which can lead to a closed solution.
Polyakov considered non-Abelian gauge theories as chiral fields determined on closed loops in the real space--time.
The Abelian lattice gauge theory can be considered similarly---the physical space of states of the theory consists of the closed loops of the electric flux.

Let us enumerate again the initial models, in which there are weak gauge interactions, deconfinement is possible, and a gauge boson and fermions appear:

(1) The lattice Z$_2$ theory, which in the case of solids is dual to the Ising model having classical phase transitions. It predicts quantum phase transitions.

(2) The Abelian O(1) model, which in the continuous limit corresponds to the Maxwell theory of the electromagnetic field.

(3) The non-Abelian SU(2) model, to which in the case of solids there corresponds a Heisenberg model.

In the first two models, there is possible a transition from the confinement to the deconfinement; they can be used for describing spin liquids.
In the third model, there is an asymptotic freedom; no deconfinement phase exists.

In Z$_2$ spin liquid and in the lattice Abelian O(1) model, strings appear, which, in turn, form networks---a string-net liquid.
Depending on the binding force, a large number or a small number of loops can appear in the phase and they can be large or small.
Strings can vibrate. The fluctuations of a string liquid are described by a vector field perpendicular to the direction of the propagation of waves, which is analogous to the Maxwell theory for the propagation of light.
Thus, there appear gauge fields---bosons, which are the only excitations in the case of the closed strings.
In the quantum electrodynamics, gapless photons correspond to these fluctuations of the string liquid.
If a string becomes torn, there appear fermions---its broken ends.
They can be identified with the electrons, since their interaction with the deformation of string network is described by the equations of electron--photon interaction \cite{Wen}.

The random appearance (``flashing'') of monopoles (see Section 3.2) leads to the confinement of fermions; this is analogous to the interaction of quarks through mesons in quantum chromodynamics.

The fermions appear as the defects of the quantum long-range entanglement, even if the initial model is purely bosonic.
The previous approaches made it possible to obtain fermions in the models of the boson field only for the U(1) gauge field in the 2 + 1 dimensionality of space.
In the approach of quantum entanglement, the fermions and bosons are obtained simultaneously for any dimensionality and any gauge group.

In the usual physics of the condensed state, the boson field arises as a result of collective oscillations of atoms: in the solids, these are phonons (two transverse and one longitudinal modes); in the liquids, longitudinal compression waves (one mode).
The collective vibrations of the extended objects (strings), which compose medium in the case of our strongly correlated system, can leads to the appearance of gauge bosons (even photons!) \cite{Wen}.
Moreover, fermions can also appear as a result of collective vibrations of a quantum liquid consisting of extended objects of a more complicated nature.

In order to avoid a field-theoretical consideration of a continuous medium, usually there are introduced lattice gauge theories, where the fields are switched along the bonds, which corresponds to a dual lattice \cite{Kogut}.
These fields are equivalent to electromagnetic field and, according to \cite{Wen}, generate loops --- gauge bosons in the phase of deconfinement.
Upon the breaking of loops, there are formed fermions, which can be considered as the bound states of a quantum of the flux of the field and of a charged particle.

Kitaev \cite{Kitaev1,Kitaev2} proposed an exactly solvable model, which confirms the results of the slave-boson theory relative to the deconfinement.
The quasiparticles in this model carry the electric and magnetic charges of the corresponding gauge group.

In the Wen approach \cite{Wen3}, the representation of qubits is used---simplest elements of quantum information.
In contrast to the approach of slave-bosons, here the primary elements are pseudo-spins, located on the lattice bonds rather than at the sites.
The properties of the electromagnetic field prove to be caused by the complex organization of qubits---nontrivial topological order.
If in the ground state (vacuum) all qubits are in the simplest ``ferromagnetic'' state with spins up, then the state with several inverted spins describes a space with several Bose particles with the zero spin, to which there corresponds a scalar field.
However, this simple approach encounters with a number of problems and requires a further development. In reality, the vacuum must have a much more complex structure.

The observed elementary particles can appear only from a nontrivial qubit ether, tangled at large distances, so that the matter consists of entangled qubits.
The variety of the properties of the matter is determined by the topological complexity of the organization of its degrees of freedom, rather than by the variety of initial fields.
In this picture, the matter can be identified with the information.

The lattice theories discussed are not local boson models, since in the lattice gauge theories the strings are continuous.
It follows from the theory of string networks that the language of strings in reality is not strictly necessary: the qubits by themselves are capable of generating gauge fields, which obey to equations of the Maxwell or Yang--Mills type.
Thus, the gauge symmetry does not have a fundamental importance for the gauge theory: a lattice gauge theory will always reproduce a gauge interaction at low energies.
Since a gauge field can be obtained from the local models of a qubit without gauge symmetry, the picture of strings is converted into the picture of a long-range entanglement, and the gauge fields become its fluctuations.
A string network is a way for describing the regularities of such a system \cite{Wen3}.

\subsection{String-Net Condensation}

In addition to the field theory of strings, a physical picture is required, which would explain how the topological phases arise from the microscopic degrees of freedom.
In \cite{Levin-Wen2005}, these problems were considered for a broad class of ``doubled'' topological phases, which are described by the sum of two topological quantum-field theories with opposite chiralities.
Physically they are characterized by parity and time-reversal invariance.

As was discussed above, the local energy limitations can lead the microscopic degrees of freedom to the organization into the effective extended objects---string nets.
If the kinetic energy of these string nets is large, then they can become condensed, giving rise to topological phases.
The condensation of string nets ensures a natural physical mechanism for the appearance of topological phases in the real systems of condensed media.

With the simplest qubits on the bonds, which form a dual lattice, we deal in the Z$_2$ lattice gauge theory.
The corresponding Hamiltonian of the Kitaev model is written as \cite{Levin-Wen2005,Sachdev4}
\begin{equation}
  \mathcal{H}=
 -U\sum_i\prod_{\langle ij\rangle} \sigma^x_{\langle ij\rangle}
 - t\sum_p\prod_{\langle ij\rangle \in p} \sigma^z_{\langle ij\rangle}, \quad
  \label{ham}
\end{equation}
where $\sigma^{x,z}$ are the Pauli matrices, $i$, $ \langle ij\rangle$ and $p$ are sites, bonds, and ``plaquettes'' of the lattice, respectively.

In the only nondegenerate ground state, we have for all bonds and plaquettes
$A_{i} =\prod_{\langle ij\rangle} \sigma^x_{\langle ij\rangle} =1$,
$F_p =\prod_{\langle ij\rangle \in p} \sigma^z_{\langle ij\rangle}=1$  (in this case, it is important that $[A_i , A_j ] = [F_p, F_{p'}] = [A_i, F_p] = 0$).
Thus, the spin liquid is a superposition of all configurations of closed strings with equal positive coefficients.

The spinon excited state in this formalism is a broken bond in the basis
$ \sigma ^ z_{\langle ij\rangle}$.
Since here there is no conservable spin quantum numbers, the spinon carries no spin, but changes the local exchange energy; in the Z$_2$ gauge theory, the spinon carries the corresponding electric charge.
The spinon state has eigenvalues $ A_i = -1 $ and $ A_j = 1$ for all $ j \neq 1 $.
In this case, $ F_p = 1 $ on all plaquettes, so that the spinon is a superposition of all configurations of strings on the lattice with the only free end at the site $i$.

A vison has an additional structure: for it,  $F_p = -1 $ on one plaquette $p$,
and $ F_{p '} = 1 $ for all $ p' \neq p $ but $ A_i = 1 $ for all sites $i$.
The wave function is, as before, a superposition of all configurations of the closed strings, as in the ground state.

The Hamiltonian (\ref{ham}) in its form resembles the Hubbard Hamiltonian (\ref{eq:G.1}).
The first term in (\ref{ham}) is the ``electric energy'' of order $U$, which is the tension of the string, whereas the ``magnetic energy'' of order $t$ is the kinetic energy of the string. Therefore, this microscopic Hamiltonian effectively describes the dynamics of strings.

\begin{figure}[tb]
\centerline{
\includegraphics[width=2.5in]{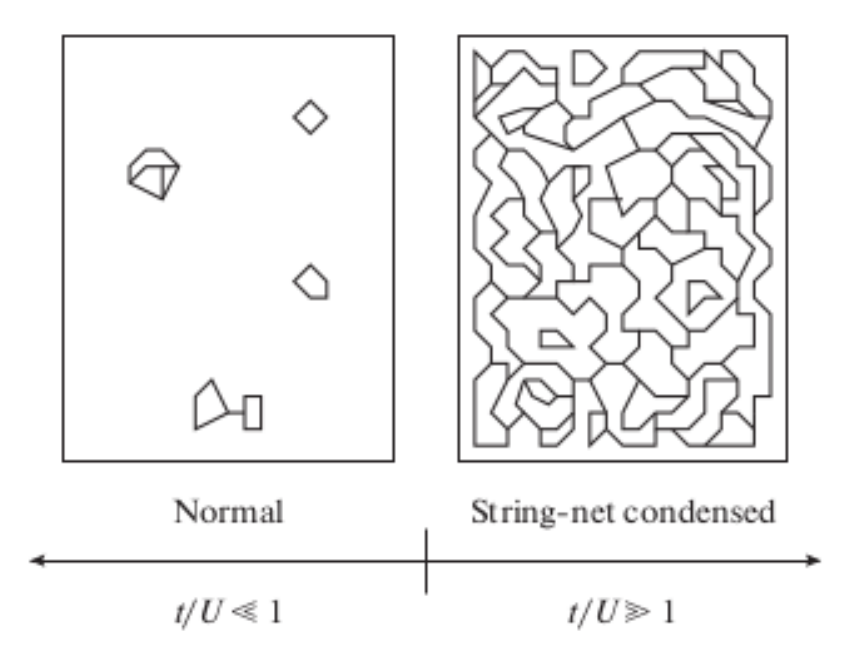}
}
\caption{
Schematic phase diagram for the Hamiltonian of string networks (\ref{ham}). When the $t/U$ (ratio of kinetic energy to the tension of the string) is small, the system is in the normal phase; the ground state is a vacuum with a small number of networks. When $t/U$ is large, the large fluctuating string networks are condensed. Thus, there is expected a phase transition between these two states at $t/U$ of the order of unity.
}
\label{fig16}
\end{figure}

The confinement phase corresponds to a high electric energy and, therefore, to a large tension of the string, $ U \gg t $.
Thus, the ground state is a vacuum configuration with several small strings.
The deconfinement phase corresponds to a high magnetic energy and, therefore, to a high kinetic energy, thereby describing the quantum liquid of large strings---a superposition of many string configurations, a string condensate.
The large string networks typically have a size of the order of the size of the system, filling the entire space (see Fig. \ref{fig16}). The transition between these states resembles the Mott--Hubbard transition.

Although these conclusions are obtained in the simplest gauge theory---on the Z$_2$ lattice, a similar, though more complex picture exists also for other deconfinement gauge theories.

\begin{figure}[tb]
\centerline{
\includegraphics[width=3in]{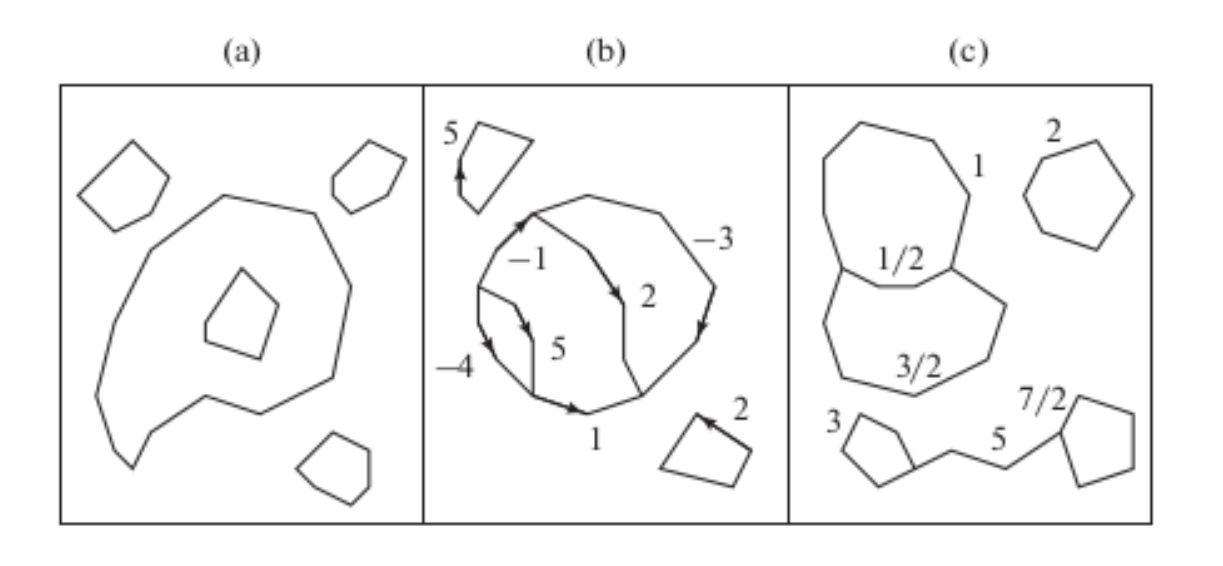}
}
\caption{
Typical configurations of string networks in the dual formulation (a) Z$_2$, (b) U(1), and (c) SU(2) of the gauge theory \cite{Levin-Wen2005}. In the case of (a), the configurations consist of closed (nonintersecting) loops. In the case (b), they are oriented graphs with the edges marked by integers. They obey to the rules of branching $ E_1 + E_2 + E_3 = 0 $ for any three edges, which meet at a point. In the case of (c), the network consist of (nonoriented) graphs with edges marked by the half-integers 1/2, 1, 3/2, etc.
}
\label{fig17}
\end{figure}

In the U(1) model, qubits are examined with $N$ integer levels (types of bonds); in the SU(2) model, half-integer values (Fig. \ref{fig17}) are possible.
A quantum of the flux is located at each site---so, the string nets arise.
Thus, the topological phases appear upon the condensation of string nets similar to how the traditional ordered phases appear through the condensation of particles.

It should be emphasized that the unification of gauge bosons and fermions in terms of string nets differs substantially from the theory of superstrings for the gauge bosons and fermions.
In the theory of string nets the intranet gauge bosons and fermions occur from qubits that form the space, and the ``string net'' is simply a model of qubits, the name, which indicates how the qubits are organized in the ground state.
In this case, the gauge bosons are the waves of collective fluctuations of string nets, and the fermion corresponds to the end of a string.
On the contrary, in the theory of superstrings the gauge bosons and fermions directly originate from the strings, corresponding to their small pieces.
Different fluctuations of these pieces of strings lead to particles of different kinds.
The fermions in the theory of superstring are obtained ``by hand,'' by the introduction of Grassman fields.

\subsection{Tensor Networks}

A new language in the theory of the condensed state comes from tensor networks (TNs) \cite{Orus,Wen}.
In particular, it makes it possible to describe the processes of entanglement and appearance of quantum topological states.
Here, the system of quantum pseudo-spins is taken as a basis, and the wave function of the system is described by a network of interconnected tensors.
It can be said that this tensor is ``the driving DNA'' for the wave function: the latter is constructed from fundamental block-pieces of the quantum state according to sufficiently simple rules \cite{Orus}.
Within the framework of this method it is possible to describe a broad class of bosonic, fermionic, and spin systems with different dimensionality, symmetry, and boundary conditions, and also phase transitions in them.

The dimensionality of the Hilbert space, which describes all connections in a many-particle system, is enormous: it increases exponentially with the number of particles (which by itself is very great, on the order of the Avogadro number (10$^{23}$)).
Fortunately, the dimensionality of the space of relevant states in a number of cases can be reduced.
Thus, the majority of Hamiltonians (including those in the solid-state theory) are local; for example, the interaction occurs only between the nearest neighbors.
It turns out that for the low-energy states of such Hamiltonians with a gapped spectrum the entropy of the entanglement of two systems is determined by the area of the boundary rather than by the volume (here, there appears an analogy with the holographic model of the universe and with the theory of black holes, according to which all information is coded on the external surface). This consideration sets limitation, also, on the physically permissible form of low-energy states. Moreover, upon the evolution of a quantum many-particle system described by a local Hamiltonian, the overwhelming majority of states proves to be unattainable in reality.

A tensor network (TN) is a set of tensors, in which some (or even all) indices are convoluted according to a certain rule (this operation is called a TN convolution).
As a result, there is obtained a new tensor with several open indices.
In this case, the complete number of operations necessary for obtaining the final result for a TN depends substantially on the order of the convolution of the indices.
The formalism of TNs makes it possible to break the wave function of a many-particle state into small fragments.

Following the survey \cite{Orus}, let us examine a quantum system consisting of $N$ particles, the degrees of freedom of each in which can be described by $p$ different states (for example, for a system of the type of the quantum Heisenberg model with spin 1/2 we have $p = 2$, so that each particle is a two-level system or a qubit). For this system, any wave function $|{\Psi}\rangle$ which describes its physical properties, can be written down as follows:
\begin{equation}
 |{\Psi}\rangle = \sum_{i_1 i_2 \ldots i_N} C_{i_1 i_2 \ldots i_N} |{i_1}\rangle\otimes |{i_2}\rangle \otimes \cdots \otimes |{i_N}\rangle ,
\label{st}
\end{equation}
where $|{i_r}\rangle$ is the basis of the states of each particle, $ r = 1, ..., N $.
In the above equation, $C_{i_1 i_2 \ldots i_N}$ are the sets $ p^N $ of complex numbers $ i_r = 1, ..., p $ for each particle $R$, the symbol $ \otimes $ designates the tensor product of separate quantum states for each of the particles in the system of many bodies.
The quantities $ C_ {i_1 i_2 \ldots i_n} $ can be considered as the coefficients of the tensor $C$ with $N$ indices $ i_1 i_2 \ldots i_n $, i.e., of the rank $N$ with the coefficients $ O (p^N) $. Hence, it can be seen that the number of parameters that describe the wave function is exponentially large (on the order of the size of the system), so that this complete description is practically ineffective.
The introduction of TN states makes it possible to decrease the complexity of the representation of the states $|{\Psi}\rangle$, ensuring the precise description of the expected characteristics of the entanglement of the state.
This is reached by replacing the ``large'' tensor $C$ by a network of ``smaller'' tensors.
This approach is reduced to the decomposition of state $|{\Psi}\rangle$ into ``fundamental blocks,'' i.e., into a network of tensors of smaller rank.
The final representation in the terms of TNs, as a rule, depends on the number of model parameters polynomially and, thus, is a computationally effective description of the quantum state of the system of many bodies.

\begin{figure}[ht]
\centering
\includegraphics[width=11cm]{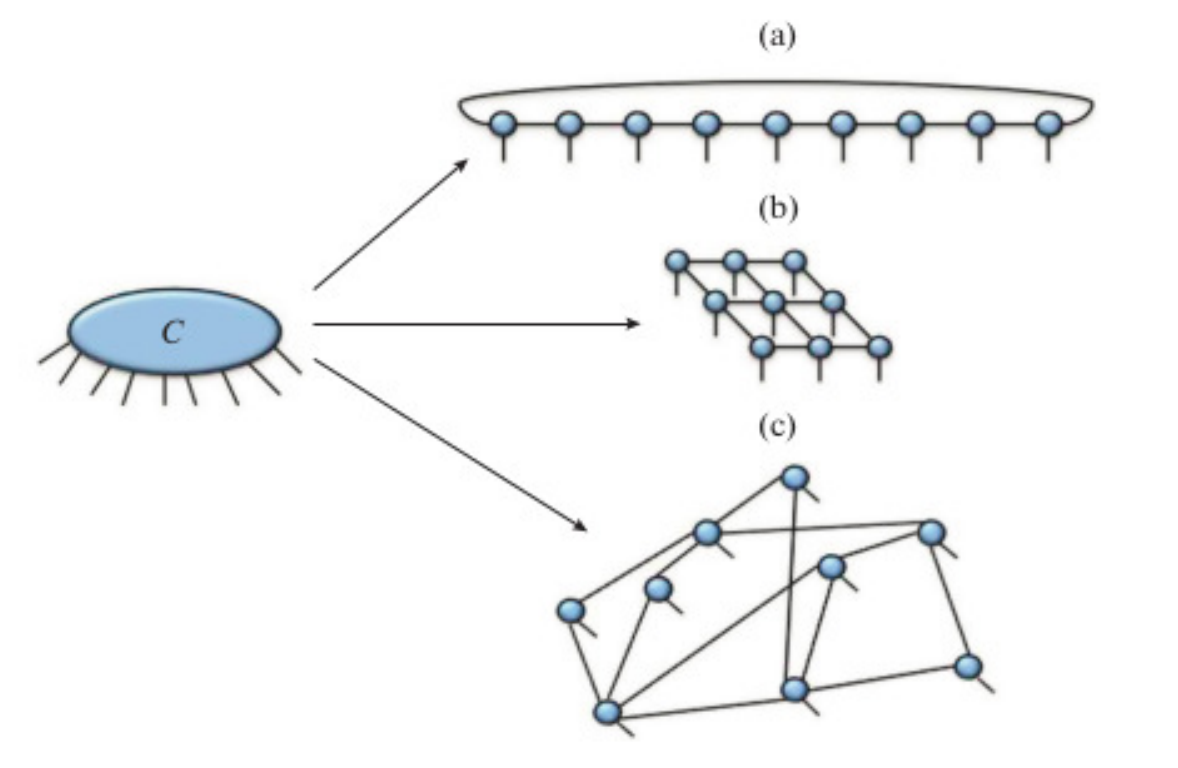}
\caption{
Decomposition of the tensor $C$ into tensor networks through (a) MPSs with periodic boundary conditions, (b) PEPSs with open boundary conditions, and (c) into an arbitrary tensor network.
\label{fig18}
}
\end{figure}

A simple example (Fig. \ref{fig18}a) is the matrix product state (MPS) with periodic boundary conditions.
Here, the number of parameters is $ O (N p D^2) $, where the open indices in TNs take on $p$ values, and the rest take on $D$ values.
Then, the convolution of TNs gives a tensor of rank $N$, and, therefore, $p^N$ coefficients. However, these coefficients are not independent variables; they are obtained from the convolution of this TN and therefore have a certain structure.

The MPS has a finite correlation length when and only when the greatest eigenvalue of its double tensor is not degenerate.
The state of the tensor product (TP) is a natural generalization of the MPS to the case of two or greater number of dimensions due to the arrangement of tensors of higher rank (instead of matrices) at each lattice site. It also satisfies the area law for the entanglement.
On the other hand, it describes not only states with a short-range entanglement, but also topologically ordered states with a long-range entanglement.

The replacement of the tensor $C$ by TNs assumes the appearance of additional degrees of freedoms in the system, which are responsible for the ``gluing'' of different blocks of our DNA.
These new degrees of freedom are designated by indices, which connect the tensors into the network.
Such indices are called the bond indices and make important physical sense: they represent the structure of the many-particle entanglement in the quantum state $|\Psi \rangle$ and the number of their different values is the quantitative measure of quantum correlations in the wave function. The corresponding maximum value $D$ is the bond dimension of the tensor network (or the internal dimensionality of the MPS).

\begin{figure}[ht]
\centering
\includegraphics[width=11cm]{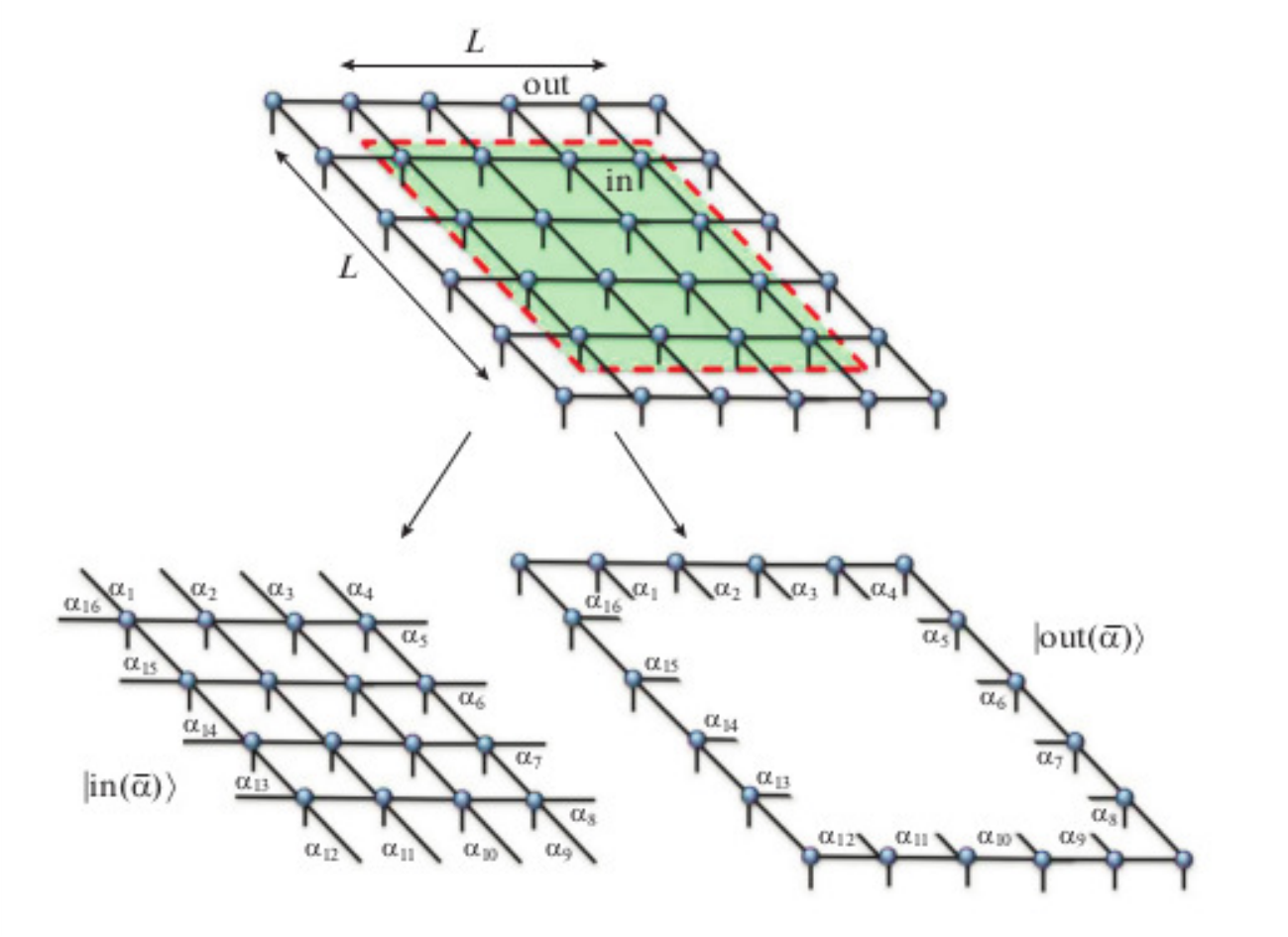}
\caption{
States $|{in(\bar{\alpha})\rangle}$ and $|{out(\bar{\alpha})}\rangle$ for the $4 \times 4$ and $6 \times 6$ blocks of PEPSs.
\label{fig19}}
\end{figure}

Let us explain how the entanglement is correlated with the bond index.
We assume that we have a TN state with a bond dimension $D$ for all indices (Fig. \ref{fig19}).
This example of the TN state is called the projected entangled pair state (PEPS).
Let us estimate the entropy of entanglement for this state in the case of a linear block of length $L$.
We introduce a joint index of all indices of TNs
$\bar{\alpha} = \{\alpha_1 \alpha_2 ... \alpha_{4L} \}$.
If the indices $\alpha_i $ take $D$ values, $\bar{\alpha}$ can change to $ D^{4L} $.
Then, in terms of the states of the internal and exterior parts of the block, we have
\begin{equation}
|{\Psi}\rangle = \sum_{\bar{\alpha}=1}^{D^{4L}} |{in(\bar{\alpha})}\rangle \otimes  |{out(\bar{\alpha})} \rangle\ .
\end{equation}

The reduced density matrix of the internal part is written as
\begin{equation}
\rho_{in} = \sum_{\bar{\alpha}, \bar{\alpha'}} X_{\bar{\alpha} \bar{\alpha'}} |{in(\bar{\alpha})}\rangle \langle{in(\bar{\alpha'})|},
\quad  \ X_{\bar{\alpha} \bar{\alpha'}} \equiv   \langle{out(\bar{\alpha'})}|{out(\bar{\alpha})}\rangle,
\end{equation}
and its rank is no more than $D^{4L}$; the reduced density matrix of the exterior part is analogous.
The entanglement entropy $S(L) = -{\rm Sp} (\rho_{in} \log \rho_{in})$ of the block is bounded from above by the logarithm of the rank of the matrix $\rho_{in}$, and we find
\begin{equation}
S(L) \le 4L \log D  \,;
\end{equation}
this is the upper boundary for the entanglement entropy according to the area law, which corresponds to the situation of deconfinement (see Section 3.2).

\begin{figure}[ht]
\centering
\includegraphics[width=12cm]{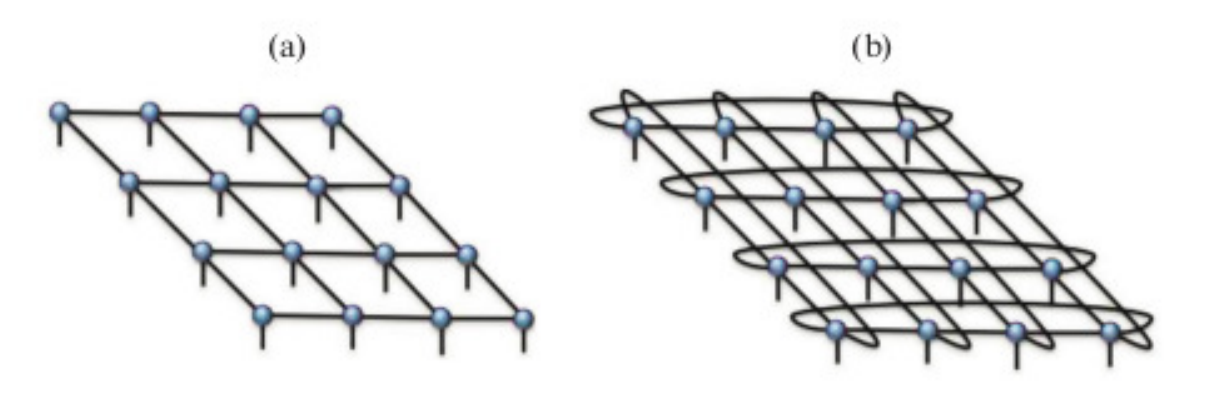}
\caption{
$4 \times 4$  PEPS states: (a) open boundary conditions, and (b) periodic boundary conditions.
\label{fig20}}
\end{figure}

The MPSs are the TN states that describe one-dimensional systems.
In them, there is one tensor for each site in the system of many bodies.
The connective indices of the bonds, which glue these tensors, assume $D$ values, and the open indices, which correspond to the physical degrees of freedom of local Hilbert spaces, assume $p$ values.
The PEPSs (natural generalization of MPSs) make it possible to describe the case of higher space dimensions. Figure \ref{fig20} shows PEPSs for a $4 \times 4$ square lattice.

In order to calculate the expected value of local observables in the state of the tensor product, we must know how to convolute two-dimensional tensor networks, which in the general case is not a simple and effective procedure.
However, for this purpose there were developed approximate methods, used in quantum calculations \cite{Deng}.

\section{Conclusions}

The modern theory of the condensed state for systems with strong correlations operates with substantially many-electron states.
In such systems, the ME operators and functions (Section 2.2) prove to be primary, and the one-electron operators, in contrast to the standard procedure of second quantization, appear upon the breaking of an ME string.
An obvious case here is the Anderson state of an RVB, where the electron is separated into a spinon and a holon (particles, which formally appear from the representation of the ME operator, Section 2.4).
These particles appear upon the breaking of the bound valence pair, and initially there is only a complete wave function of the crystal.

In contrast to the usual quantum field theory on the one hand and the classical topology on the other hand, the quantum topology operates with the whole spaces of many-electron operators; moreover, the form and writing of the latter proves to be ambiguous (as an example, the Kotliar--Ruckenshtein representation can be mentioned (\ref{eq:133}) and the introduction of a Higgs field \cite{Chowdhury}).
Each ME operator is an elaborate dynamic complex: in it, as from a seed, there germinate different operator products corresponding to processes of recombination of ME states.
Thus, there are described numerous quantum phases, to which there correspond different excitation spectra and physical models (mean-field theories), and the fluctuations are described by gauge fields, which determine the stability of these phases.

The picture of the quantum long-range entanglement changes our representations about the structure of space--time and correlation, making the latter nonlocal; this is substantially connected with the topological properties of the system (see article \cite{Kauffman}, where the generalized formalism of Feynman path integral is discussed in detail, which describes the topology of space–time).

This situation resembles the Einstein--Podolsky--Rosen effect, in which there occur noncausal quantum correlations at the infinitely large distance.
An additional factor, which favors quantum coherence, is the topological invariants, which stabilize the quantum state (in the usual macroscopic systems of the type of the ``Schr\"odinger cat,'' the weak external actions lead to decoherence of such states).

The new description in the language of collective excitations of the extended structures---string networks---makes it possible in a united manner to introduce both the gauge bosons and fermionic particles \cite{Wen,Wen3}.
On the contrary, the classical geometric point of sight (``fiber bundles'') can lead only to gauge interactions, so that the quantum entanglement opens a new chapter in physics.

In the theory of the condensed state, there are possible languages based on the concepts of spins, bosons, or bits of information--qubits (the simplest pseudo-spin representation of the latter is the occupied or vacant site of the lattice, or two directions of arrows).
According to \cite{Wen3}, on the deep level, the matter can be considered as a totality of qubits. This space is a dynamic medium; it is an ocean of qubits, a ``qubit ether.''
Then, the substance, i.e., the elementary particles, are excitations--waves, vortices, ``bubbles,'' and other defects in this ether.
In this case, according to the formulas of quantum physics and theory of relativity, there is established an equivalence of mass, energy, frequency, and information (entropy).

In the topological approaches, the ground state is considered within the framework of the nonperturbative approach.
The description of the ground state in the ordered phases is not too difficult problem.
As a rule, it is constructed rather simply and is characterized by weak coupling and by Goldstone excitations.
The most interesting problem is that of the description of a ``paramagnetic'' (formally, disordered) state, whose structure (e.g., the topological structure, including a series of spin liquids) can be in reality very complicated, and of a fundamental importance here can be effects of frustration.
In the problems discussed, we encounter with the search for the ground state of the system, which is formed by quantum fluctuations and exhibits a strong topological degeneracy (the simplest example is the singlet ground state of a Heisenberg antiferromagnet, Section 2.3).
Of fundamental importance is the problem of entropy of this state and of information contained in it, which can be colossal.
Indeed, the liquid state is a totality (superposition) of all possible positions of sites, and in a string liquid all the positions are themselves matrices or involve tensors. In such a system, a long-range entanglement can naturally appear.

The topological phases, just as any other quantum systems, are characterized by an enormous space of states, and the addition of only one element can substantially change the state of the entire system (see, e.g. the Anderson's catastrophe of orthogonality \cite{6331a,Zou1}).

The topological order and entanglement lead to a number of new states of a quantum matter and to new physical phenomena, such as fractional charge, fractional and non-Abelian statistics, etc.
If we could realize a quantum liquid consisting of oriented strings in real materials, this would allow us to create artificial elementary particles, an artificial world in artificial vacuum \cite{Wen,Wen3}.
This subject matter also brings us to philosophical questions of the nonlocality and collective quantum nature of consciousness \cite{mensky}, to the problems of virtual reality, etc.
There is also an opportunity of applying the concept of exotic topological states to the problem of transfer and storage of information in the living matter (here, there appear analogies with the structure of DNA, see, for example, \cite{Orus} and the survey \cite{Nikiforov}).

Although the theory of superstrings in application to physics of elementary particles and cosmology met with serious difficulties \cite{Smolin}, similar string approaches prove to be highly useful in the physics of the condensed state.
In this context, different string concepts are proposed by the different authors \cite{Wen,Weng20071}.
Deep connections with quantum electrodynamics \cite{Kogut} and chromodynamics \cite{Kogut1,Kogut11} appear here; the formalism of electric and magnetic components of the gauge field proves to be convenient.
Besides the physics of elementary particles, at present there are widely used formal methods, which unite the theory of topological states with the theory of gravity and structure of the universe, including holographic models (AdS/CFT theory) \cite{end1,gravity}.
It turns out that the theory of loop quantum gravity can be reformulated in terms of a specific type of string networks, where the strings are marked by positive integers.
Thus, the condensation of string networks in the spin model can lead to gravity \cite{Levin-RMP}.

The modern picture of the excitation spectrum in the condensed media gives a new sight also on a number of old problems of the physics of solids, in particular on the problem of the itinerant magnetism and especially on the description of the paramagnetic state of both strongly correlated and usual systems.
Their nature proves to be extremely complicated: it includes a complex topology and enormous hidden information.
The topology proves to be essential also upon the description of the correlation (Hubbard) splitting of the spectrum in many-electron systems (see Section 3.6).

As to practical applications, the formalism of tensor networks finds use in quantum neuron networks \cite{Deng}, and the topological phases in the linear and flat structures and even in three-dimensional materials can be used in the new generations of electronics and have a potential in quantum computers.

\section*{Acknowledgments}

The work was performed within the framework of the state task of the FANO of Russia (theme ``Flux'' no. AAAA-A18-118020190112-8; and theme ``Quantum'', no. AAAA-A18-118020190095-4). Section 4 was supported by the Russian Scientific Foundation (project 17-12-01207).

%\bibliography{x1}

\newpage


\section*{Supplementary material}

# Современная физика конденсированного состояния: сильные корреляции и квантовая топология

В. Ю. Ирхин, Ю. Н. Скрябин

15 ноября 2019 г.

В пространстве материи твердой,
под властью ее законов
и вихревых потоков,
в сомнениях фрустрационных
тайно хранит электрон,
будучи несвободным
в дебрях дуальной решетки,
тройственную природу,
включая бозон, фермион и фотон -
агент калибровочного поля.
Пускай конфайнмент препоной,
свирепствуют монополи,
не потеряет в сумерках он
мелодию скрытых струн,
послание мира иного,
память квантовой топологии.

**Аннотация**

Темой обзора является применение новых идей необычных квантовых состояний к физике конденсированного состояния, в особенности твердого тела, в контексте современной теории поля. Проведено сопоставление с классическими работами по многоэлектронной теории, включая формализм многоэлектронных операторов. Обсуждаются принципиально многочастичная природа основного состояния, индивидуальные и коллективные возбуждения и квантовые флуктуации в рассматриваемых системах, а также квантовые фазовые переходы с упором на топологические аспекты и с учетом эффектов фрустрации. Рассмотрены вариационные подходы и представления вспомогательных частиц, соответствующие приближения среднего поля и теория калибровочных полей, проблема конфайнмента–деконфайнмента, нарушение картины ферми-жидкости и экзотические неферми-жидкостные состояния. Дан обзор современной теории запутанных топологических состояний, формирования спиновой жидкости, струн и струнных сетей.

# 1 Введение

Нобелевская премия 2016 года была присуждена за теоретические открытия топологических фазовых переходов и топологических фаз материи, за работы, сделанные в 1970-80х гг., которые "раскрыли секреты экзотической материи". За время, истекшее после пионерских работ Д. Халдейна, М. Костерлица и Д. Таулеса, был достигнут значительный прогресс в физике конденсированного состояния, связанный с применением новых, существенно квантовых топологических концепций. Сюда относятся топологические фазы вещества, в том числе защищенные симметрией, спиновые жидкости, экзотические возбуждения, струны, тензорные сети... Эти достижения были связаны с работой нескольких групп исследователей: Андерсона [1, 2], Вена [3, 4, 5, 6], Сачдева [7], Коулмена [8]... Большой вклад здесь внесли и современные российские ученые - А.М. Поляков и А.Ю. Китаев.

## 1.1 Квантовые фазы и понятие квантовой топологии

Кратко остановимся на формальных математических определениях и подходах. Как известно из современной математики [9], топологическая теория поля и теория узлов тесно связаны. Действительно, интуитивно понятно, что топологическое запутывание ассоциируется с запутыванием (entanglement) квантовых состояний (в частности, это дает красивую аналогию с фейнмановским интегралом по траекториям). Таким образом, с практической точки зрения квантовая топология оказывается тесно связанной с проблемой квантовых вычислений [10].

Соответственно, была предложена математическая основа для классификации топологических фаз: каждая из них связана с математическим объектом, известным как "тензорная категория" [11, 12], который удовлетворяет некоторым алгебраическим уравнениям. Этот объект характеризует различные топологические фазы

и определяет универсальные свойства квазичастичных возбуждений, как и аппарат обычных групп симметрии в теории Ландау. Таким образом, математическая структура тензорных категорий вместе с физической картиной струнной конденсации обеспечивает общую теорию топологических фаз.

Данный подход также предоставляет точно решаемые модели и волновые функции основного состояния для каждой из топологических фаз. Это локальные бозонные (или спиновые) модели, которые реализуют все дискретные калибровочные теории в любой размерности и теории Черна-Саймонса (в размерности 2+1). Топологический порядок характеризуется устойчивым вырождением основного состояния (квантовое состояние стабилизируется топологическими инвариантами) и нетривиальной статистикой частиц. В отличие от ситуации нарушения симметрии, возникающие (emergent) частицы в топологически упорядоченных состояниях включают калибровочные бозоны, а также фермионы либо энионы (частицы с дробной статистикой), которые могут возникать как коллективные возбуждения чисто бозонных моделей.

С точки зрения физики, квантовые фазы вещества являются фазами материи при нуле температур и соответствуют основным состояниям квантового гамильтониана системы [7]. В этом смысле изоляторы, магнетики и сверхпроводники существуют при нулевой температуре $T$ и являются примерами квантовых фаз.

После открытия дробного квантового эффекта Холла [13] и высокотемпературной сверхпроводимости (ВТСП) [14] было осознано, что существует новый тип порядка – топологический (элементарные возбуждения с энергетической щелью) или квантовый порядок (в более общем случае бесщелевых возбуждений). С тех пор изучение и классификация новых типов порядка в конденсированных средах стало весьма интенсивной областью исследования.

Согласно [4], возможна следующая классификация. Если квантовое состояние имеет щелевой спектр, то соответствующий квантовый порядок будет называться топологическим порядком. Низкоэнергетическая эффективная теория топологически упорядоченного состояния будет топологической квантовой теорией поля (TQFT [15]). Второй класс квантовых порядков соответствует ферми-жидкостям или свободным фермионным системам; при этом различные квантовые порядки классифицируются согласно топологии поверхности Ферми [16]. Третий, самый интересный класс квантового порядка описывается в терминах конденсации струнных сетей [12]. Такие состояния имеют некоторое сходство с обычной ситуацией нарушенной симметрии при конденсации отдельных частиц: они аналогичны бозе-конденсированным состояниям, но конденсат формируется из протяженных объектов. Коллективные возбуждения в струнном конденсате не являются обычными скалярными бозонами, – колебания замкнутых струн порождают калибровочные бозоны, а при разрыве струн их концы могут давать фермионы.

Таким образом, квантовая топология, рассматривающая внутренние степени свободы, может быть переформулирована на теоретико-полевом языке замкнутых струн, которые построены из локальных спинов либо псевдоспинов (кубитов – понятие, широко используемое в квантовых вычислениях). В свою очередь, от этих струн можно перейти к представлению "электрических" и "магнитных" калибровочных полей [17].

Само название "топологический порядок" исторически заимствовано из низко-

энергетической эффективной теории киральных спиновых состояний в TQFT [15]. Термин "топологический" здесь означает дальнодействующее запутывание, и поэтому он относится к квантовой топологии [5], – необходимо его отличать от обычной классической топологии, оперирующей с вихрями в сверхтекучей жидкости, различием между сферой и тором и т. д.

Киральное спиновое состояние как особый тип квантового состояния было предложено в 1987 году для объяснения высокотемпературной сверхпроводимости [18, 19]. В отличие от известных в то время квантовых состояний, оно содержит новые экзотические частицы – спиноны и голоны – в состоянии деконфайнмента и соответствует стабильной фазе при нулевой температуре. Сначала, в предположении справедливости теории нарушенной симметрии Ландау, киральное состояние рассматривалось как состояние, которое нарушает симметрии обращения времени и четности, но сохраняет симметрию относительно вращения спинов и трансляций. Однако было быстро обнаружено, что существует множество различных киральных состояний, которые имеют ту же самую симметрию. Таким образом, одной симметрии недостаточно, чтобы характеризовать такие состояния. Это означает, что они представляют новый тип порядка, который и был назван топологическим порядком.

С макроскопической точки зрения топологический порядок характеризуется сильным вырождением основного состояния и неабелевыми геометрическими фазами [20]. С микроскопической же точки зрения топологический порядок определяется состоянием квантовой спиновой жидкости (QSL – quantum spin liquid) с щелевым энергетическим спектром [21, 22], которое не может быть представлено произведением одночастичных собственных состояний, если не рассматривать фазовых переходов с закрытием энергетической щели. Такие QSL проявляют дальнодействующее запутывание (entanglement), которое и есть внутренняя природа топологического порядка [23, 24, 25].

Формирование корреляций в топологических системах весьма необычно. Обычных "силовых" корреляций (предвестников дальнего порядка) здесь нет, но возникают особые, обусловленные топологией образца. При этом основное состояние вырождено, но не в силу симметрии, а благодаря топологическим характеристикам. Так, система "чувствует" сворачивание в тор, радикальным может оказаться влияние одной особой (выколотой) точки [26]...

## 1.2 Квантовые фазовые переходы

Традиционно считалось, что феноменологическая теория фазовых переходов Ландау [27, 28] может описать все возможные фазы вещества и фазовые переходы. Казалось, что все непрерывные фазовые переходы ассоциируются с нарушенной симметрией в рамках так называемого подхода Ландау-Гинзбурга-Вильсона. Долгое время была уверенность, что теория нарушенной симметрии Ландау также описывает все возможные квантовые фазы и непрерывные квантовые фазовые переходы (quantum phase transitions, QPT) между ними, то есть фазовые переходы при $T = 0$.

В отличие от классических фазовых переходов, QPT учитывают флуктуации параметра порядка во мнимом времени. Отличительной чертой QPT является наличие нуль-температурной квантовой критической точки, разделяющей квантово-упорядоченную и квантово-разупорядоченную фазы на фазовой диаграм-

ме температура–параметр (константа связи, давление, концентрация и т. д.), описывающий близость к квантовой критической точке. При конечной $T$ эти фазы разделены квантовой критической областью, ограниченной линиями кроссоверов, сходящимися в критическую точку. Наличие такой области на фазовой диаграмме приводит ко многим необычным явлениям, в частности, к образованию неферми-жидкости в металлических системах.

В частности, были предложены так называемые проективные группы. Как известно, симметрия обычных упорядоченных фаз позволяет классифицировать их по 230 пространственным группам (в трех измерениях), а также приводит к появлению коллективных бесщелевых возбуждений – голдстоуновских бозонов. Проективные группы симметрии (Projective symmetry groups, PSG) для квантовых порядков имеет аналогичные приложения, позволяя классифицировать более 100 различных двумерных $Z_2$-спиновых жидкостей, причем все они имеют одинаковую обычную симметрию [5]. Проективная группа симметрии также может приводить к возникновению бесщелевых калибровочных бозонов и фермионов. В методе PSG волновые функции не конкретизируются, а классифицируются по симметрии путем анализа с помощью локальных унитарных преобразований. Преобразования соответствующей ренормгруппы позволяют уловить простую суть возникающих сложных явлений. Метод PSG позволяет определить фазовые диаграммы, включающие фазы как с дальним, так и с ближним топологическим порядком и запутыванием. Кроме того, в теории топологических систем могут быть использованы термины и аналогии из теории информации и электрических цепей, а также нейронных сетей [29, 30].

После открытия ВТСП появилась надежда, что экзотические сильнокоррелированные состояния и квантовые спиновые жидкости могут играть решающую роль в ее понимании [1]. В таких состояниях возможно разделению спина и заряда: электрон распадается на две квазичастицы — спинон (спин $1/2$, заряд $0$) и голон (спин $0$, заряд $e$). Бозе-конденсация голонов приводит к сверхпроводимости, и этот ее новый механизм объясняет бурный интерес к изучению различных спиновых жидкостей.

Однако, несмотря на определенные успехи, теория фаз с нарушенной симметрией не может объяснить существование QSL в основном состоянии из-за конфайнмента (взаимного удержания) спинонов и голонов, обусловленного возникающим калибровочным полем [31]. Действительно, было обнаружено (см., напр., [32]), что в $U(1)$-калибровочной теории основное состояние является либо антиферромагнитным, либо сверхпроводящим, либо ферми-жидкостью – все эти фазы находятся в состоянии конфайнмента.

Тем не менее, на фазовой диаграмме высокотемпературных сверхпроводников при конечных температурах могут существовать области ("псевдощелевые" фазы), где фермионы и бозоны находятся в состоянии деконфайнмента. Следует отметить, что здесь фермионы и бозоны не могут рассматриваться как свободные из-за их сильного взаимодействия с калибровочным полем. В основном состоянии область деконфайнмента как правило редуцируется в "деконфайнментную" квантовую критическую точку [33, 34]. С этой точкой связан ряд важных физических явлений, включая разделение спиновых и зарядовых степеней свободы электрона.

В настоящем обзоре будут рассмотрены как феноменологические подходы в квантовой топологии, так и микроскопические модели для систем с сильными корреляциями. Мы постараемся проследить связи современных концепций, использующих

последние достижения квантовой теории поля, с классическими представлениями многоэлектронной теории, начатой в работах Шубина и Вонсовского, Гутцвиллера, Хаббарда и других авторов.

## 2 Развитие представлений и методов многоэлектронной теории

Первые успехи теории металлов в 20-е–30-е годы XX века были связаны с созданием квантовой механики и открытием статистики Ферми. В рамках приближения свободных электронов, а затем одноэлектронной зонной теории — в работах Паули, Блоха, Вильсона, Пайерлса, Зоммерфельда – было дано объяснение парамагнетизма, поведения теплоемкости и кинетических свойств [35]. Однако для описания ферромагнетизма и ряда других явлений (например, перехода металл-изолятор) эти представления оказались недостаточными. С другой стороны, попытки использовать для магнитных металлов модель Дирака-Гейзенберга, основанную на атомной картине локализованных спинов, также не дали хороших результатов (в частности, она была не в состоянии объяснить дробные значения магнитных моментов). Таким образом, потребовался определенный синтез многоэлектронной модели Гейзенберга и одноэлектронной зонной модели.

В 1934 г. была предложена полярная модель Шубина и Вонсовского [36, 37], а в 1946 г. — $s-d$ обменная модель. Обе эти модели сыграли исключительно важную роль в теоретическом описании $d$– и $f$–металлов и их соединений. Дальнейшее развитие многоэлектронных подходов в основном пошло по направлениям диаграммных методов и феноменологической теории ферми-жидкости Ландау.

Новый импульс многоэлектронной теории кристалла придали идеи Хаббарда [38, 39], выделившего в своей модели наиболее существенную часть кулоновского взаимодействия – сильное отталкивание электронов на одном узле $U$. В случае невырожденной зоны ее гамильтониан запишется как

$$\mathcal{H} = \sum_{ij\sigma} t_{ij} c_{i\sigma}^\dagger c_{j\sigma} + U \sum_i c_{i\uparrow}^\dagger c_{i\uparrow} c_{i\downarrow}^\dagger c_{i\downarrow},  \quad (1)$$

где $t_{ij}$ – интегралы переноса. Начиная с работ Хаббарда, эта модель широко использовалась для рассмотрения ферромагнетизма коллективизированных электронов, перехода металл-изолятор и других физических явлений, включающих сильную корреляцию и не описываемых методами теории возмущений.

В своей лекции "Многочастичная физика: незаконченная революция" [8] П. Коулмен говорит о трех эрах физики конденсированного состояния: первых успехах после открытия квантовой механики (свободные фермионы), многочастичной физике середины XX века (использование теории возмущений и фейнмановской диаграммной техники, начало изучения коллективных явлений) и современной эре физики сильнокоррелированной материи – исследования "экзотических" систем, в которых коллективные явления выходят на первый план.

В настоящее время активно обсуждается роль квантовых корреляций, где на первый план выходят топологическое вырождение, приводящее к радикальному усложнению основного состояния системы.

## 2.1 Вторичное квантование: фермионы и многоэлектронные операторы

При переходе к стандартному представлению вторичного квантования [40] волновые функции кристалла $\Psi(x_1\ldots x_N)$ ($x = \{\mathbf{r}_i s_i\}$, $s_i$ — спиновые координаты) выбираются в виде линейных комбинаций слэтеровских определителей. Они составляются из одноэлектронных волновых функций $\psi_\lambda(x)$ ($\lambda = \{\nu\gamma\}$, $\nu$ — индексы ячеек в решетке, а $\gamma$ — одноэлектронные наборы квантовых чисел):

$$\Psi(x_1\ldots x_N) = \sum_{\lambda_1\ldots\lambda_N} f(\lambda_1\ldots\lambda_N)(N!)^{-1/2} \sum_P (-1)^P P \prod_i \psi_{\lambda_i}(x_i), \qquad (2)$$

где $P$ пробегает всевозможные перестановки $x_i$. Представление вторичного квантования вводится путем использования одноэлектронных чисел заполнения $n_\lambda$ в качестве новых переменных:

$$\Psi(x_1\ldots x_N) = \sum_{\{n_\lambda\}} f(\ldots n_\lambda \ldots)\Psi_{\{n_\lambda\}}(x_1\ldots x_N). \qquad (3)$$

Тогда $f(\ldots n_\lambda \ldots)$ играет роль новой волновой функции. Одноэлектронные операторы рождения и уничтожения Ферми определяются следующим образом:

$$c_\lambda f(\ldots n_\lambda \ldots) = (-1)^{\eta_\lambda} n_\lambda f(\ldots n_\lambda - 1 \ldots),$$
$$c_\lambda^\dagger f(\ldots n_\lambda \ldots) = (-1)^{\eta_\lambda}(1 - n_\lambda) f(\ldots n_\lambda + 1 \ldots), \qquad (4)$$

причем

$$\eta_\lambda = \sum_{\lambda' > \lambda} n_{\lambda'}, \qquad c_\lambda^\dagger c_\lambda = \hat{n}_\lambda.$$

Следует отметить, что само введение одноэлектронных операторов, ставшее для нас привычным, является весьма нетривиальным шагом. Альтернативно операторы фермионов в решеточных (псевдо)спиновых моделях могут быть сконструированы из операторов псевдоспина [3]

$$c(i_a) = \sigma^+(i_a) \prod_{b<a} \sigma^z(i_b) \qquad (5)$$

(произведение идет по узлам решетки, упорядоченным определенным образом по контуру), что обобщает преобразование Йордана-Вигнера для одномерной цепочки на решетки высших размерностей. Таким образом, фермион предстает как существенно многочастичное и нелокальное образование. Это дает другое направление развития теории многочастичных систем, связанное со струнной картиной, использующей язык кубитов (см. раздел 5).

Теперь обобщим метод вторичного квантования, вводя квантовые числа электронных групп $\Lambda = \{\nu\Gamma\}$, $\Gamma$ — многоэлектронные (МЭ) уровни. В частности, можно объединить электроны на каждом узле в решетке ($\Lambda = \{\nu\Gamma\}$, $\Gamma_i$ — МЭ уровни), получая

$$\Psi(x_1\ldots x_N) = \sum_{\{N_\lambda\}} f(\ldots N_\lambda \ldots)\Psi_{\{N_\lambda\}}(x_1\ldots x_N). \qquad (6)$$

7

В случае атомной конфигурации эквивалентных электронов $l^n$ МЭ волновая функция электронной группы определяется рекуррентным соотношением (см. [41]):

$$\Psi_{\Gamma_n}(x_1 \ldots x_N) = \sum_{\Gamma_{n-1},\gamma} G^{\Gamma_n}_{\Gamma_{n-1}} C^{\Gamma_n}_{\Gamma_{n-1},\gamma} \Psi_{\Gamma_{n-1}}(x_1 \ldots x_{n-1}) \psi_\gamma(x_n), \qquad (7)$$

где $C$ — коэффициенты Клебша—Гордана. В случае $LS$-связи

$$C^{\Gamma_n}_{\Gamma_{n-1},\gamma} \equiv C^{L_n M_n}_{L_{n-1} M_{n-1}, lm} C^{S_n \mu_n}_{S_{n-1} \mu_{n-1}, \frac{1}{2}\sigma}, \qquad (8)$$

суммирование по $\gamma = \{lm\sigma\}$ стоит вместо суммирования по одноэлектронным орбитальным проекциям $m$ и спиновым проекциям $\sigma$, но не по $l$; $G^{\Gamma_n}_{\Gamma_{n-1}} \equiv G^{S_n L_n}_{S_{n-1} L_{n-1}}$ — генеалогические коэффициенты. Если добавленный электрон принадлежит другой оболочке, можно просто записать

$$\Psi_{\Gamma_n}(x_1 \ldots x_n) = n^{-1/2} \sum_{i,\Gamma_{n-1},\gamma} (-1)^{n-i} C^{\Gamma_n}_{\Gamma_{n-1},\gamma} \times$$
$$\times \Psi_{\Gamma_{n-1}}(x_1 \ldots x_{i-1}, x_{i-1} \ldots x_{n-1}) \psi_\gamma(x_i) \quad (9)$$

(в отличие от случая эквивалентных электронов, здесь необходима антисимметризация). Волновая функция всего кристалла (6) может быть теперь получена как антисимметризованное произведение МЭ функций для электронных групп, что в принципе позволяет ввести МЭ состояния для всего кристалла.

По аналогии с (7), (9) можно ввести МЭ операторы рождения для электронных групп [42]. Для эквивалентных электронов

$$A^\dagger_{\Gamma_n} = n^{-1/2} \sum_{\Gamma_{n-1},\gamma} G^{\Gamma_n}_{\Gamma_{n-1}} C^{\Gamma_n}_{\Gamma_{n-1},\gamma} a^\dagger_\gamma A^\dagger_{\Gamma_{n-1}}. \qquad (10)$$

При добавлении электрона из другой оболочки

$$A^\dagger_{\Gamma_n} = \sum_{\Gamma_{n-1},\gamma} C^{\Gamma_n}_{\Gamma_{n-1},\gamma} a^\dagger_\gamma A^\dagger_{\Gamma_{n-1}}. \qquad (11)$$

Операторы (10), (11) удобны только для работы с конфигурациями с фиксированным числом электронов. Чтобы избежать проблем "неортогональности" при различных $n$, следует ввести проекционные множители [43, 44]:

$$\tilde{A}^\dagger_\Gamma = A^\dagger_\Gamma \prod_\gamma (1 - \hat{n}_\gamma). \qquad (12)$$

Формально произведение в (12) идет по всем допустимым одноэлектронным состояниям $\gamma$. Однако из-за тождества $c^\dagger_\gamma \hat{n}_\gamma = 0$ достаточно сохранить только те $\gamma$, которые не входят в соответствующие произведения операторов в $A_\Gamma$. Вводя МЭ операторы, которые зависят от всех одноэлектронных квантовых чисел (как занятых, так и свободных состояний), мы делаем следующий шаг в квантово-полевом описании после обычного вторичного квантования. Таким образом, можно прийти к представлению МЭ чисел заполнения $N_\Gamma = \tilde{A}^\dagger_\Gamma \tilde{A}_\Gamma$ на данном узле.

Теперь мы можем вести обобщенные проекционные операторы $X$-операторы Хаббарда [39]:
$$X(\Gamma, \Gamma') = \tilde{A}^\dagger_\Gamma \tilde{A}_{\Gamma'}, \tag{13}$$
которые переводят состояние $\Gamma'$ в состояние $\Gamma$.
$$X(\Gamma, \Gamma') = |\Gamma\rangle\langle\Gamma'|, \quad X(\Gamma, \Gamma')X(\Gamma'', \Gamma''') = \delta_{\Gamma'\Gamma''}X(\Gamma, \Gamma'''), \tag{14}$$
где $|\Gamma\rangle$ — точные собственные состояния гамильтониана. В простейшем случае $s$-электронов $\gamma = \sigma = \pm(\uparrow, \downarrow)$, $\Gamma = 0$, $\sigma$, $2$ и $|0\rangle$ — свободное состояние (дырка), а $|2\rangle$ — дважды занятое синглетное состояние на узле (двойка).

Произвольный оператор $\hat{O}$, действующий на электроны на данном узле $i$, выражается через $X$-операторы:
$$\hat{O} = \sum_{\Gamma\Gamma'} \langle\Gamma|\hat{O}|\Gamma'\rangle X(\Gamma, \Gamma'). \tag{15}$$

Тогда, находя матричные элементы одноэлектронных ферми-операторов из (10), получаем представление [45]:
$$c^\dagger_\gamma = \sum_n n^{1/2} \sum_{\Gamma_n \Gamma_{n-1}} G^{\Gamma_n}_{\Gamma_{n-1}} C^{\Gamma_n}_{\Gamma_{n-1},\gamma} X(\Gamma_n, \Gamma_{n-1}). \tag{16}$$

В частности, для $s$-электронов
$$c^\dagger_\sigma = X(\sigma, 0) + \sigma X(2, -\sigma). \tag{17}$$

В действительности МЭ состояния вообще можно трактовать как первичные, получающиеся при решении полной МЭ задачи, а не строить из одноэлектронных состояний (как было сделано выше на примере атомного подхода). При этом важную роль играют нелокальность, а природа коллективных состояний может быть очень сложной и включать дальнодействие.

Существование нетривиальных многочастичных состояний подтверждается обобщенной теоремой Либа-Шульца-Маттиса [46]. Она утверждает, что в системе с полуцелым спином на ячейку и глобальной U(1)-симметрией спектр возбуждения в термодинамическом пределе не может одновременно подчиняться двум требованиям: (а) основное состояние единственно и (б) существует конечная щель для всех возбуждений. Это означает, что щелевое состояние с ненарушенной симметрией должно иметь вырождение основного состояния, которое по своей природе является топологическим.

При увеличении силы корреляций в МЭ системах происходит смена статистики элементарных возбуждений с зонной на атомную, которая проявляется как корреляционное хаббардовское расщепление и переход металл-изолятор (для наполовину заполненной зоны). Это превращение сопровождается формированием локальных магнитных моментов [47] и изменением числа состояний под поверхностью Ферми, если таковая вообще имеется (см. раздел 3.6).

Как указал Андерсон [48], образование хаббардовских подзон связано с катастрофой ортогональности, в результате которой исчезает квазичастичный вычет и нарушается картина ферми-жидкости Ландау. Исчезновение вычета означает, что

добавление одной частицы в систему вызывает глобальное изменение ее состояния, что характерно для МЭ систем. В такой ситуации, в частности, можно ожидать формирования экзотических частиц и состояния типа спиновой жидкости. Все эти вопросы будут обсуждаться ниже.

## 2.2 Вариационные подходы

Чтобы работать с МЭ состояниями, на протяжении многих лет развивались различные приближенные методы. В работах Шубина и Вонсовского было использовано простейшее квазиклассическое приближение [36, 49]. Оно по существу состоит в замене X-операторов $c$-числовыми функциями, определяющими амплитуду вероятности пребывания узла в состоянии однократно занятого узла, двойки или дырки:

$$X_i(+, 0) \to \varphi_i^* \Psi_i, \ X_i(2, -) \to \Phi_i^* \psi_i, \ X_i(2, 0) \to \Phi_i^* \Psi_i \tag{18}$$

с дополнительным условием $|\varphi_i|^2 + |\psi_i|^2 + |\Phi_i|^2 + |\Psi_i|^2 = 1$ Это соответствует определению полной энергии системы из вариационного принципа с волновой функцией

$$\phi = \prod_i [\varphi_i^* X_i(+, 0) + \psi_i^* X_i(-, 0) + \Phi_i^* X_i(2, 0\ ) + \Psi_i^*]|0\rangle. \tag{19}$$

Она смешивает возбуждения бозевского и фермиевского типа, а потому не удовлетворяет принципу Паули. Тем не менее, квазиклассическое приближение позволяет грубо описать переход металл-изолятор, что и было проделано впоследствии рядом авторов (см. обзор [49]). Так, в работе Карона и Пратта [50] было даже рассмотрено среднее поле для фермионов, так что фактически на с-числа заменялись не X-операторы, а обычные фермиевские операторы.

Предвосхищая работы Хаббарда, уравнения квазиклассического приближения дают изменение ширины энергетической полосы, а также относительного расположения различных полос в зависимости от числа двоек и магнитного момента.

Более строгим и последовательным оказался вариационный метод Гутцвиллера, см. [51]. Соответствующая волновая функция может быть записана как

$$\Psi = \prod_i [1 - (1 - \hat{n}_{i\uparrow}\hat{n}_{i\downarrow})]|\psi_0\rangle, \tag{20}$$

где вариационный параметр $g$ $(0 < g < 1)$ учитывает уменьшение вероятности состояний с большим числом двоек, $|\psi_0\rangle$ – волновая функция некоррелированного состояния.

Другой вариант вариационного подхода, учитывающий дальнодействие, дает экситонный подход, в котором рассматривается конденсация электрон-дырочных или электрон-электронных пар [52, 53]. Проще всего он может быть проиллюстрирован на описании сверхпроводимости. Стандартная волновая функция БКШ записывается как

$$|\Psi_0\rangle = \prod_{\mathbf{k}}(u_{\mathbf{k}} + v_{\mathbf{k}} c_{\mathbf{k}\uparrow}^\dagger c_{-\mathbf{k}\downarrow}^\dagger)|0\rangle \tag{21}$$

и является суперпозицией состояние с различным числом частиц $N$, т.е. вводятся боголюбовские аномальные средние. В работе [54] были проведены вычисления с

волновой функцией с фиксированным $N$

$$|\Psi_0\rangle = \left[\sum_{\mathbf{k}} \phi_{\mathbf{k}} c^{\dagger}_{\mathbf{k}\uparrow} c^{\dagger}_{-\mathbf{k}\downarrow}\right]^{N/2}|0\rangle, \qquad (22)$$

явно вводящей конденсат куперовских пар, причём использовалась теория перечисления графов. В работе [55] было аналогично построено основное состояние хаббардовского антиферромагнетика без введения аномальных средних (магнитных подрешёток):

$$|\Psi_0\rangle = \left[\sum_{\mathbf{k}} \phi_{\mathbf{k}} c^{\dagger}_{\mathbf{k}+\mathbf{Q}\downarrow} c_{\mathbf{k}\uparrow}\right]^{N/2}|F\rangle, \qquad (23)$$

где $|F\rangle$ – ферромагнитное состояние, $\mathbf{Q}$ – волновой вектор магнитной структуры. Это описание по наблюдаемым результатам соответствует слэтеровскому антиферромагнетизму. Таким образом, состояние с дальним порядком описать нетрудно, однако для описания коррелированного парамагнитного состояния требуется решение очень сложной задачи – учёт экситонов с разными импульсами.

Аналогичное описание дальнего порядка в антиферромагнетике может быть дано в рамках модели Гейзенберга с помощью функции $|\Psi_0\rangle = (S^+_{\mathbf{Q}})^{N/2}|F\rangle$ [56].

Вариационное описание состояния резонирующих валентных связей (RVB – resonating valence bond), сочетающее упомянутые выше подходы, было развито Андерсоном [1]. В случае половинного заполнения зоны здесь может быть использована пробная функция

$$\Psi = P_G|\psi_0\rangle, \qquad (24)$$

где $P_G = \prod_i(1 - \hat{n}_{i\uparrow}\hat{n}_{i\downarrow})$ – проекционный оператор Гутцвиллера в пределе сильных корреляций, полностью устраняющий дважды занятые состояния на узле. Поскольку $\phi(\mathbf{k}) = v_{\mathbf{k}}/u_{\mathbf{k}} \simeq \mathrm{sign}(t_{\mathbf{k}} - E_F)$ ($t_{\mathbf{k}}$ – зонная энергия), функция (24) может быть записана и в виде

$$|\Psi_0\rangle = \prod_{k<k_F} c^{\dagger}_{\mathbf{k}\uparrow} c^{\dagger}_{-\mathbf{k}\downarrow}|0\rangle. \qquad (25)$$

Именно под действием проектора функция БКШ, описывающая сверхпроводящее состояние, превращается в состояние RVB – появляется поверхность Ферми, поскольку коэффициенты преобразования Боголюбова принимают значения 0 и 1. Таким образом, функция основного состояния должна строиться именно на МЭ возбуждениях. Однако, как будет проиллюстрировано в разделе 3.1, на практике часто работают с особой функцией среднего поля, а правило сумм на узле может быть учтено через топологические вихревые возбуждения. Частный выбор функции $\phi(\mathbf{k})$ (например, в приближении ближайших соседей) позволяет учесть особое короткодействующее спаривание (образование синглетов).

Неелевское состояние обладает дальним порядком и бесконечным вырождением основного состояния, что приводит к появлению голдстоуновских мод – магнонов. С другой стороны, состояние RVB является невырожденным синглетным состоянием с ближним антиферромагнитным порядком. Возникновение подрешёток может быть описано как декогеренция (нарушение квантовой когерентности в синглетном состоянии) [57].

Согласно подходу [32], пробная функция для половинного заполнения в случае антиферромагнетика может быть записана в виде

$$|\Psi_0\rangle = P_G \left[\sum_{\mathbf{k}}(A_{\mathbf{k}}a^\dagger_{\mathbf{k}\uparrow}a^\dagger_{-\mathbf{k}\downarrow} + B_{\mathbf{k}}b^\dagger_{\mathbf{k}\uparrow}b^\dagger_{-\mathbf{k}\downarrow})\right]^{N/2}|0\rangle \qquad (26)$$

где операторы $a^\dagger_{\mathbf{k}\sigma}$ и $b^\dagger_{\mathbf{k}\sigma}$ соответствуют верхней и нижней слэтеровской подзонам с энергиями $\pm\xi_{\mathbf{k}}$, $A_{\mathbf{k}} = (E_{\mathbf{k}} + \xi_{\mathbf{k}})/\Delta_{\mathbf{k}}$, $B_{\mathbf{k}} = (-E_{\mathbf{k}} + \xi_{\mathbf{k}})/\Delta_{\mathbf{k}}$, причем $E_{\mathbf{k}} = \sqrt{\xi_{\mathbf{k}}^2 + \Delta_{\mathbf{k}}^2}$ и $\Delta_{\mathbf{k}} = (3/8)J\Delta(\cos k_x - \cos k_y)$. За счет включения синглетного RVB-спаривания, определяемого величиной $\Delta$, эта волновая функция дает существенно лучшее значение энергии по сравнению с обычным слэтеровским антиферромагнетизмом.

Волновую функцию (24) также можно записать через многоэлектронные X-операторы [1]:

$$|\Psi_0\rangle = \prod_{\mathbf{k}}(u_{\mathbf{k}} + v_{\mathbf{k}}X^{+0}_{\mathbf{k}}X^{-0}_{-\mathbf{k}}|0\rangle). \qquad (27)$$

Идеи экситонного конденсата были также применены к системам с промежуточной валентностью типа $SmB_6$ и $SmS$ в рамках двухзонных моделей Фаликова-Кимбалла с гибридизацией или решетки Андерсона [58]. В зависимости от значений параметров в этих задачах могут возникать экситоны большого и малого радиуса.

В последнее время такие подходы к проблеме промежуточной валентности развивались и на современном языке экзотических частиц – спинонов и голонов [59, 60]. Согласно [59], нейтральный фермионный композитный экситон формируется как связанное состояние электрона проводимости и голона.

## 2.3 Представления вспомогательных частиц

Непосредственная работа с многоэлектронными X-операторами затруднительна (в частности, отсутствуют обычные разложения теории возмущений и диаграммная техника, хотя могут быть использованы формальные разложения п обратному координационному числу $1/z$ [44]). Поэтому часто используются их представления через вспомогательные ("auxiliary", "slave" – слейв) бозевские и фермиевские операторы.

Исходя из конкретной физической задачи и ситуации, используются различные версии таких представлений. В связи с теорией двумерных высокотемпературных сверхпроводников Андерсон [1] предложил представление

$$c^\dagger_{i\sigma} = f^\dagger_{i\sigma}e_i + \sigma d^\dagger_i f_{i-\sigma}. \qquad (28)$$

Выражение (28) соответствует разбиению одноэлектронного оператора на два многоэлектронных X-оператора (17), т.е. формированию хаббардовских подзон. Здесь $f^\dagger_{i\sigma}$ — операторы рождения для нейтральных фермионов (спинонов), $e^\dagger_i$, $d^\dagger_i$ — операторы рождения для заряженных бесспиновых бозонов. Физический смысл таких возбуждений можно объяснить следующим образом. Рассмотрим решетку с одним электроном на узле с сильным хаббардовским отталкиванием, так что каждый узел нейтрален. В основном состоянии RVB каждый узел принимает участие в одной связи. Когда связь нарушается, появляются два неспаренных узла, которые обладают спинами, равными 1/2. Соответствующие возбуждения (спиноны) не заряжены.

Вместе с тем, пустой узел (дырка) в системе несет заряд, но не спин. Такая картина описывает состояние спиновой жидкости, где имеет место разделение спиновых и зарядовых степеней свободы электрона и существенную роль играют калибровочные поля.

В работе [61] было предложено расширение физического пространства через введение четырех бозонов $p_{i\sigma}$, $e_i$, $d_i$, которые осуществляют проектирование на однократно занятые состояния, дырки и двойки соответственно. В результате гамильтониан Хаббарда принимает вид

$$\mathcal{H} = \sum_{ij\sigma} t_{ij} f_{i\sigma}^\dagger f_{j\sigma} z_{i\sigma}^\dagger z_{j\sigma} + U \sum_i d_i^\dagger d_i, \qquad (29)$$

где $f_{i\sigma}, f_{i\sigma}^\dagger$ – вспомогательные ферми-операторы,

$$z_{i\sigma}^\dagger = (1 - e_i^\dagger e_i - p_{i-\sigma}^\dagger p_{i-\sigma})^{-1/2} (p_{i\sigma}^\dagger e_i + d_i^\dagger p_{i-\sigma})(1 - d_i^\dagger d_i - p_{i\sigma}^\dagger p_{i\sigma})^{-1/2}, \qquad (30)$$

причем накладываются дополнительные ограничения

$$\sum_\sigma p_{i\sigma}^\dagger p_{i\sigma} + d_i^\dagger d_i + d_i^\dagger d_i = 1, \quad f_{i\sigma}^\dagger f_{i\sigma} = p_{i\sigma}^\dagger p_{i\sigma} + d_i^\dagger d_i.$$

Здесь имеется определенный произвол; дополнительные операторные множители имеют собственные значения 1 в физическом подпространстве, но при расцеплении позволяют перейти к пределу свободных электронов, что важно для построения теории среднего поля. Это представление позволило качественно воспроизвести ряд прежних результатов (например, как и квазиклассическое приближение (18), получить описание перехода металл–изолятор [62]). Теория среднего поля для представления бозонов Котляра–Рукенштейна (29) дает результаты, идентичные теории Гутцвиллера в пределе бесконечного координационного числа. В рамках этого представления может быть рассмотрено и магнитное упорядочение [63].

Существует также вращательно-инвариантная версия [64]

$$c_{i\sigma}^\dagger = \sum_{\sigma'} f_{i\sigma}^\dagger z_{i\sigma\sigma'}^\dagger, \ \hat{z}_i = e_i^\dagger \hat{L}_i M_i \hat{R}_i \hat{p}_i + \widehat{\widetilde{p}}_i^\dagger \hat{L}_i M_i \hat{R}_i d_i, \qquad (31)$$

где множители $L$, $M$, $R$ аналогичны (30), скалярное и векторное вспомогательные бозонные поля $p_{i0}$ и $\mathbf{p}_i$ введены как $\hat{p}_i = \frac{1}{2}(p_{i0}\sigma_0 + \mathbf{p}_i\boldsymbol{\sigma})$ и $\widetilde{\widehat{p}}_i$ – обращенный по времени оператор $\hat{p}_i$. Эта версия удобна в случае магнитоупорядоченных фаз, позволяя учесть поправки от спиновых флуктуаций. В частности, она может просто описывать неквазичастичные состояния, обусловленные электрон-магнонным рассеянием, которые ранее рассматривались в многоэлектронном представлении операторов Хаббарда (см. [65]); при этом операторы $p^-$ описывают магноны, а $f$ – спиноны.

Отметим также представление вспомогательных роторов (slave rotor) [66]

$$c_{i\sigma}^\dagger = f_{i\sigma}^\dagger b_i = f_{i\sigma}^\dagger \exp(i\theta_i), \qquad (32)$$

в котором исходное взаимодействие Хаббарда заменяется простым кинетическим членом для фазового поля $(U/2)\hat{L}_i^2$ с моментом количества движения $\hat{L} = -i\partial/\partial\theta$.

Это представление удобно для описания перехода металл-диэлектрик в парамагнитной фазе, а также было обобщено на случай допирования [67, 68, 69].

В работе [70] было предложено представление, содержащее два сорта фермиевских операторов – голонов $e_i$ и дублонов $d_i$, которые соответствуют дыркам и двойкам:

$$\begin{aligned} X_i(+,0) &= e_i(1 - d_i^\dagger d_i)(1/2 + s_i^z), \ X_i(-,0) = e_i(1 - d_i^\dagger d_i)s_i^-, \\ X(2,-) &= d_i^\dagger(1 - d_i^\dagger d_i)s_i^+, \ X(2,+) = d_i^\dagger(1 - d_i^\dagger d_i)(1/2 + s_i^z). \end{aligned} \quad (33)$$

При этом операторы физических спинов связаны с псевдоспиновыми операторами $s_i^\alpha$ соотношением $\mathbf{S}_i = \mathbf{s}_i(1 - d_i^\dagger d_i - e_i^\dagger e_i)$. Ранее рассматривались различные частные случаи этого представления, соответствующие пределу больших $U$ ($t-J$ модели, см. обзор [71]).

В дальнейшем были предложены также суперсимметричные представления [72].

## 2.4 $t$—$J$ модель и $s$—$d$ обменная модель с сильными корреляциями

При рассмотрении сильнокоррелированных соединений, например, медь-кислородных высокотемпературных сверхпроводников, широко используется $t$—$J$ модель (модель Хаббарда с $U \to \infty$ и учетом гейзенберговского обмена). Ее гамильтониан в МЭ представлении имеет вид

$$\mathcal{H} = -\sum_{ij\sigma} t_{ij} X_i(0,\sigma) X_j(\sigma,0) + \sum_{ij} J_{ij} \mathbf{S}_i \mathbf{S}_j. \quad (34)$$

При выводе $t - J$ модели из модели Хаббарда с большим $U$ устранение двоек каноническим преобразованием [73] приводит к андерсоновскому антиферромагнитному сверхобмену $J = -2t^2/U$; впрочем, иногда удобно считать $J$ независимой переменной. В простой модели (34) возникает богатая фазовая диаграмма, включающая спиральные магнитные структуры и неоднородные состояния (см. обзор [71]).

По утверждению Андерсона [74], $t - J$ гамильтониан – не просто удобная альтернатива модели Хаббарда; он отражает физический факт, что низкоэнергетические состояния находятся в подпространстве, которое является переполненным и описывается одной полосой электронных состояний, поскольку антисвязанные состояния двоек из верхней части зоны выброшены. Не существует такого сходящегося пертурбативного метода, которые бы связал низкоэнергетические состояния с первоначальной зоной модели Хаббарда, поскольку они существуют в гильбертовых пространствах разных размерностей.

Первопринципной моделью для описания купратов является трехзонная $p - d$ модель

$$\begin{aligned} \mathcal{H} &= \sum_{\mathbf{k}a\sigma} \left[ \varepsilon p_{\mathbf{k}a\sigma}^\dagger p_{\mathbf{k}a\sigma} + \Delta d_{\mathbf{k}\sigma}^\dagger d_{\mathbf{k}\sigma} + V_\mathbf{k}(p_{\mathbf{k}\sigma}^\dagger d_{\mathbf{k}\sigma} + h.c.) \right] \\ &+ U \sum_i d_{i\uparrow}^\dagger d_{i\uparrow} d_{i\downarrow}^\dagger d_{i\downarrow}, \end{aligned} \quad (35)$$

где $\varepsilon$ и $\Delta$ – положения $p$- и $d$-уровней для O- и Cu-ионов, $V_{\mathbf{k}} = 2V_{pd}\left(\sin^2 k_x + \sin^2 k_y\right)^{1/2}$ – матричные элементы $p$–$d$ гибридизации (ср. [75]). В пределе больших $U$ мы можем использовать представление слейв-бозонов $d_{i\sigma}^\dagger \to X_i(\sigma 0) = f_{i\sigma}^\dagger e_i$.

При $|V_{pd}| \ll \varepsilon - \Delta$ (большая щель для переноса заряда) гамильтониан (35) снова сводится каноническим преобразованием к $t-J$ модели с $t_{\text{eff}} = V_{pd}^2/(\varepsilon - \Delta)$. Интересно, что $t-J$ модель, полученная из однозонной модели Хаббарда, также формально сводится к аналогичной структуре в представлении (30) со вспомогательными, а не физическими частицами $p_i$. Таким образом, модель Хаббарда и модель (35) могут рассматриваться параллельным образом [75].

Модель (34) описывает взаимодействие носителей тока с локальными моментами. Для того, чтобы продемонстрировать явно разделение этих степеней свободы, укажем на ее эквивалентность $s-d$ обменной модели с параметром обмена $|I| \to \infty$. Действительно, после перехода к МЭ представлению гамильтониан последней запишется в виде [76, 77]

$$\mathcal{H} = \sum_{ij\sigma} t_{ij} g_{i\sigma\alpha}^\dagger g_{j\sigma\alpha} + \mathcal{H}_d, \quad \alpha = \text{sign} I, \tag{36}$$

где

$$g_{i\sigma\alpha}^\dagger = \sum_M \{[S + \sigma\alpha M + (1+\alpha)/2]/(2S+1)\}^{1/2} X_i\left(M + \frac{\sigma}{2}, \alpha; M\right). \tag{37}$$

Модель (36) соответствует случаю, когда носители тока не принадлежат той же энергетической зоне, где формируются магнитные моменты. Такая ситуация имеет место в некоторых магнитных полупроводниках и изоляторах. Однако легко видеть, что при $S = 1/2$, $\alpha = -$ гамильтониан (36) совпадает с (34), причем $t_{ij}$ заменяется в (36) на $2t_{ij}$ (множитель 2 возникает из-за эквивалентности перескоков электронов с обоими противоположными спинами в модели Хаббарда).

Гамильтониан (36) может быть выражен через фермиевские операторы и операторы локализованных спинов. Используя (13), (11), можно выделить операторы электронов проводимости из $X$-операторов:

$$g_{i\sigma\alpha}^\dagger = \sum_{\sigma'} c_{i\sigma'}^\dagger (1 - n_{i,-\sigma'}) \left[P_\alpha \delta_{\sigma\sigma'} + \frac{\alpha}{2S+1}(\mathbf{S}_i \boldsymbol{\sigma}_{\sigma'\sigma})\right], \tag{38}$$

$$P_+ = \frac{S+1}{2S+1}, \qquad P_- = \frac{S}{2S+1}. \tag{39}$$

Далее, использование свойств матриц Паули дает

$$\mathcal{H} = \sum_{ij\sigma\sigma'} t_{ij} \left\{ \left[P_\alpha^2 + \frac{\mathbf{S}_i \mathbf{S}_j}{(2S+1)^2}\right] \delta_{\sigma\sigma'} + \frac{\alpha}{(2S+1)^2} P_\alpha (\mathbf{S}_i + \mathbf{S}_j) \boldsymbol{\sigma}_{\sigma\sigma'} + \right.$$
$$\left. + \frac{i}{(2S+1)^2} \boldsymbol{\sigma}_{\sigma\sigma'} [\mathbf{S}_i \times \mathbf{S}_j] \right\} c_{i\sigma}^\dagger (1-\hat{n}_{i,-\sigma})(1-\hat{n}_{j,-\sigma'}) c_{j\sigma'} + \mathcal{H}_d. \tag{40}$$

Для спиновых операторов имеем

$$\mathbf{S}_i = \frac{1}{2} \sum_{\sigma\sigma'} a_{i\sigma}^\dagger \boldsymbol{\sigma}_{\sigma\sigma'} a_{i\sigma'}, \tag{41}$$

15

причем, в зависимости от задачи, можно использовать представление швингеровских бозонов ($a_{i\sigma}^\dagger = b_{i\sigma}^\dagger$) или фермиевских спинонов ($a_{i\sigma}^\dagger = f_{i\sigma}^\dagger$). Члены в (40), линейные по спиновым операторам, обеспечивают возможность эффективной гибридизации электронов со спинонами.

Представление гамильтониана (40) (разумеется, оно справедливо и в $t-J$ модели) было получено в [43, 44]. В дальнейшем оно было использовано для построения особой теории среднего поля в применении к фазовой диаграмме ВТСП-купратов как "новая формулировка $t-J$ модели" [78]. В теории среднего поля для $t-J$ модели этих авторов электронный спектральный вес происходит из двух зон: низкоэнергетической спинонной полосы и высокоэнергетической электронной зоны. Спектральный вес от спинонной зоны представляет собой острый когерентный пик. Широкий спектральный вес от электронной зоны соответствует некогерентному фону.

При рассмотрении спектра купратов возникает нодально-антинодальная дихотомия [32]: природа спектра различна на разных участках поверхности Ферми. Спектр является бесщелевым вблизи нодальных точек ($\pm\pi/2, \pm\pi/2$) (где возбуждения описываются как дираковские фермионы, см. раздел 3.3) и щелевым вблизи антинодальной точки $(0, \pi)$. Вблизи нодальной точки возникает сильное гибридизационное перемешивание между спинонами и электронами (допонами в терминологии [78]), а вблизи антинодальной точки смешивание отсутствует. По существу описанная картина близка к гибридизационной двухзонной модели решеток Кондо, где локализованные и коллективизированные состояния исходно разделены, но смешиваются в приближении среднего поля для вспомогательных f-псевдофермионов, так что последние принимают участие в формировании поверхности Ферми [79] (см. также раздел 3.5).

Таким образом, использование концепции двухзонной модели помогает получить физическую картину экзотических состояний сильно коррелированных систем.

Члены, содержащие векторные произведения, описывают анизотропное рассеяние электронов и могут быть важны при рассмотрении кинетических явлений в узких зонах, например аномального эффекта Холла. Гамильтониан (40) может быть полезен при исследовании состояний с аномальными "киральными" параметрами порядка, которые исследовались в рамках двумерной модели Гейзенберга и $t-J$ модели (см., напр., [80]).

# 3 Ферми-жидкость и неферми-жидкостные фазы

Концепция ферми-жидкости предполагает взаимно однозначное соответствие низкоэнергетических состояний между рассматриваемой взаимодействующей системой и системой невзаимодействующих электронов. Эта подразумевает, в частности, существование только квазичастичных возбуждений с зарядом $\pm e$ и спином $1/2$. Отсюда следует существование поверхности Ферми, определяемой полюсами одночастичной функции Грина при энергии $E = 0$. Эта поверхность Ферми подчиняется теореме Латтинджера, т. е. окружает объем импульсного пространства, определяемый полной плотностью электронов.

Нарушение ферми-жидкостного поведения при низких температурах может иметь различные источники. В чистых системах эффекты взаимодействия могут

давать стабильные неферми-жидкостные фазы. Один из сценариев состоит в том, что низкоэнергетические возбуждения приобретают квантовые числа, отличные от обычных электронов или дырок, что приводит к необычным низкотемпературным свойства. В размерности $d = 1$ такое поведение носит общий характер и хорошо понято (латтинджеровские жидкости с разделением спина и заряда). Аналогичное разделение ожидается в спиновых жидкостях в более высоких размерностях.

В случае полузаполненной полосы при уменьшении кулоновского отталкивания моттовский изолятор переходит в металл. Вблизи этого перехода металлическое состояние будет сильно коррелированным, включая хаббардовские подзоны и локальные моменты, которые подвержены фрустрации. Совместное действие этих факторов может приводит к значительно уменьшенной шкале когерентности, а следовательно к тяжелофермионному поведению. Для рассмотрения перехода металл-диэлектрик удобно использовать роторное представление (32) электронного оператора уничтожения как произведение заряженного бозона $b_i$ и нейтрального фермиона со спином (спинона) $f_{i\sigma}$ [81, 82]. При нечетном заполнении зоны система бесспиновых бозонов переходит с увеличением хаббардовского параметра $U$ в сверхтекучий моттовский изолятор. В приближении среднего поля спиноны являются свободными (невзаимодействующими), несмотря на сильные корреляции в электронной системе. Если бозон $b_i$ конденсируется, т.е. $\langle b \rangle \neq 0$, мы получаем фазу ферми-жидкости физических электронов: при замене $b$ на его $c$-числовое среднее $\langle b \rangle$ фермионы $f_\sigma$ приобретают те же квантовые числа, что и исходные электроны, так что их ферми-поверхность описывает обычный металл. Если бозон является щелевым и, следовательно, несконденсирован, образуется спин-жидкостной изолятор Мотта, в котором выживает поверхность Ферми нейтральных фермионных возбуждений (спинонов). Изолятор Мотта для бозонов также является диэлектрическим состоянием для электронов с щелью для всех заряженных возбуждений, и происходит непрерывный переход к изолятору с "призрачной" (ghost) поверхностью Ферми спинонов.

Аналогичный сценарий перехода металл-изолятор может быть рассмотрен в представлении бозонов Котляра-Рукенштейна (30) [62, 83]. При этом, когда $\langle b \rangle \neq 0$ или мало (квазичастичный вычет мал), недостающий спектральный вес определяется флуктуациями – некогерентным вкладом хаббардовских подзон [83].

Помимо уменьшения взаимодействия, еще один способ превратить фрустрированный изолятор Мотта в металл – допирование. При малых уровнях допирования имеются локальные моменты, и, следовательно, важными могут быть эффекты фрустрации. Поэтому были высказаны ряд идей о возникновении необычных металлических фаз с неферми-жидкостным поведением; в частности, такой металл может быть фракционализованным состоянием со спинонными и голонными возбуждениями (раздел 3.6).

Можно предположить, что вдали от половинного заполнения зоны, особенно при малых плотностях электронов, возникает ферми-жидкостное состояние, поскольку акты рассеяния частиц становятся редкими, так что эффекты даже сильного кулоновского взаимодействия $U$ слабы. Однако ситуация не столь проста. Действительно, согласно спектральному представлению функций Грина имеем для числа двоек,

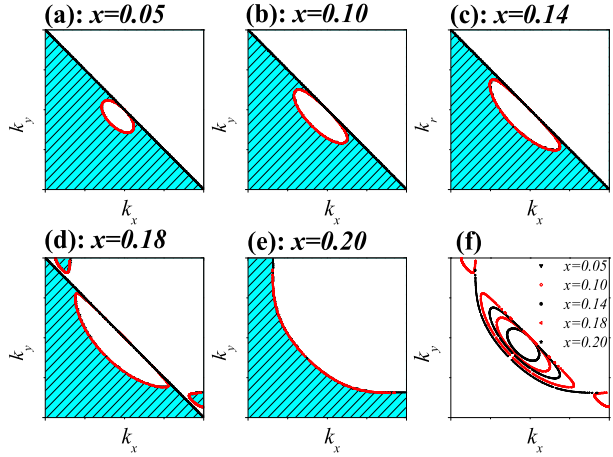

Рис. 1: Эволюция поверхности Ферми с концентрацией дырок $x$ в модели [84]. В затененной области $G(\mathbf{k}, 0) > 0$. Толстые линии на (a) – (d) являются линиями нулей Латтинджера; вблизи критического допирования $x_c = 0.18$ (d) имеется дополнительный электронный карман.

определяющего локальный момент [47]:

$$N_2 = \langle c^\dagger_{i\uparrow} c_{i\uparrow} c^\dagger_{i\downarrow} c_{i\downarrow} \rangle = -\frac{1}{\pi U} \int \operatorname{Im} \sum_{\mathbf{k}} \frac{\Sigma_{\mathbf{k}}(E)}{E - t_{\mathbf{k}}} f(E)\, dE. \qquad (42)$$

Поскольку собственная энергия $\Sigma_{\mathbf{k}}(E)$ конечна в пределе $U \to \infty$ (в частности, в T-матричном приближении Канамори), $N_2$ ведет себя как $1/U$. Тогда теорема Гелмана–Фейнмана $N_2 = \partial \mathcal{E}/\partial U$ дает расходимость энергии основного состояния $\mathcal{E}$:

$$\mathcal{E}(U) - \mathcal{E}(0) = \int\limits_0^\infty N_2(U)\, dU \sim \ln U. \qquad (43)$$

Эта расходимость указывает на неадекватность одноэлектронной картины ферми-жидкости при больших $U$ и формирование хаббардовских подзон. С другой стороны, расчет в многоэлектронном представление X-операторов [47] дает правильную асимптотику $N_2 \sim 1/U^2$. Верный результат получается и в представлении бозонов Котляра-Рукенштейна (29).

Авторы [84] предложили феноменологическое представление для одноэлектронной функции Грина, которая удовлетворяет теореме Латтинджера по-другому: электронная плотность связана не с объемом поверхности Ферми, а с числом особенностей логарифма функции Грина при нулевой энергии. Таким образом, учитываются как обычные полюса, так и нули функции Грина (т.е. квазичастичные полюса электронной собственной энергии); последняя поверхность называется латтинджеровской.

Из феноменологической формы функции Грина получаются дисперсия квазичастиц и спектральный вес, характеризующие когерентную часть одноэлектрон-

ной функции Грина. При этом существует критическая концентрация дырок, выше которой спин-жидкостная аномальная собственная энергия исчезает. Как следствие, существует квантовая критическая точка, которая разделяет две топологически различные фазы. Ниже критической плотности дырок $G(\mathbf{k}, \mathbf{0})$ характеризуется сосуществующими поверхностью Латтинджера и дырочными карманами ферми-поверхности, а выше ее имеется только замкнутая поверхность Ферми, как в теории Ландау.

Как показано на рис.1, в узком диапазоне $x \lesssim x_c$ возникает новый набор электронных ферми-карманов вблизи седловых точек, вне латтинджеровской поверхности нулей. Эти карманы сливаются с дырочными карманами внутри поверхности Латтинджера при $x = x_c$, чтобы образовать единую поверхности Ферми. Таким образом, имеется два близко расположенных топологических изменения: переход типа Лифшица, при котором форма ферми-поверхность полюсов претерпевает топологические изменения, и квантовая критическая точка $x = x_c$, связанная с открытием одночастичной щели при возникновении RVB-спиновой жидкости для $x < x_c$, что ведет к образованию поверхности Латтинджера.

Форма функции Грина [84] близка к результату [85], полученному на основе слейв-бозонной средней теории поля в $t - J$ модели. В этом подходе спиноны и голоны взаимодействуют друг с другом через калибровочное поле, а для описания калибровочных флуктуаций вводится феноменологическое притяжение между ними, который является константой в некотором диапазоне расстояний. Такое притяжение и приводит к связыванию спинона и голона, а следовательно к квазичастичному полюсу собственной энергии.

## 3.1 Теория среднего поля и калибровочные поля

Обсудим картину спиновой жидкости в рамках представления (28). Для наполовину заполненной зоны присутствуют только спинонные возбуждения с кинетической энергией порядка $|J|$. При допировании системы дырками возникают носители заряда, которые описываются операторами голонов. Таким образом, возникает состояние спиновой жидкости с подавленным дальним магнитным порядком. При этом в простейшем бесщелевом варианте теории RVB в чисто спиновых (недопированных) системах появляются малый энергетический масштаб $J$ и большой линейный член в удельной теплоемкости с $\gamma \sim 1/|J|$, связанный с существованием ферми-поверхности спинонов [1].

При дальнейшем развитии теории было предложен другой вариант теории RVB, в котором на поверхности Ферми возникает щель, так что образуется структура типа d-волны с максимумом щели вблизи $(0, \pi)$; это согласуется с фотоэмиссионными экспериментами (см. [2]). Были развиты и более сложные версии теории RVB, которые используют топологическое рассмотрение и аналогии с дробным квантовым эффектом Холла (см., напр., [86]). Эти идеи привели к необычным и красивым результатам. В частности, спиноны могут подчиняться дробной статистике, т. е. волновая функция системы приобретает комплексный коэффициент при перестановке двух квазичастиц.

Рассмотрим приближение среднего поля в $t - J$ модели в представлении вспомогательных бозонов $c_{i\sigma}^\dagger \to X_i(\sigma 0) = f_{i\sigma}^\dagger b_i$, соответствующее теории U(1). Здесь можно

19

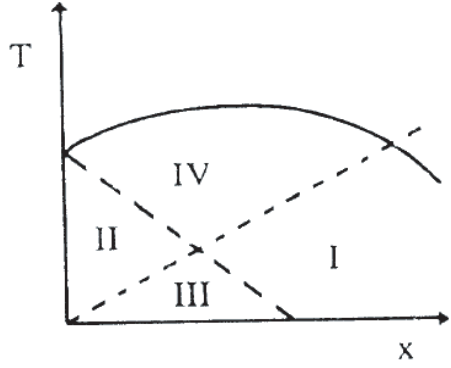

Рис. 2: Схематическая фазовая диаграмма теории среднего поля U(1). Сплошная линия обозначает начало однородного состояния RVB ($\chi \neq 0$), пунктирная линия – возникновение фермионного спаривания ($\Delta \neq 0$), а короткий пунктир – бозе-конденсацию в подходе среднего поля ($b \neq 0$). Возникает четыре области: (I) ферми-жидкость с $\chi \neq 0$, $b \neq 0$; (II) состояние спиновой щели с $\chi \neq 0$, $\Delta \neq 0$; (III) $d$-волновой сверхпроводник с $\chi \neq 0$, $\Delta \neq 0$, $b \neq 0$ и (IV) странный металл с $\chi \neq 0$ [87].

ввести спаривания

$$\chi_{ij} = \sum_\sigma \langle f^\dagger_{i\sigma} f_{j\sigma}\rangle, \ \Delta_{ij} = \langle f_{i\uparrow} f_{j\downarrow} - f_{i\downarrow} f_{i\uparrow}\rangle \tag{44}$$

В однородной RVB-фазе (uRVB) $\chi_{ij} = \chi$ для всех связей и вещественно, а щель $\Delta_{ij} = 0$, так что спектр $f$-фермионов имеет вид $E_\mathbf{k} = -2J\chi(\cos k_x + \cos k_y)$. Однако имеются фазы с более низкой энергией, включая $d$-волновой сверхпроводник. Фазовая диаграмма с учетом последнего состояния показана на рис. 2.

Следует еще раз отметить, что описание ферми-жидкостной фазы в $t-J$ модели является в некотором смысле внутренне противоречивым, поскольку часть состояний уже исключена благодаря проектированию на однократно занятые состояния. В этом смысле более последователен подход, учитывающий формирование двоек и включающий оба вспомогательных бозона $e$ и $d$ в рамках представлений (28) либо (29). Последнее представление позволяет также учесть формирование верхней и нижней хаббардовской подзоны как некогерентного вклада – связанного состояния спинона и голона [83].

Более сложные спин-жидкостные фазы получаются при учете SU(2)-инвариантности $t-J$ гамильтониана, что позволяет устранить ряд трудностей теории U(1) [32]. SU(2) слейв-бозонная теория, описывающая допированные спиновые жидкости, включает два SU(2)-дублета $\psi_i$ и $h_i$ с двумя бозонными полями $b_{1i}$ и $b_{2i}$ со спином 0, которые формируют SU(2) синглеты. Тогда имеем представление электронного оператора

$$\begin{aligned} c_{\uparrow i} &\to \frac{1}{\sqrt{2}} h_i^\dagger \psi_i = \frac{1}{\sqrt{2}}\left(b_{1i}^\dagger f_{\uparrow i} + b_{2i}^\dagger f_{\downarrow i}^\dagger\right) \\ c_{\downarrow i} &\to \frac{1}{\sqrt{2}} h_i^\dagger \bar\psi_i = \frac{1}{\sqrt{2}}\left(b_{1i}^\dagger f_{\downarrow i} - b_{2i}^\dagger f_{\uparrow i}^\dagger\right) \end{aligned} \tag{45}$$

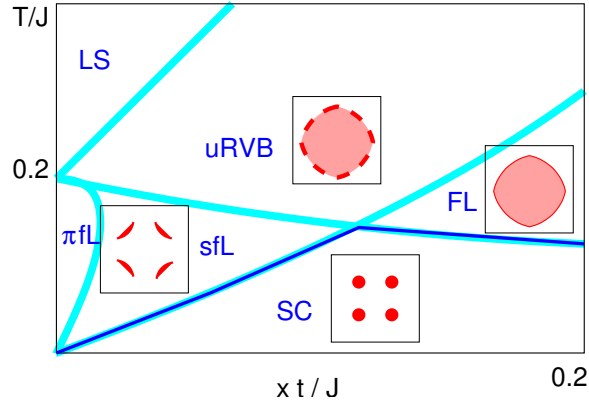

Рис. 3: SU(2) фазовая диаграмма среднего поля для $t/J = 1$ [87]. Показаны поверхность Ферми, дуги Ферми или точки Ферми в различных фазах.

В состоянии sfL (staggered flux liquid) получаем

$$E_{\mathbf{k}} = \pm \frac{3}{4} J \sqrt{\chi^2 (\cos k_x + \cos k_y)^2 + \Delta^2 (\cos k_x - \cos k_y)^2}.$$

Эта фаза имеет дираковский линейный спектр в точках $(\pm\pi/2, \pm\pi/2)$. В фазе πfL

$$E_{\mathbf{k}} = \pm \frac{3}{4} J |\chi| \sqrt{\sin^2 k_x + \sin^2 k_y}$$

Помимо этих двух бесщелевых фаз, может возникать также щелевое $Z_2$-состояние.

Фазовая диаграмма в SU(2) слэйв-бозонной теории допированного моттовского изолятора показана на Рис. 3. Эта теория также не свободна от трудностей. В частности, площадь поверхности Ферми приближается к $1 - x$ очень медленно с ростом допирования $x$ и понижением температуры.

Интуитивная картина спиновой жидкости с симметрией SU(2) как раз дается андерсоновским состоянием RVB, которое включает синглетное спаривание спинов и формирование квантовой суперпозиции многих различных пар, т. е. различные димерные покрытия решетки, причем симметрия гамильтониана сохраняется. Эта картина включает дробные возбуждения, поскольку разрушение димера приводит к двум возбуждениям с независимой динамикой: получающиеся мономеры представляют собой объекты, несущие заряд 0 и спин $1/2$, то есть спиноны. В спиновой жидкости $Z_2$ они связаны с возникающими $Z_2$ возбуждениями, которые представляют собой $Z_2$-вихри (или потоки), называемые визонами (см. также раздел 5.1).

Волновая функция $|\Psi_{\mathrm{MF}}(\chi_{ij})\rangle$ гамильтониана среднего поля

$$\mathcal{H}_{\mathrm{MF}} = J \sum_{ij\sigma} (\chi_{ij} f_{i\sigma}^\dagger f_{j\sigma} + h.c.) + \sum_{i\sigma} a_0 (f_{i\sigma}^\dagger f_{i\sigma} - 1), \qquad (46)$$

не является вполне удовлетворительной функцией спиновой системы, поскольку удовлетворяет условию однократного заполнения узлов лишь в среднем. Чтобы учесть его точнее, необходимо ввести U(1)-калибровочные флуктуации фазы величины $a_0$, соответствующие флуктуациям киральности, $\chi_{ij} \to \bar{\chi}_{ij} \exp(-i a_{ij})$. Выполняя проектирование на однократного занятые состояния, получаем физическую волновую

21

функцию [32]
$$\Psi_{\text{spin}}(\chi_{ij}, \{\sigma_i\}) = \langle 0_f | \prod_i f_{i\sigma_i} | \Psi_{\text{MF}}(\chi_{ij}) \rangle. \qquad (47)$$

где $|0_f\rangle$ – состояние без $f$-фермионов: $f_{\sigma_i}|0_f\rangle = 0$. Данная конструкция позволяет явно построить волновую функцию, соответствующую калибровочным флуктуациям $a_{ij}$:

$$\Psi_{\text{spin}}(a_{ij}) = \langle 0_f | \prod_i f_{i\sigma_i} | \Psi_{\text{MF}}(\bar{\chi}_{ij} \exp{(i a_{ij})}) \rangle, \qquad (48)$$

так что естественным образом возникает калибровочное поле.

Обсудим топологическую природу квазичастичных возбуждений и калибровочного поля в киральном спиновом состоянии [3, 32]. Простейшее возбуждение в приближении среднего поля может быть получено путем добавления спинона в зону проводимости. Однако это возбуждение еще не является физическим, поскольку спинон в зоне проводимости нарушает ограничение $\sum_\sigma f_{i\sigma}^\dagger f_{i\sigma} = 1$. Дополнительную плотность спинонов можно устранить, вводя вихревой поток калибровочного поля

$$\Phi = -\pi \sum_{i\sigma} (\langle f_{i\sigma}^\dagger f_{i\sigma} \rangle - 1/2).$$

Поэтому физические квазичастицы в киральном спиновом состоянии являются спинонами, одетыми $\pi$-вихрем. Одетые спиноны несут спин $1/2$, поскольку вихрь не может индуцировать произвольные спиновые квантовые числа. Однако спинон, который несет единичный заряд калибровочного поля, имеет дробную статистику, будучи связанным состоянием заряда и вихря. Уровни Ландау, возникающие в "электромагнитном" калибровочном поле, соответствуют хаббардовским подзонам.

В целом теория среднего поля оказывается чрезвычайно богатой. В рамках калибровочной теории возможно возникновение большого числа спиновых жидкостей – как щелевых, так и бесщелевых. Кроме спин-бозонного представления, описывающего фазу деконфайнмента с бесщелевыми спинонами, может быть использовано спин-фермионное представление, которое однако описывает только щелевые фазы [3].

## 3.2 Конфайнмент и деконфайнмент

Формально представление (28) справедливо всегда, однако в реальности картина спинонов и голонов может реализоваться лишь при определенных условиях (например, в точно решаемых моделях типа Китаева [88, 89]). Здесь принципиален вопрос о конфайнменте и деконфайнменте [3, 32]. В фазе конфайнмента вся картина спиновой жидкости разрушается при учете флуктуационных эффектов вследствие взаимодействия спинона с калибровочным бозоном: размножаются топологические дефекты, описываемые соответствующими гомотопическими группами [26]. В результате спиноны взаимодействуют друг с другом через линейно возрастающий с расстоянием потенциал, а для бозонов имеется большая энергетическая щель, так что все эти экзотические частицы никак не проявляются в низкоэнергетическом спектре. В итоге спиноны снова "склеиваются" с бозонами в обычные электроны и мы возвращаемся к тривиальному состоянию обычной ферми-жидкости – эффективной одночастичной

теории. Таким образом, используемый сложный формализм согласуется с привычными физическими представлениями одноэлектронной теории [3].

Взаимодействие калибровочных полей на классическом уровне не приводит к конфайнменту. Однако, если мы выйдем за пределы классического приближения и рассмотрим взаимодействие между калибровочными флуктуациями, картина полностью изменится. В размерности $2+1$ такое взаимодействие трансформирует потенциал $g \ln R$ в линейный, независимо от константы связи $g$. Таким образом, среднеполевые SU(2)-состояния с щелевыми фермионами нестабильны, а теории среднего поля нельзя доверять.

Иная картина возникает в фазе деконфайнмента (бесщелевой топологической фазе). Здесь топологические дефекты отсутствуют, калибровочное поле некомпактно, сохраняются поток электрического поля и топологический заряд. В результате спиноны и калибровочные бозоны являются хорошо определенными частицами, возникают струны электрического потока и топологический порядок (струнные сети), что означает многочастичное квантовое запутывание.

Взаимодействия и механизмы, ведущие к деконфайнменту, могут быть разными: фрустрации обменных параметров, эффект Кондо (подавление магнитных моментов электронами проводимости), туннелирование, псевдоспины... Часто их описание можно свести к эффективной изотропной или анизотропной модели Гейзенберга [90].

В качестве параметра порядка для перехода конфайнмент–деконфайнмент можно использовать петлю Вильсона (см. [32]). Она связана с калибровочным потенциалом $V(R)$ между двумя статическими калибровочными зарядами $\pm q$ с противоположным знаком, помещенными на расстояние $R$ как $W(C) = \exp[-V(R)T]$, причем петля $C$ состоит из путей длины $T$ по времени направления и длины $R$ вдоль пространственного направления. Существует два типа поведения $W(C)$: закон площади с $W(C) \sim \exp[-\alpha RT]$ и закон периметра с $W(C) \sim \exp[-\beta(R+T)]$, где $\alpha$ и $\beta$ являются константами. В первом случае потенциал $V(R)$ линейно возрастает в $R$, и, следовательно, два калибровочных заряда не могут быть свободны. Поэтому он соответствует конфайнменту в то время как второй случай – деконфайнменту.

Компактная квантовая электродинамика (чистая калибровочная модель) в размерности $2+1$ конфайнментна при сколь угодно малой константе связи из-за возникновения инстантонов [91]. При этом инстантон ведет себя как положительный магнитный заряд, а антиинстантон – как отрицательный. Кулоновский газ в $2+1$ измерениях всегда находится в фазе экранирования: кулоновское взаимодействие становится короткодействующим из-за экранирования заряда облаком окружающих противоположных зарядов заряд. Поэтому энергия рождения инстантона конечна и освобождаются магнитные заряды, которые разупорядочивают калибровочное поле, что приводит к закону площадей и конфайнменту. Однако конфайнмент может исчезать выше некоторой температуры перехода.

При наличии материальных полей вопрос о конфайнменте еще более усложняется. Пары частиц-античастиц могут возбуждаться из вакуума, экранируя статические заряды. При конечных температура переход к деконфайнменту заменяется постепенным кроссовером в плазменную фазу. Поэтому можно ожидать, что слейв-бозонная теория среднего поля без конфайнмента описывает физику на промежуточной энергетической шкале, хотя основное состояние является конфайнментным (ан-

тиферромагнитным, сверхпроводящим, либо ферми-жидкостным). Примером может служить "деконфайнментное" псевдощелевое состояние, обсуждаемое для ВТСП-систем, которое существует только при конечных температурах и описывается фермионами и бозонами, связанными с некомпактными калибровочными полями.

Еще один способ получения фазы деконфайнмента – выход в высшие измерения. В размерности 3 + 1 бесщелевые U(1) флуктуации не создают конфайнментных взаимодействий. В размерности 4 + 1 и выше даже неабелева калибровочная теории может быть в фазе деконфайнмента. Таким образом, можно построить бозонные модели на кубической решетке, которые имеют бесщелевые фотоны – U(1) калибровочные бозоны [32].

Явление деконфайнмента в последнее время активно изучается в расчетах квантовым методом Монте Карло для специфических моделей типа Китаева, включающих фрустрацию [92]; эти расчеты достаточно надежны, поскольку, в отличие от стандартных фермионных задач, проблема знака отсутствует. Картина деконфайнмента может быть достаточно сложной: вовлеченными в него оказываются не все частицы. Размер спинонов получается достаточно большим – сравнимым с размером их связанного состояния, что отличается от "классического" RVB с короткими связями.

Деконфайнмент в двумерных системах также исследуется экспериментально для металлоорганических соединений с квадратной решеткой. Он проявляется как возникновение спинонов в высокоэнергетическом спектре возбуждений даже в отсутствие фрустраций [93]. При удалении от соответствующего аномального волнового вектора эти дробные частицы связываются, образуя обычные магноны.

## 3.3 Дираковские фермионы и алгебраическая спиновая жидкость

Существует еще один путь к деконфайнментному состоянию – взаимодействие компактного калибровочного поля U(1) с бесщелевыми фермионами, в частности с дираковскими, возникающим в фазе sfL. В этом случае коррелированное металлическое состояние представляет собой полуметалл, содержащий бесщелевые электронные возбуждения только в изолированных фермиевских точках зоны Бриллюэна, что и определяет особую топологию проблемы [16]. Эти точки суть поверхность Ферми электронов. Такая модель может быть применима к ВТСП купратам [32] либо двухслойному графену [94].

Поскольку состояния вблизи ферми-точек имеют дираковский спектр, проблема может быть проанализирована в рамках соответствующего релятивистского формализма. Низкоэнергетическое действие для нейтральных дираковских спинонов $\Psi$ в изоляторной фазе имеет структуру

$$S = \int d^3x \sum_\mu \sum_{\sigma=1}^N \bar{\Psi}_\sigma(\partial_\mu - ia_\mu)\gamma_\mu\Psi_\sigma \qquad (49)$$

где интегрирование выполняется в 2 + 1 измерениях, $\gamma_\mu$ – матрицы Дирака, $\bar{\Psi}_\sigma \equiv \Psi_\sigma^\dagger \gamma_0$, а $a_\mu$ – возникающее (emergent) калибровочное поле.

В зависимости от деталей решетки, $a_\mu$ может быть калибровочным полем типа U(1) или SU(2). Для большого количества "цветов" $N$ (определяемого числом точек Дирака в зоне Бриллюэна) действие $S_D$ описывает конформную теорию поля (CFT). Таким образом, мы имеем масштабно-инвариантное квантовое состояние с сильным взаимодействием и степенным спектром для всех возбуждений, причем хорошо определенные квазичастицы отсутствуют. Это состояние именуется как алгебраическая спиновая жидкость (см. [82, 32]).

Итак, бесщелевые возбуждения, несущие калибровочные заряды, могут экранировать калибровочное взаимодействие, делая его менее конфайнментным [32]. В отличие от истинных деконфайнментных фаз, где невзаимодействующие квазичастицы становятся свободными при низких энергиях, здесь деконфайнмент означает только то, что бесщелевые заряженные частицы остаются бесщелевыми, но не вполне свободны. Соответствующие бесщелевые спиновые жидкости получаются из фазы экзотической жидкости sfL и uRVB.

На уровне среднего поля электронная функция Грина состояния uRVB в координатном пространстве является произведением фермионной и бозонной функции Грина, а в импульсно-частотном представлении – их сверткой. Спектральная функция состоит из уширенного квазичастичного пика с весом порядка допинга и некогерентного фона. При допировании фазы uRVB возникает состояние странного металла с большой поверхностью Ферми, так что картина изолированных спинонных дираковских точек оказывается несколько измененной [32].

Среднеполевая фаза sfL приводит к бесщелевой U(1) спиновой жидкости. Она дает пример алгебраической спиновой жидкости, в которой корреляционные функции имеют разрезы вместо полюсов, а спиновые корреляции убывают по алгебраическому степенному закону. Для других широко известных бесщелевых состояний, таких как твердые вещества, сверхтекучие жидкости, ферми-жидкости и т. д., бесщелевые возбуждения всегда описываются свободными бозонами или свободными фермионами. Единственным исключением является одномерная латтинджеровская жидкость; алгебраическую спиновую жидкость можно рассматривать как ее обобщение за пределами одного измерения. Важно, что алгебраическая спиновая жидкость является фазой материи, а не только возникает в критической точке между двумя фазами.

Ферми-жидкостная фаза включает бозонную конденсацию, которая восстанавливает квазичастичную картину. Поэтому низкоэнергетические возбуждения в этой фазе описываются электронно-подобными квазичастицами, и эта фаза соответствует ферми-жидкости электронов.

Динамика для калибровочного поля U(1) возникает из-за экранирования бозонами и фермионами, причем и те, и другие несут калибровочный заряд. В случае малого допирования можно учитывать только экранирование фермионами. После интегрирования по $\Psi$ в (49) эффективная статистическая сумма для калибровочного поля U(1) имеет вид [95]

$$\begin{aligned}\mathcal{Z} &= \int Da_\mu \exp\Big(-\frac{1}{2}\int \frac{d^3q}{(2\pi)^3} a_\mu(q)\Pi_{\mu\nu}a_\nu(-q)\Big),\\ \Pi_{\mu\nu} &= \frac{N}{8}\sqrt{q^2}\Big(\delta_{\mu\nu} - \frac{q_\mu q_\nu}{q^2}\Big).\end{aligned} \qquad (50)$$

Поляризуемость $\Pi$ делает калибровочную связь $a_\mu j^\mu$ ($j^\mu$ – ток) маргинальным воз-

25

мущением в фиксированной точке свободных фермионов.

Рассмотрим электронную функцию Грина. В первом порядке в $1/N$ было получено

$$G(x) = \langle b^\dagger(x)b(0)\rangle_0 \langle f(x)f^\dagger(0) \exp(i\int_0^x dx \cdot a)\rangle, \tag{51}$$

где $\int_0^x dx$ – интеграл по прямому пути возврата и $\langle...\rangle$ означает интеграл по калибровочным флуктуациям [32]. Тогда имеем

$$G(x) \propto (x^2)^{-(2-\alpha)/2} \tag{52}$$

с показателем $\alpha \sim 1/N$, являющимся аномальной размерностью; для антиферромагнетика с квадратной решеткой $\alpha = 32/(3\pi^2 N)$. Эти результаты описывают частичный конфайнмент спинонов и бозонов, связанных калибровочным полем.

## 3.4 Модель фазовых струн

Поскольку при наличии проекторов Хаббарда статистика Ферми-Дирака неприменима, Венг предложил "статистику фазовых струн", описывающую процесс допирования в $t-J$ модели [96, 97, 98]. Вся фермионная знаковая структура на простых решетках для полузаполненной зоны в моттовском изоляторе полностью приводима (устранима), так как проблема взаимодействующих электронов превращается в проблему взаимодействия локализованных спинов. Когда решетка фрустрирована, спиновая задача может по-прежнему включать проблему знака, что затрудняет расчеты типа Монте-Карло, но характер спина в моттовском состоянии подразумевает локализацию, так что электроны становятся различимыми.

Инжектированная дырка остается стабильной из-за конфайнмента спинонов и голонов полем фазового сдвига, несмотря на то, что фон представляет собой спинон-голонное море. Точный деконфайнмент возникает только в пределе нулевого допинга, когда дырка теряет свою целостность и распадается на голон и спинон. При допировании моттовского изолятора дырками возникает неприводимая знаковая структура, но при низком допинге эта структура должна быть очень разреженной по сравнению с эквивалентной системой свободных фермионов с той же плотностью.

Простой способ сформулировать статистику Венга – систематически посчитать неприводимые знаки на языке мировых линий [98]. Для невзаимодействующего ферми-газа эмпирические знаки входят в статсумму как

$$Z_{FG} = \sum_c (-1)^{N_{\text{ex}}[c]} Z_0[c], \tag{53}$$

где $Z_0[c]$ обозначает статсумму для мировой линии одной частицы, сумма идет по конфигурациям линий $c$, $Z_0[c] > 0$ и $N_{\text{ex}}[c] = \sum_w w C_w[c] - \sum_w C_w[c]$ – целое число, подсчитывающее количество обменов (оборотов $w$ раз вокруг временной оси), $C$ – всевозможные циклические разложения. Напротив, неприводимые знаки, входящие в статсумму $t-J$ модели, можно посчитать как

$$Z_{t-J} = \sum_c (-1)^{N^h_{\text{ex}}[c] + N^\downarrow_h[c]} \mathcal{Z}[c]. \tag{54}$$

Сумма теперь идет по конфигурациям мировых линий спиновых частиц ("спинонов") и дырок. Относительно друг друга дырки ведут себя как фермионы, а $N_{ex}^h[c]$ подсчитывает обмены в конфигурации $c$. Спиноны здесь представляют собой бозоны с твердой сердцевиной (спины вверх рассматриваются как фон). Новый аспект статистики Венга заключается в том, что знаковая структура определяется четностью числа дырочно-спинонных столкновений, связанных с конкретной конфигурацией мировых линий.

Формально, спроектированный оператор уничтожения электронов записывается в слейв-фермионном представлении, причем явно учитывается знаковый множитель $(-\sigma)^j$: $c_{j\sigma} = (-\sigma)^j f_j^\dagger b_{j\sigma}$, где $f_j^\dagger$ – фермионный голон, а для спиновой системы используются швингеровские бозоны $b_{j\sigma}^\dagger$. Из-за знакового множителя межспиновый суперобмен приобретает общий отрицательный знак, поэтому волновая функция основного состояния чисто спиновой системы при половинном заполнении не имеет узлов.

Венг предложил еще одну формулировку, которая строго эквивалентна представлению мировых линий. Она включает бозонные операторы $h_{i\sigma}^\dagger$ и $b_{i\sigma}$ для "голонов" и "спинонов", удовлетворяющие ограничению $h_i^\dagger h_i + \sum_\sigma b_{i\sigma}^\dagger b_{i\sigma} = 1$, которое может интерпретироваться на псевдоспиновом языке. Используя формализм калибровочных полей, можно записать спроектированный электронный оператор как

$$c_{i\sigma} = h_i^\dagger b_{i\sigma} \exp(\frac{1}{2}\left[\Phi_i^s - \Phi_i^0 - \sigma\Phi_i^h\right])(\sigma)^{N_h}(-\sigma)^i, \quad (55)$$

где $N_h$ – полный оператор числа голонов. Данное представление типа Йордана-Вигнера включает фазы $\Phi^{s,h,0}$, нелокальным образом связанные с положением всех других частиц. Этот нелокальный эффект учитывается путем введения "взаимной статистики" между спинами и дырками.

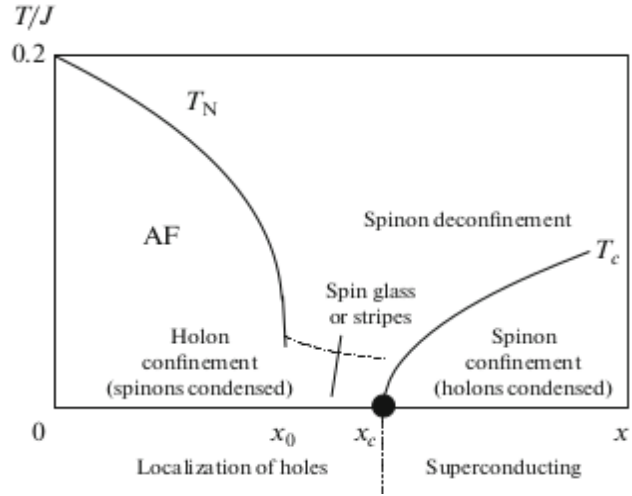

Рис. 4: Фазовая диаграмма модели фазовых струн при малом допировании [99]. Дуальная топологическая калибровочная структура определяет различные фазы посредством конфайнмента-деконфайнмента.

Согласно работе [100], антиферромагнитная и сверхпроводящая фазы в картине Венга дуальны: в первой голоны находятся в конфайнменте, а спиноны в деконфайн-

менте, во втором – наоборот. Эти две фазы разделены новой фазой – так называемым бозе-изолятором, где и голоны, и спиноны деконфайнментны.

## 3.5 Фрустрации и спиновая жидкость в решетках Кондо

За счет подавления (экранирования) магнитного момента эффект Кондо может приводить к возникновению экзотических состояний – тяжелой ферми-жидкости или спиновой жидкости. Выигрыш в энергии немагнитного состояния по сравнению с однозонной моделью Хаббарда или моделью Гейзенберга определяется температурой Кондо $T_K$ [101]. В свою очередь, тенденция к состоянию спиновой жидкости (образование RVB-синглетов) дает дополнительный выигрыш для кондовского состояния по сравнению с магнитоупорядоченными фазами.

Для описания основного состояния кондо-решетки в режиме сильной связи было разработано специальное приближение среднего поля, которое использует псевдофермионное представление для операторов локализованных спинов $S = 1/2$. В рамках этого приближения Коулмен и Андрей [101], учитывая SU(2)-симметрию в формализме континуального интеграла, рассмотрели формирование состояния спиновой жидкости в двумерной ситуации. Затем это приближение было применено к случаю ферро- и антиферромагнитного упорядочения [102, 79, 103]. При этом смешивание между локальными моментами и электронами проводимости может быть представлено через вспомогательный бозон $b_i \sim \sum_\sigma c^\dagger_{i\sigma} f_{i\sigma}$. Его конденсация и означает формирование тяжелой ферми-жидкости: в области, где электронные состояния имеют преимущественно $f$-характер, зона является почти плоской.

Роль фрустраций в решетках Кондо была детально исследована в работе [104]. Как видно из Рис. 5, кондовское экранирование уменьшает величину локального момента, уменьшая тем самым критическое значение $Q_c$, необходимое для формирования спиновой жидкости. Граница антиферромагнитной фазы простирается от $K = K_c$ на кондовской оси до $Q = Q_c$ на оси фрустраций. При больших $Q$ и малых $K$ реализуется спин-жидкостной металл с локализованными $f$-электронами и малой поверхностью Ферми, а при больших $K$ возникает тяжелая ферми-жидкость с большой ферми-поверхностью и делокализованными $f$-электронами. Поскольку объем ферми-поверхности сохраняется, она не может плавно эволюционировать от малой до большой, поэтому спин-жидкостной металл и тяжелая ферми-жидкость должны быть разделены квантовым фазовым переходом при нулевой температуре.

В [104] также был рассмотрен простой пример фрустрированной спиновой модели – решеточная модель Шастри-Сазерленда-Кондо, гамильтониан которой имеет вид $\mathcal{H}_{SSK} = \mathcal{H}_K + \mathcal{H}_{SS}$, причем слагаемые

$$\begin{aligned}
\mathcal{H}_K &= \sum_{\mathbf{k}\sigma} t_\mathbf{k} c^\dagger_{\mathbf{k}\sigma} c_{\mathbf{k}\sigma} + 2J_K \sum_i \mathbf{s}_c(i)\mathbf{S}_i \\
\mathcal{H}_{SS} &= J_1 \sum_{\boxtimes,\boxdot} \mathbf{S}_k \mathbf{S}_l + J_2 \sum_{\langle i,j \rangle} \mathbf{S}_i \mathbf{S}_j
\end{aligned} \quad (56)$$

содержат кондовское и гейзенберговское взаимодействия. Здесь $\boxtimes, \boxdot$ означает сумму по плакетам с чередующимися диагональными связями. Отношение диагонального взаимодействия к взаимодействию ближайших соседей $Q = J_1/J_2$ играет роль параметра фрустрации.

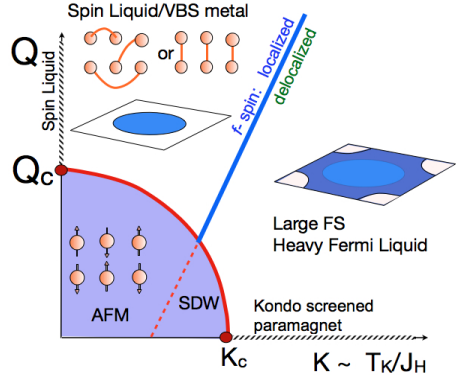

Рис. 5: Схематическая диаграмма Кондо-решетки в пространстве параметров квантовой фрустрации $Q$ и $K = T_K/J_H$ ($J_H$ – гейзенберговское взаимодействие) [104]. Переход между малой и большой поверхностью Ферми может также происходить через промежуточную фазу "странного металла".

Решетка Шастри-Сазерленда имеет то преимущество что основное состояние при большой фрустрации является состоянием валентной связи со спиновыми синглетами, расположенными по диагональным связям, и четко определенной волновой функции. В отличие от других типов спиновой жидкости можно показать, что это димерное состояние является *точным* основным состоянием модели при условии $J_1 \gtrsim 2J_2$.

## 3.6 Фракционализованная ферми-жидкость

Еще один сценарий формирования нефермиевской жидкости был назван фракционализованной ферми-жидкостью FL$^*$. В такой фазе заряженные возбуждения имеют обычные квантовые числа (заряд $\pm e$ и спин $1/2$), но они сосуществуют с дополнительными дробными степенями свободы. Важно отметить, что эта фаза имеет поверхность Ферми (ПФ) с объемом, нарушающим теорему Латтинджера.

В работах Сачдева и др. [103, 105] концепция фракционализованной ферми-жидкости была введена первоначально для $s-d(f)$ обменной модели (решетки Кондо). В фазе FL$^*$ отсутствует кондовское спаривание, т.е. конденсация хиггсовского бозона $b$ ($\langle b_i \rangle = \langle f_i^\dagger c_i \rangle = 0$), однако существует аномальное среднее RVB-типа $\chi_{ij} = \langle f_i^\dagger f_j \rangle$. Таким образом, локализованные моменты не принимают участие в формировании ПФ (они образуют отдельную спинонную ферми-поверхность), но адиабатически связаны со спиновой жидкостью, описываемой калибровочной теорией и обладающей соответствующими экзотическими возбуждениями в фазе деконфайнмента. В пространственных размерностях $d \geq 2$ стабильна спиновая жидкость типа Z$_2$, а при $d \geq 3$ существует спиновая жидкость U(1). В этой фазе коэффициент электронной теплоемкости $C/T$ логарифмически расходится. На фоне такого состояния может возникать магнитная неустойчивость для спинонной ПФ – металлическое магнитное состояние волны спиновой плотности SDW$^*$, которое может характеризоваться малым моментом.

Упомянем трудности теории: в работе [103] пришлось искусственно ввести поле,

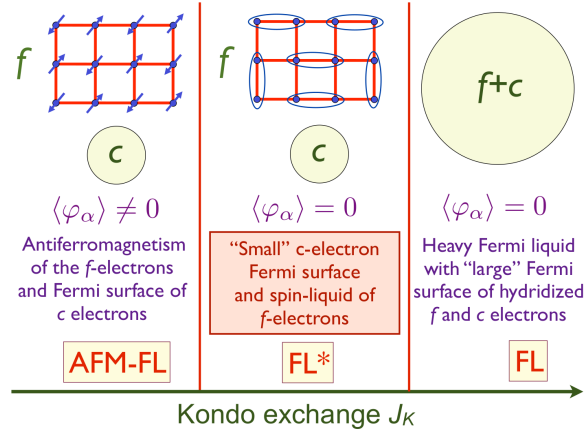

Рис. 6: Ферми-жидкостные фазы модели решетки Кондо [107]. В состоянии FL* спиновая жидкость локальных $f$-моментов отделена от малой ПФ $c$-электронов проводимости, которая не подчиняется обычной теореме Латтинджера.

стабилизирующее ненасыщенное состояние с малым моментом. В работе [106] такого введения удалось избежать, и в численных расчетах было получено состояние, где эффект Кондо сосуществует с антиферромагнетизмом, однако спин-жидкостной параметр порядка исчезает при переходе к малой ПФ. Отметим, что в простом ренормгрупповом рассмотрении квантовых фазовых переходов в магнитных решеток Кондо может возникать промежуточная область с неферми-жикостным поведением и частично подавленными магнитными моментами [79].

В последующих работах [105] фаза FL* как основное состояние уже рассматривалась только в квантовой критической точке, а за ее пределами – как неустойчивая относительно фазы с локальными моментами (см. раздел 4.1).

В дальнейшем, следуя идее Рибейро и Вена [78] о возникновении двух гибридизованных подсистем (спинонов и допонов) в $t-J$ модели (раздел 2.5), представление о фракционализованной ферми-жидкости было обобщено на однозонную модель Хаббарда [108, 107]. При этом в антиферромагнитной и спин-жидкостной (неупорядоченной) фазах спиноны описываются как швингеровские бозоны, а в ферми-жидкостной фазе – как фермионы (см. (41)).

Изменение статистики спинонов происходит в точке квантового фазового перехода между двумя конфайнментными фазами – магнитной и фазой с большой ПФ. Описание такого перехода требует дальнейших исследований. Здесь могут оказаться полезными суперсимметричные представления (см., напр., [72]). Упомянем также полностью фракционализованное представление электрона, где он разделяется на квазичастицы, несущие его спин и заряд, а также майорановские фермионы, несущие статистику Ферми [109].

При сравнении рисунков 6 и 7 интересно отметить, что в отличие от модели Хаббарда, где кулоновское взаимодействие приводит к разрушению ферми-жидкостного состояния (формированию хаббардовских подзон с малой ПФ), эффект $s-d$ обмена противоположен: с ростом $s-d$ обменного параметра возникает тяжелая ферми-жидкость и большая ПФ. Причина состоит в том, что эти взаимодействия ведут к разным типам спаривания электронов и спинонов – диагональному и недиагональ-

ному (гибридизационному).

Представление швингеровских бозонов может быть введено не только в $t-J$ модели [108], но и в антиферромагнитной фазе спин-фермионной модели – модификации однозонной модели Хаббарда с конечным взаимодействием [107], что позволяет говорить о реализации в последней состояния необычной ферми-жидкости FL*.

Спин-фермионная модель [110], используемая при интерполяционном описании коллективизированного магнетизма (ее аналогии с $s-d$ обменной моделью обсуждаются в [111]), также позволяет разделить спиновые и электронные степени свободы.

Механизм перехода и описание квантовой критической точки между фазами FL и FL* обсуждается в работе [105] (рис. 6). Когда мы движемся к фазе FL* в пределах фазы FL, две поверхности Ферми с $k_{F1}$ и $k_{F2}$ ведут себя по-разному. ПФ с $k_{F1}$ эволюционирует в ПФ с $k_{F*}$ – в состояние FL* и остается "холодной" вблизи квантовой критической точки: сильного рассеяния этих квазичастиц нет, а квазичастичный вычет электронов при переходе остается конечным. Напротив, ПФ с $k_{F2}$ становится "горячей": время жизни квазичастиц сокращается, а вычет уменьшается и в конечном счете исчезает в квантовой критической точке, так что динамика квазичастиц проявляет неферми-жидкостное поведение. Это аномальное поведение распространяется на всю горячую ПФ, в отличие от изолированных "горячих точек", которые появляются в теории волн спиновой плотности.

В состоянии FL* имеются дополнительные низкоэнергетические возбуждения локальных моментов, которые дают добавочный топологический вклад в изменение импульса кристалла. Действительно, действие вихревого потока аналогично преобразованию Либа-Шульца-Маттиса [46], так что спин-жидкостное состояние в размерности $d=2$ приобретает это изменение импульса. Обычно объем поверхности Ферми определяется общим числом электронов в системе. Однако имеется общий топологический анализ [112], основанный на "протаскивании" кванта потока (благодаря циклическим граничным условиям система рассматривается как тор, в конту-

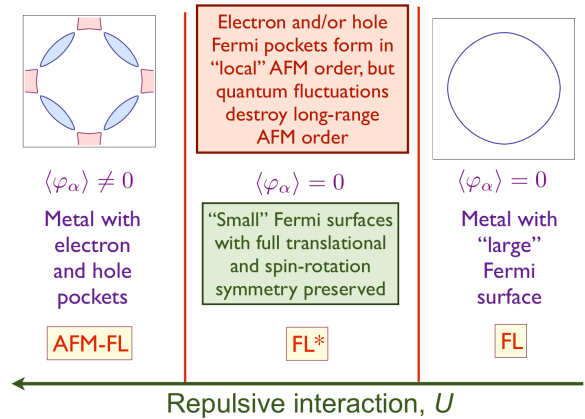

Рис. 7: Фазовая диаграмма основного состояния модели Хаббарда согласно [107], $\varphi_\alpha$ – параметр антиферромагнитного порядка. В фазе AFM-FL удвоение элементарной ячейки АФМ порядком гарантирует выполнение теоремы Латтинджера. Фаза FL* не имеет антиферромагнитного порядка ($\langle\varphi_\alpha\rangle = 0$), однако наследует малые карманы ПФ от фазы с $\langle\varphi_\alpha\rangle \neq 0$, причем теорема Латтинджера не выполняется.

31

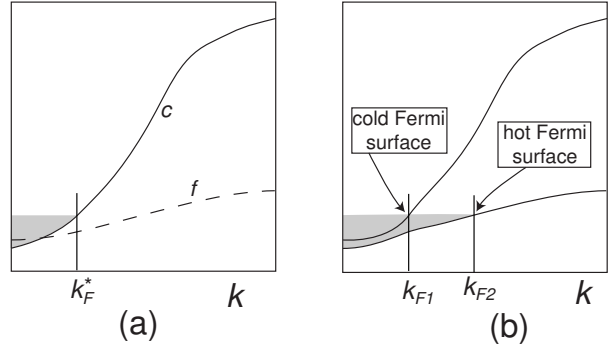

Рис. 8: (а) Состояние FL$^*$ с холодной ПФ и импульсом Ферми $k_F^*$. Фермионы $f$, образующие спиновую жидкость, схематически представлены пунктирной линией. (б) Состояние FL, которое подчиняется обычной теореме Латтинджера, но имеет две ПФ.

рах которого возникает кристаллический импульс, что аналогично появлению силы Фарадея с изменением потока) и глобальной калибровочной симметрии U(1) (сохранение заряда). Из него следует, что существование немагнитного состояния FL$^*$ с другим объемом ПФ разрешено, если мы допускаем глобальные топологические возбуждения. Такие возбуждения естественно возникают в калибровочных теориях для FL$^*$ состояния. Таким образом, нарушение теоремы Латтинджера должно сопровождаться топологическим порядком [103]. Следовательно, формирование малой ПФ, хаббардовского расщепления (отметим, что теорема Ошикавы [112] справедлива и для изолятора) и состояния с неупорядоченными локальными моментами можно связать с топологическим порядком в состоянии спиновой жидкости.

В многозонной системе существует возможность для парциального перехода Мотта между двумя металлическими фазами, где поверхность Ферми претерпевает квантованное изменение. В простейшем случае одна подполоса (или орбиталь) меняет свой характер с металлического на моттовский, в то время как другие полосы остаются металлическими. (орбитально-селективный переход Мотта, см. [113]).

Будучи стабильной в низкотемпературном пределе, парциальная фаза Мотта нарушает теорему Латтинджера и, следовательно, является неферми-жидкостным металлом. Согласно концепции Сачдева, это и есть фракционализованная ферми-жидкость (FL$^*$), а переход между FL и FL$^*$ является орбитально-селективным переходом Мотта.

Парциальный переход Мотта был получен из численных результатов для допированной однозонной хаббардовской модели купратных сверхпроводников. Это поведение, по-видимому, является импульсно-селективным, то есть некоторые области в пространстве импульсов ведут подобно ферми-жидкости, а другие – как моттовский изолятор.

# 4 Теория квантовых фазовых переходов. Спиновая жидкость и топологический порядок

С точки зрения квантовых фазовых переходов, особый интерес представляют фрустрированные магнитные системы, в которых дальний порядок подавлен (наиболее простой пример здесь – треугольная решетка, где согласно численным расчетам имеется конечная область деконфайнмента, обусловленная фрустрацией). При учете квантовых флуктуаций в них могут реализоваться состояния с локальными моментами без нарушения симметрии – вплоть до самых низких температур существует только ближний порядок. В случае половинного заполнения зоны эти системы могут демонстрировать переход металл-изолятор в парамагнитной фазе.

Такие квантовые спиновые жидкости (QSL) заметно отличаются от их классических аналогов по ряду позиций [113]. Квантовые колебания обычно устраняют сильное вырождение основного состояния фрустрированных систем за счет квантового туннелирования, что приводит к единственному основному состоянию (с точностью до глобальных преобразований симметрии или топологического вырождения). QSL – термодинамически стабильные фазы вещества, характеризующиеся динамическими калибровочными полями и топологическим порядком. Это подразумевает существование дробных возбуждений, связанных с калибровочным полем. Несмотря на такую связь, дробные возбуждения асимптотически свободны (деконфайнмент). Волновые функции в QSL могут характеризоваться дальнодействующим запутыванием. При этом QSL следует отличать от "тривиальных" квантовых парамагнетиков без топологического порядка и фракционализации (например, синглетные магнетики типа $TlCuCl_3$, где связываются в пары спины $1/2$).

## 4.1 Деконфайнментная квантовая критичность

Обратимся теперь к вопросу о том, как подход калибровочной теории может быть применен к проблеме высокотемпературной сверхпроводимости. В основном состоянии последняя возникает при допировании моттовского изолятора, когда антиферромагнитный порядок пропадает, причем при температурах выше точки сверхпроводящего перехода нормальное состояние описывается псевдощелевой картиной. Чтобы увязать одновременное исчезновение магнитного порядка с возникновением псевдощелевого поведения и сверхпроводимости, был предложен подход, позволяющий описать появление сначала немагнитного моттовского изолятора при нулевом допинге (спиновая жидкость), который при допировании превращается в сверхпроводник (см. рис. 9).

Рассмотрим некоторое спин-жидкостное моттовское состояние. При изменении химического потенциала в нем произойдет нуль-температурный моттовский фазовый переход, обусловленный не изменением взаимодействия, а допированием. Квантовая критическая точка будет определять поведение системы в конечном интервале параметров. Линии кроссоверов, возникающих вблизи этого перехода, показаны на рис.10.

Интересно отметить, что роторное представление также позволяет описать промежуточную псевдощелевую фазу, давая переход к металлическому состоянию только при конечной концентрации дырок [69]. Похожую картину дает и модель фазовых

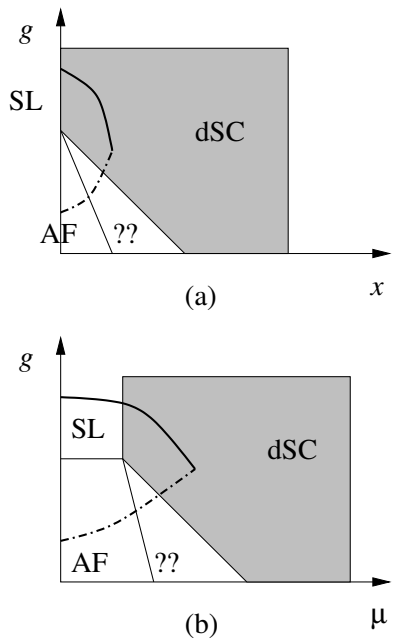

Рис. 9: (a) Схема нуль-температурной фазовой диаграммы, показывающая маршрут между антиферромагнитным изолятором Мотта и $d$-волновым сверхпроводником как функцию допирования $x$. Вертикальная ось помечена параметром $g$, который означает меру фрустрации спиновой системы в изоляторной фазе. AF – антиферромагнитное упорядоченное состояние, SL – изоляторная спиновая жидкость, которая получеается путем увеличения фрустрации. б) То же, что на (a), но как функция химического потенциала.

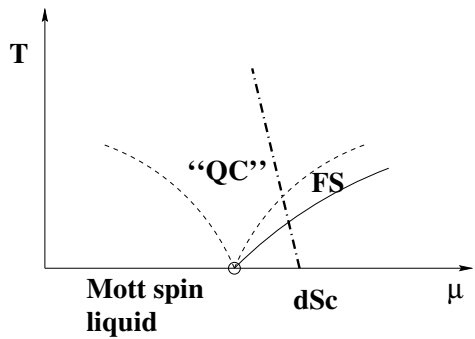

Рис. 10: Схематическая фазовая диаграмма для индуцированного допированием перехода Мотта между спин-жидкостным изолятором и $d$-волновым сверхпроводником. Полужирная линия – путь, принятый системой с плотностью дырок $x$, которая имеет сверхпроводящее основное состояние. Область, обозначенная FS, представляет собой флуктуационный режим сверхпроводящего перехода. QC является квантовой критической областью, связанной с критической точкой Мотта. Эта область может быть сопоставлена с высокотемпературной псевдощелевой фазой.

струн (раздел 3.4), а также вычисления в представлении МЭ Х-операторов (некогерентные состояния при малом допировании) [65].

Среди возможных кандидатов на спин-жидкостное состояние в литературе упоминаются димерная фаза (валентно-связанное состояние, VBS), $Z_2$ и U(1) спиновые жидкости. Поскольку наиболее вероятно, что трансляционно-инвариантное спин-жидкостное состояние может при допировании привести к $d$-волновой сверхпроводимости с нодальными квазичастицами, наиболее подходящей является фаза U(1) [32]. Более того, даже если физическая система не обладает стабильным состоянием U(1) спиновой жидкости, последняя может существовать как критическое состояние, разделяющее антиферромагнитную фазу и $Z_2$ спиновую жидкость, что способствует возникновению псевдощелевой и сверхпроводящей фаз [114].

Простейшей ситуацией, выходящей за рамки теории Ландау, оказывается QPT при нулевой температуре, когда между различными квантовыми фазами с нарушенными симметриями в квантовой критической точке происходит переход, запрещенный по Ландау. Примером является фрустрированный антиферромагнетик [33, 34], где деконфайнментная квантовая критическая точка разделяет упорядоченные фазы с нарушенными вращательной и трансляционной симметрией. Для таких точек лучшим отправным пунктом при описании критического поведения является не обычный параметр порядка, а набор дробных возбуждений, которые специфичны для критической точки, но не присутствуют ни в одной из фаз рядом с ней. Более того, существует дополнительная топологическая структура, присутствующая в критической точке и связанная с топологическим законом сохранения. Дополнительную сохраняющуюся величину удобно интерпретировать как полный поток калибровочного поля, возникающего в критической точке. Он сохраняется только асимптотически — при низких энергиях в квантовой критической точке. Этот закон сохранения обеспечивает резкое различие между деконфайнментными и обычными критическими точками.

В континуальном пределе топологический поток определяется как

$$Q = \frac{1}{4\pi} \int dxdy \epsilon_{\alpha\beta\gamma} \phi_\alpha \partial_x \phi_\beta \partial_y \phi_\gamma$$

и связан с числом скирмионов в спиновой конфигурации. Сохранение $Q$ означает появление дополнительного глобальной U(1) симметрии в критической точке и наличие дополнительного набора бесщелевых калибровочных возбуждений. Оно естественно приводит к большим значениям критического индекса корреляционной функции $\eta$ для деконфайнментной критической точки.

Известно, что компактная U(1) калибровочная теория в 2 + 1-мерном пространстве-времени допускает наличие точечных дефектов (монополей), которые меняют топологический заряд (число скирмионов). Следуя [115], рассмотрим траектории двумерной ренормализационной группы с управляющим параметром $s$ и параметром $\lambda_4$, описывающим эффективность монополей (monopole fugacity); они показаны на рис. 11.

На линии $\lambda_4 = 0$ расположена нестабильная неподвижная точка $s = s_c$, отделяющая линии ренормгрупповых траекторий в направлении магнитоупорядоченного неелевского состояния от линий в направлении парамагнитной спиновой жидкости с бесщелевым U(1) фотоном и щелевыми спинонными возбуждениями.

35

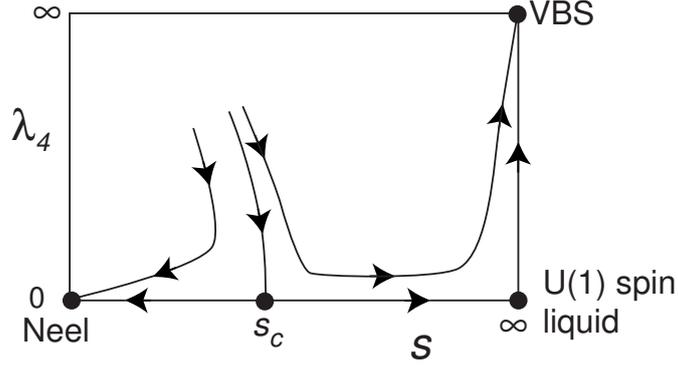

Рис. 11: Траектории ренормализационной группы для антиферромагнетика со спином $S = 1/2$ и квадратной решеткой.

Обратимся теперь к случаю $\lambda_4 \neq 0$. При малых $s$ величина $\lambda_4$ оказывается несущественной (иррелевантной) для неподвижной точки $\lambda_4 = 0, s_c = 0$, Однако для $s > s_c$ траектории отклоняются и притягиваются в сторону больших значений $\lambda_4$. Это связано с тем, что параметр $\lambda_4$ становится опасно релевантным возмущением для неподвижной точки при больших $s$, описывающей спин-жидкостную фазу. Вблизи $s_c$ значение $\lambda_4$ очень мало на масштабах порядка корреляционной длины $\xi_{\text{spin}}$, но оно становится порядка единицы и более на втором масштабе длины, который обозначается как $\xi_{\text{RVB}}$. Тогда мы имеем связь $\xi_{\text{RVB}} \sim \xi_{\text{spin}}^{\lambda}$, где $\lambda > 1$, поэтому $\xi_{\text{RVB}} \gg \xi_{\text{spin}}$.

Второй пример деконфайнментной квантовой критической точки представляет собой QPT в тяжелой ферми-жидкости; рассмотрим ее неустойчивости, следуя [105] (рис.12). В изоляторных магнетиках существует прямой переход второго рода между неелевским состоянием и состоянием с другим порядком — валентно-связанным твердым состоянием (VBS). Его можно рассматривать как аналог прямого перехода между тяжелой ферми-жидкостью и металлом с локальными магнитными моментами (LMM), где происходит потеря "кондовского" порядка одновременно с появлением магнитного. Действительно, неелевское состояние и состояние FL (решетка Кондо) являются устойчивыми хиггсовскими фазами компактной U(1) калибровочной теории. Они нестабильны относительно "деконфайнментного" перехода в U(1) спиновую жидкость и фазу FL* соответственно. Однако такие деконфайнментные фазы являются довольно лабильными состояниями вещества и могут быть неустойчивым по отношению к конфайнментным фазам с обычным порядком — например, неустойчивость U(1) спиновой жидкости к переходу в VBS. Естественно рассмотреть неустойчивость состояния FL* по отношению к состоянию с LMM, но таким образом, чтобы квантовая критичность оставалась деконфайнментной. В изоляторных магнетиках переход между двумя конкурирующими порядками (неелевским и VBS) происходит не как функция некоторого параметра настройки, а динамически, как функция масштаба длины-времени. Действительно, в парамагнетике потеря корреляций Нееля происходит на одном масштабе $\xi$, тогда как пиннинг VBS-порядка появляется на гораздо большем масштабе $\xi_{\text{VBS}}$, который расходится как степень $\xi$. Аналогично, прямой переход второго рода между конкурирующими порядком Кон-

36

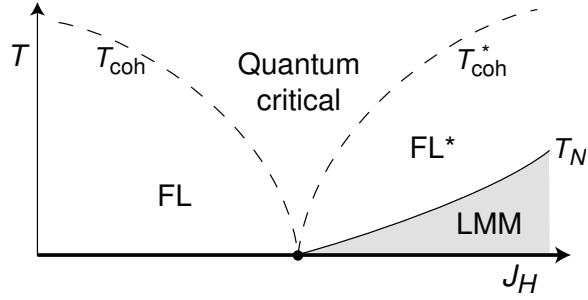

Рис. 12: Фазовая диаграмма вблизи квантового перехода между ферми-жидкостью FL и состоянием с локальными моментами LMM [105]. Две различные энергетические шкалы приводят к двум отличающимся критическим показателям, согласно которым точка Нееля $T_N$ и температура когерентности $T_{\text{coh}}$ (или $T_{\text{coh}}^*$) приближаются к квантовой критической точке.

до в FL и магнитным порядком в металле с локальными моментами возможен, но требует по крайней мере двух расходящихся масштабов длины-времени в квантовой критической точке, причём деконфайнмент обнаруживается только на более коротких масштабах.

## 4.2 Хиггсова критичность

Как обсуждалось выше, начиная с работы [103] переходы между ферми-жидкостью и неферми-жидкостными состояниями обсуждаются в терминах хиггсовского бозона $b$. Как и в теории элементарных частиц, такой формализм удобен при описании нарушенных симметрий.

В фазе Хиггса калибровочное поле в кулоновском газе (раздел 3.2) станоится щелевым по механизму Андерсона-Хиггса, так что для заряда $q = 1$ фаза Хиггса и конфайнментные фазы плавно связаны друг с другом. Ситуация резко отличается при $q = 2$, т.е. когда поле бозонов соответствует спариванию, например, в сверхпроводнике. В этом случае может существовать граница между конфайнментной и хиггсовской фазами, так что последняя (фаза спаривания) деконфайнментна и имеет остаточную калибровочную симметрию $Z_2$, поэтому параметр порядка инвариантен при изменении знака зарядов $q = 1$, составляющих пару [32].

Следуя [116], обсудим эволюцию фазы U(1)–FL$^*$ с её малой поверхностью Ферми к обычной "большой" ПФ при сильном допировании. Существует обычный путь преобразования, связанный с исчезновением антиферромагнитного порядка. Это переход между двумя состояниями ферми-жидкости, который описывает стандартная теория Герца-Миллиса [117, 118].

Простая SU(2)-теория поля [116], допускает конденсаты двух бозонных полей: калибровочного бозона $R$ и хиггсовского бозона $H^a$. Соответствующие фазы получают путём изменения управляющих параметров $s$ и $g$ (рис. 13) и обозначаются по их конденсатам.

1) Фаза Хиггса, обозначенная как (A) на рис. 13, где обе симметрии SU(2)$_{\text{spin}}$ и SU(2)$_{\text{gauge}}$ нарушены, что приводит к $\langle R \rangle \neq 0, \langle H^a \rangle \neq 0$. Здесь калибровочные возбуждения-фотоны являются щелевыми. Эта фаза описывает AFM-металл, где

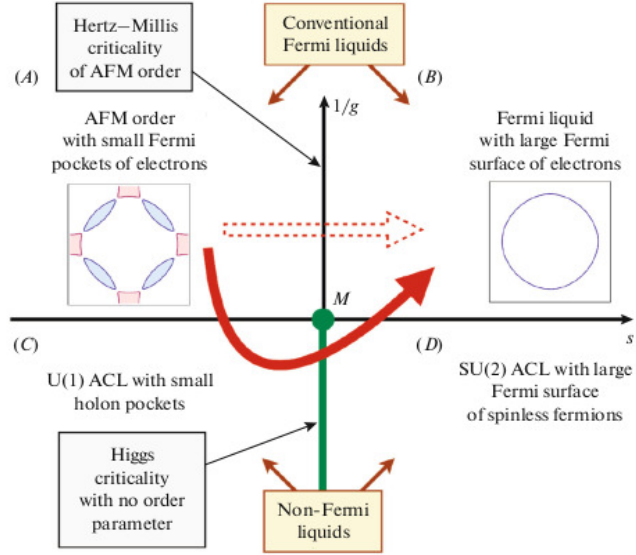

Рис. 13: Схема металлических фаз по теории [116] в координатах обратная константа связи $1/g$ – параметр хиггсовской критичности $s$. Только фаза A имеет нарушенную глобальную симметрию, связанную с наличием дальнего антиферромагнитного (AFM) порядка. Обычные ферми-жидкостные фазы сверху претерпевают переход от малых к большим поверхностям Ферми, сопровождаемый потерей AFM порядка. Пунктирная стрелка – прямой маршрут между этими фазами, который может дать описание купратов с электронным допированием. Стрелка вокруг точки M – маршрут с увеличением дырочного допирования. Фаза $U(1)$-$FL^*$ спускается из фазы $U(1)$ ACL и имеет малую поверхность Ферми электронов из-за наличия топологического порядка, тогда как фаза A выше нее имеет "малую" ПФ электронов из-за нарушения трансляционной симметрии.

большая поверхность Ферми реконструируется в дырочные и электронные карманы из-за конденсации бозона $H^a$ вдоль вектора неелевского параметра порядка $\mathbf{n}$.

2) Конфайнментная фаза SU(2), обозначенная как (B) на рис. 13. Симметрия $SU(2)_{\text{spin}}$ здесь не нарушена. Мы имеем $\langle R \rangle \neq 0$, $\langle H^a \rangle = 0$, что необходимо для сохранения спиновой вращательной инвариантности, так как $\mathbf{n} = 0$. Это обычная ферми-жидкость с большой ПФ.

3) В фазе Хиггса (C) симметрия $SU(2)_{\text{gauge}}$ нарушена, а $SU(2)_{\text{spin}}$ – нет, что ведет к $\langle R \rangle = 0, \langle H^a \rangle \neq 0$. Это означает развитые локальные AFM флуктуации без какого-либо дальнего порядка. Если выбрать $H^a$ вдоль направления $(0,0,1)$, проведя калибровочное преобразование, симметрия U(1)-подгруппы $SU(2)_{\text{gauge}}$ не нарушается, так что калибровочный $A^z$-фотон остается бесщелевым. Таким образом, эта фаза описывает U(1) алгебраическую зарядовую жидкость (ACL), или голонный металл [119]. Однако из-за выраженного локально AFM порядка ПФ реконструируется в карманы голонов $\psi_p$, которые минимально связаны с калибровочным полем U(1). В зависимости от температуры может происходить непрерывный кроссовер из U(1) ACL в U(1) $FL^*$ (или "голон-дырочный" металл), где некоторые из голонов ($\psi_\pm$) начинают формировать связанные состояния с щелевыми спинонами.

4) Последняя фаза (D) имеет полную симметрию, так что ни одно из полей не сконденсировано: $\langle R \rangle = \langle H^a \rangle = 0$. В отличие от U(1) ACL, где только $A^z$-фотон был бесщелевым, в этой фазе имеется триплет бесщелевых SU(2)-фотонов, связанных с большой ПФ. Эта фаза может быть описана как SU(2) алгебраическая зарядовая жидкость. При достаточно низких температурах она неустойчива по отношению к сверхпроводимости.

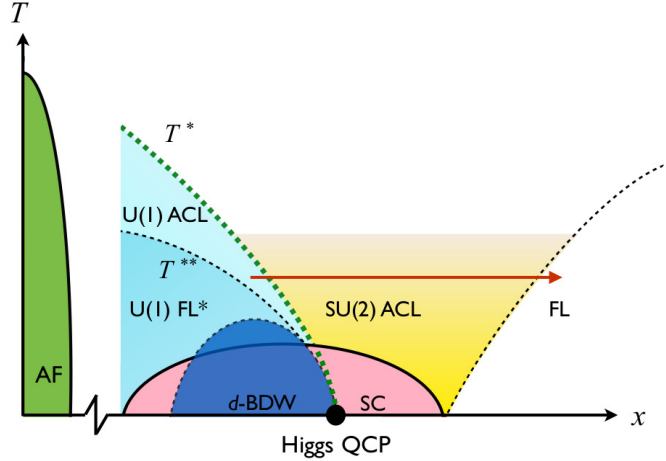

Рис. 14: Фазовая диаграмма для купратов, основанная на теории критичности Хиггса. Алгебраические зарядовые жидкости (ACL) обладают ферми-поверхностями бесспиновых $\psi$-фермионов, несущих электромагнитный заряд: в SU(2) ACL поверхность Ферми большая и связана с SU(2) калибровочным полем, тогда как в U(1) ACL поверхность Ферми малая и связана с калибровочным полем U(1). Ферми-жидкость FL* опускается из U(1) ACL при связывании $\psi$-фермионов с нейтральными спинонами. $d$-BDW – волна плотности связей, SC – $d$-волновой сверхпроводник,

Сравнение вычислений электронной функции Грина в топологически упорядоченной хиггсовской фазе SU(2) калибровочной теории флуктуирующего антиферромагнетизма на квадратной решетке с результатами первопринципных расчетах в динамической теории среднего поля (DMFT) и квантовом методе Монте-Карло проведено в работе [120].

## 4.3 Фазовые переходы и некогерентные состояния в проводящих магнетиках

В проводящих магнетиках при изменении параметров взаимодействия возникают дополнительные квантовые фазовые переходы по сравнению с моделью Гейзенберга – появление магнитного упорядочения в основном состоянии, качественное изменение поведения магнитного момента и т.д. В пределе сильных корреляций такие переходы не описываются простыми подходами типа Стонера или Герца-Миллиса [117, 118]. В частности, в ферромагнетике может происходить переход из насыщенной в ненасыщенную фазу. В насыщенном (полуметаллическом) состоянии поверхность Ферми исчезает для одной из проекций спина, так что этот переход может

считаться топологическим. Отметим, что строгое доказательство теоремы Нагаока о насыщенном ферромагнитном основном состоянии модели Хаббарда с бесконечным $U$ имеет топологический характер; теория фазовых струн Венга в $t-J$ модели может рассматриваться как обобщение этого подхода на случай конечных $J$.

Состояния со спином вниз в модели Хаббарда в насыщенном ферромагнетике дают самый простой пример неквазичастичных состояний. Формально в пределе $U = \infty$ удобно факторизовать оператор Хаббарда $X_i(-,0) = X_i(-,+)X_i(+,0)$, что аналогично разделению голона и спинона. Тогда вычисление дает [65]

$$\langle\langle X_{\mathbf{k}}(-,0)|X_{-\mathbf{k}}(0,-)\rangle\rangle_E = \left\{E - t_{\mathbf{k}} + \left[G^0_{\mathbf{k}\downarrow}(E)\right]^{-1}\right\}^{-1}, \; G^0_{\mathbf{k}\downarrow}(E) = \sum_{\mathbf{q}} \frac{n_{\mathbf{k}+\mathbf{q}}}{E - t_{\mathbf{k}+\mathbf{q}} + \omega_{\mathbf{q}}}. \tag{57}$$

где $t_{\mathbf{k}}$ и $\omega_{\mathbf{q}}$ – затравочный спектр дырок и магнонов. Функция Грина нулевом приближении $G^0_{\mathbf{k}\downarrow}(E)$ является сверткой функций Грина свободного носителя со спином вверх и магнона и описывает полностью полностью неквазичастичные состояния. При увеличении концентрации дырок выражение (57) приобретает полюс на уровне Ферми, что и означает разрушение насыщенного ферромагнетизма.

Интересно, что исторически понятие спинонов было введено Андерсоном в связи с проблемой спиновой поляризации в ферромагнитных металлах [121] (см. также [122]). Спинонными были названы некогерентные вклады в электронную спектральную плотность, обусловленные взаимодействием с магнонами. Соответствующий формализм описания сильного зонного ферромагнетизма, основанный на тождестве Уорда, был разработан в работах Эдвардса и Герца [123]. Он был также применен для описания перехода металл-изолятор в парамагнитной фазе и трактовки неферми-жидкостного состояния [124]. Парамагнитное состояние с неупорядоченными магнитными моментами, полученное в [124] в аналогичном приближении, может быть отождествлено с андерсоновским состоянием RVB.

Дальнейшее развитие теории неустойчивости ферромагнетизма связано с анализом трехчастичных вкладов [125].

В $t-J$ модели антиферромагнетика большая часть состояний является некогерентной, хотя для конечных $J$ состояния вблизи дна зоны образуют узкую когерентную полосу с малым вычетом порядка $|J/t| \ll 1$ и тяжелой массой $\sim |t/J|$ [126].

Конкуренция ферро- и антиферромагнетизма при допировании моттовского изолятора приводит к переходам первого рода, формированию неоднородных состояний и спиральных магнитных структур см. (рис.15). Соответствующие вычисления в обобщенном приближении Хартри-Фока и при учете корреляций в методе вспомогательных бозонов Котляра–Рукенштейна были проведены в работах [127, 63, 128]. При этом возникают различные антиферромагнитные фазы, в том числе характеризующиеся малым вычетом [128]. Методы типа Хартри-Фока переоценивают устойчивость магнитных фаз, в то время как учет корреляций позволяет успешно описать парамагнитную фазу при больших $U$, т.е. фазу типа спиновой жидкости [63].

## 4.4 Сверхпроводимость и топологический порядок

Экспериментальное открытие сверхпроводящего порядка Камерлинг-Оннесом исторически привело к теории нарушенной симметрии, однако последовательное кванто-

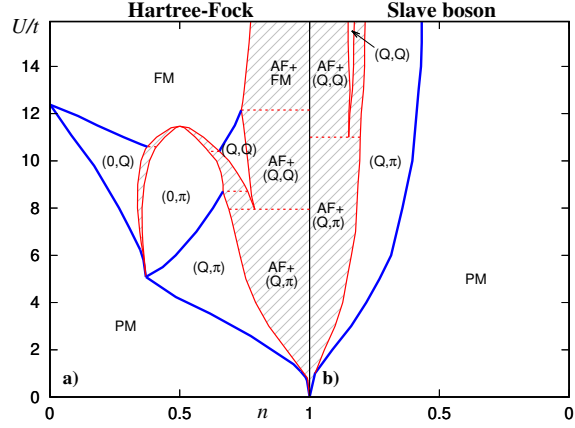

Рис. 15: Магнитная фазовая диаграмма основного состояния двумерной модели Хаббарда в рамках приближения Хартри-Фока (a) и при учете корреляций (b) для квадратной решетки при концентрации электронов $n < 1$ [63]. Спиральные фазы обозначены в соответствии с формой их волнового вектора. Жирные линии обозначают фазовые переходы второго рода. Штриховка показывает области фазового расслоения, пунктирные линии – границы между различными парами фаз. Сплошные линии соответствуют границам между однородной фазой и областями фазового расслоения, $\mathbf{Q}_{\text{AFM}} = (\pi, \pi)$, PM и FM – пара- и ферромагнитная фазы.

вое топологическое рассмотрение усложняет эту картину. В действительности сверхпроводимость может быть описана теорией Гинзбурга-Ландау с динамическим U(1) калибровочным полем [5]. Конденсация электронной пары зарядов $2e$ нарушает калибровочную теорию U(1) до калибровочной теории $Z_2$ при низких энергиях. Последняя является эффективной теорией топологического порядка $Z_2$, так что реальный сверхпроводник имеет именно такой топологический порядок. Часто сверхпроводимость описывается теорией Гинзбурга-Ландау без динамического калибровочного поля U(1); вместо него рассматривается нарушенная симметрия U(1) [4]. Однако реальные сверхпроводники в электромагнитном калибровочном поле не являются состояниями с нарушенной симметрией, а представляют собой топологически упорядоченные состояния [129, 130]. Решающее значение здесь имеет теорема Элицура [131] о невозможности существования калибровочно-инвариантного локального параметра порядка для состояния в электромагнитном калибровочном поле.

Сам по себе сверхпроводящий порядок *в поле* не есть порядок, нарушающий симметрию, но является топологическим порядком, который выходит за рамки теории симметрии Ландау (хотя, разумеется, без поля эта теория дает полное описание); по иронии судьбы, именно он был обнаружен в 1911 году. При условии, что калибровочное U(1) электромагнитное поле является динамическим, сверхпроводник имеет струнные возбуждения, характеризующие топологический порядок, – петли потока $hc/2e$.

Оказывается, что разрушение сверхпроводящего порядка происходит за счет обычного механизма Березинского-Костерлица-Таулеса, включающего размножение топологических вихрей. В спиновой жидкости инстантоны, разрушающие топологический порядок, иррелевантны. Тогда U(1) калибровочный поток является сохраня-

ющейся величиной, и мы имеем деконфайнментное состояние.

Теория флуктуаций фазы в двух измерениях хорошо описывается подходом Березинского-Костерлица-Таулеса (БКТ), поэтому разрушение сверхпроводящего порядка идет по обычному механизму БКТ – за счет распространения вихрей и теплового "развязывания" пар вихрь-антивихрь.

Одним из следствий феноменологического подхода Гинзбурга-Ландау является конкуренция между переходом БКТ для фермионного спаривания и конденсацией бозонов. Поэтому линии фазового перехода для фермионного спаривания и бозонная конденсация на фазовой диаграмме становятся линиями кроссовера, и только сверхпроводящий переход остается реальным БКТ-переходом.

Во внешнем электромагнитнои поле вихри выталкиваются. Это поле при конденсации приобретает массу (отсюда возникает эффект Мейсснера), симметрия U(1) понижается до $Z_2$ (т.е возникает спиновая жидкость). Таким образом, в поле механизм Ландау нарушения симметрии, который имеет место в сверхтекучести, не работает – возникает топологический переход.

Теперь обратимся к вихревым структурам в сверхпроводящем состоянии. Самая интригующая проблема здесь – квантование магнитного потока. Поскольку бозон имеет заряд $e$, в то время как фермионная пара – $2e$, вопрос состоит в том, может ли вихрь $hc/e$ быть более стабильным, чем обычный вихрь $hc/2e$. Как оказывается, вихрь типа B с квантованием потока $hc/e$ будет иметь более низкую энергию, чем вихри $hc/2e$, особенно при низком допировании. Этот вывод является общей особенностью калибровочной теории U(1). Экспериментально стабильные $hc/e$–вихри до сих пор не обнаружены; вихри $hc/2e$ естественно объясняются в SU(2) модели с двумя бозонами [32].

Статистика Венга приводит к изменению скрытой топологической структуры по сравнению с ферми-жидкостным БКШ-сверхпроводником: сверхпроводник представляет собой конденсат бозонов с зарядом $e$ и на первый взгляд кажется, что вихри несут поток $hc/e$. Однако, если рассмотреть влияние взаимной статистики Черна-Саймонса на топологию сверхпроводника, голон с зарядом $e$ приобретает скачок фазы $\pi$ при обходе изолированного спинона. Таким образом, связывание спинона с вихрем приводит к уменьшению кванта потока вдвое до значения $hc/2e$, как и для обычного конденсата куперовских пар [96, 98].

# 5  Решеточные калибровочные теории и струны

В теории поля часто используется переход к решеточным моделям, включающим дискретность пространства и времени. Такой переход является еще более естественным в квантовой теории твердого тела при использовании мнимого времени $t = i\tau$, поэтому ряд трудностей калибровочной теории разрешается легче. В пределе нулевой температуры временной интервал становится бесконечным, так что возникает дополнительное пространственное измерение.

Физическое пространство состояний абелевой решеточной калибровочной теории состоит из замкнутых петель электрического потока. Обсудим их происхождение, следуя обзору [17]. Удобно использовать гамильтонов подход в приближении непре-

рывного времени. Рассмотрим действие с анизотропными связями

$$S = \beta_\tau \sum_{nk}[1 - \cos\theta_{0k}(n)] - \beta \sum_{n,ik}\cos\theta_{ik}(n), \qquad (58)$$

где $\theta(n)$ – угловая непрерывная переменная, пространственные связи обозначены латинскими индексами, а временное направлением $\tau$ – индексом 0 (данная теория строится аналогично калибровочной теории модели Изинга). В классическом непрерывном пределе электрическое поле выражается через оператор углового момента как $E_k(\mathbf{n}) = (g/a^2) L_k(\mathbf{n})$, где $a$ – параметр решетки, $g$ – калибровочный заряд, играющий роль параметра взаимодействия (здесь имеется аналогия с роторным представлением (32)). Тогда находим

$$\mathcal{H} = (a^3/2)\sum_{nk} E_k^2(\mathbf{n}) - (1/g^2 a) \sum_{n,ik}\cos\theta_{ik}(\mathbf{n}). \qquad (59)$$

После разложения в непрерывном пределе имеем

$$\mathcal{H} = (a^3/2)\sum_{nk}[E_k^2(\mathbf{n}) - B_k^2(\mathbf{n})], \qquad (60)$$

причем $\theta_{jk} = a^2 g B_i$, индексы $ijk$ определены на цикле. Первый член в (59) – решеточная форма объемного интеграла от квадрата электрического поля, а второй – от магнитного.

Физическое пространство состояний является локально калибровочно-инвариантным и удовлетворяет закону Гаусса. Поэтому допустимы только замкнутые петли электрического потока: теория не имеет источников или стоков. При сильной связи важен только электрический член в уравнении (59), причем энергия замкнутой петли потока пропорциональна ее длине. Магнитный член в (59) позволяет петлям флуктуировать, но всегда оставляет их замкнутыми.

Потоки электрического заряда соответствуют потенциальной энергии, магнитная часть – кинетической. Сохранение потока в топологической фазе означает сохранение заряда, а следовательно наличие бозонов. Линии электрического потока определяют струны (strings; либо, в другой терминологии – петли, loops). Детали струнной картины определяются величиной константы взаимодействия $g$. Для электрического поля константа связи равна $g^2$, а для магнитного – $1/g^2$. Поэтому при больших $g$ доминирует потенциальная энергия, замкнутые петли электрического потока стягиваются, их мало. Напротив, при малых $g$ (в слабой связи) доминирует кинетическая энергия, роль которой играет магнитное поле, так что петли легко возбуждаются, их много.

А. М. Поляков установил глубокие соответствия решеточных калибровочных теорий в трех и четырех измерениях двумерным спиновым системам [132]. Двумерные O(n) гейзенберговские спиновые системы допускают точное решение в континуальном пределе для $n \geq 3$. Причина этого заключается в том, что эти теории имеют "скрытую симметрию", что приводит к бесконечному числу законов сохранения. В двух измерениях эти законы сохранения запрещают образование частиц в процессах рассеяния. Такие законы сохранения не могут существовать в нетривиальных

четырехмерных теориях. Однако Поляков предположил, что существует нетривиальное обобщение законов сохранения на неабелевы калибровочные поля в четырех измерениях, что может привести к замкнутому решению. Поляков рассмотрел неабелевы калибровочные теории как киральные поля, определенные на замкнутых петлях в реальном пространстве-времени. Абелеву решеточную калибровочную теорию можно рассматривать аналогичным образом – физическое пространство состояний теории состоит из замкнутых петель электрического потока.

Перечислим еще раз исходные модели, в которых есть слабые калибровочные взаимодействия и возможен деконфайнмент, а также возникают калибровочный бозон и фермионы:

1. Решеточная $Z_2$ теория, которая в твердом теле дуальна модели Изинга, имеющей классические фазовые переходы. Она предсказывает квантовые фазовые переходы.

2. Абелева O(1) модель, которая в непрерывном пределе соответствует максвелловой теории электромагнитного поля.

3. Неабелева SU(2) модель, которой в твердом теле соответствует модель Гейзенберга.

В первых двух моделях возможен переход от конфайнмента к деконфайнменту, они могут быть использованы для описания спиновой жидкости. В третьей модели имеет место асимптотическая свобода, фазы деконфайнмента нет.

В $Z_2$ спиновой жидкости и решеточной абелевой O(1) модели возникают струны, которые в свою очередь образуют сети – жидкость струнных сетей. В зависимости от силы связи, в фазе возникает много или мало петель, они будут маленькими или большими. Струны могут колебаться. Колебания струнной жидкости описываются векторным полем, перпендикулярным направлению распространения волн, что аналогично теории Максвелла для распространения света. Таким образом, появляются калибровочные поля – бозоны, которые являются единственными возбуждениями в случае замкнутых струн. В квантовой электродинамике этим колебаниям струнной жидкости соответствуют бесщелевые фотоны. Если разорвать струну, возникают фермионы – ее оборванные концы. Они могут быть отождествлены с электронами, поскольку их взаимодействие с деформацией струнной сети описывается уравнениями электрон-фотонного взаимодействия [3].

Случайное возникновение ("вспыхивание") монополей (см. раздел 3.2) приводит к конфайнменту фермионов; это аналогично взаимодействию кварков через мезоны в квантовой хромодинамике.

Фермионы возникают как дефекты квантовой дальнодействующей запутанности (entanglement), даже если исходная модель – чисто бозонная. Предыдущие подходы позволяли получить фермионы в моделях бозонного поля лишь для U(1) калибровочного поля в размерности пространства 2 + 1. В подходе же квантовой запутанности фермионы и бозоны одновременно получаются для любой размерности и калибровочной группы.

В обычной физике конденсированного состояния бозонное поле возникает в результате коллективных колебаний атомов: это фононы в твердом теле (две поперечных и одна продольная мода) и продольные волны сжатия в жидкости (одна мода). Коллективные колебания протяженных объектов (струн), составляющих среду в случае нашей сильно коррелированной системы, могут приводить к возникновению

калибровочных бозонов (даже фотонов!) [3]. Более того, фермионы могут также появиться в результате коллективных колебаний квантовой жидкости, состоящей из протяженных объектов более сложной природы.

Чтобы избежать теоретико-полевого рассмотрения сплошной среды, обычно вводятся решеточные калибровочные теории, где поля включаются по связям, что соответствует дуальной решетке [17]. Эти поля эквивалентны электромагнитному полю и, согласно [3], порождают петли – калибровочные бозоны в фазе деконфайнмента. При разрыве петель образуются фермионы, которые можно рассматривать как связанные состояния кванта потока поля и заряженной частицы.

Китаев [88, 89] предложил точно решаемую модель, которая подтверждает результаты слейв-бозонной теории относительно деконфайнмента. Квазичастицы в этой модели несут электрические и магнитные заряды соответствующей калибровочной группы.

В подходе Вена [4] используется представление кубитов – простейших элементов квантовой информации. В отличие от подхода слейв-бозонов, здесь первичными элементами являются псевдоспины, расположенные не в узлах, а на связях решетки. Свойства электромагнитного поля оказываются обусловленными сложной организацией кубитов – нетривиальным топологическим порядком. Если в основном состоянии (вакууме) все кубиты находятся в простейшем "ферромагнитном" состоянии со спином вверх, то состояние с несколькими перевернутыми спинами описывает пространство с несколькими бозе-частицами нулевого спина, которым соответствует скалярное поле. Однако такой простой подход сталкивается с рядом проблем и требует дальнейшего развития. В действительности вакуум должен иметь гораздо более сложное строение.

Наблюдаемые элементарные частицы могут возникать только из нетривиального кубитного эфира, запутанного на больших расстояниях, так что материя состоит из запутанных кубитов. Многообразие свойств материи определяется топологической сложностью организации ее степеней свободы, а не исходным многообразием полей. В этой картине материю можно отождествить с информацией.

Обсуждаемые решеточные теории не являются локальными бозонными моделями, так как в решеточных калибровочных теориях струны неразрывны. Из теория струнных сетей следует, что язык струн на самом деле не обязателен: кубиты сами способны генерировать калибровочные поля, подчиняющиеся уравнениям типа Максвелла или Янга-Миллса. Таким образом, калибровочная симметрия не имеет принципиального значения для калибровочной теории: решеточная калибровочная теория всегда будет воспроизводить калибровочное взаимодействие при низких энергиях. Поскольку калибровочное поле может получаться из локальных моделей кубита без калибровочной симметрии, картина струн превращается в картину дальнодействующей запутанности, а калибровочные поля становятся ее флуктуациями. Струнная сеть есть способ описания закономерностей такой системы [4].

## 5.1 Конденсация струнных сетей

В дополнение к полевой теории струн необходима физическая картина, которая разъясняет, как возникают топологические фазы из микроскопических степеней свободы. В работе [12] эти вопросы рассмотрены для широкого класса "удвоенных"

топологических фаз, которые описываются суммой двух топологических квантово-полевых теорий с противоположными киральностями. Физически они характеризуются четностью и инвариантностью относительно обращения времени.

Как обсуждалось выше, локальные энергетические ограничения могут приводить микроскопические степени свободы к организации в эффективные протяженные объекты – струнные сети. Если кинетическая энергия этих струнных сетях велики, то они могут конденсироваться, давая начало топологическим фазам. Конденсация струнных сетей обеспечивает естественный физический механизм для возникновения топологических фаз в реальных системах конденсированных сред.

С простейшими кубитами на связях, образующих дуальную решетку, мы имеем дело в $Z_2$ решеточной калибровочной теории. Соответствующий гамильтониан модели Китаева запишется в виде [12, 82]

$$\mathcal{H} = -U \sum_i \prod_{\langle ij \rangle} \sigma^x_{\langle ij \rangle} - t \sum_p \prod_{\langle ij \rangle \in p} \sigma^z_{\langle ij \rangle}, \tag{61}$$

где $\sigma^{x,z}$ – матрицы Паули, $i$, $\langle ij \rangle$ и $p$ обозначают узлы, связи и плакеты решетки.

В единственном невырожденном основном состоянии мы имеем для всех связей и плакетов $A_i = \prod_{\langle ij \rangle} \sigma^x_{\langle ij \rangle} = 1$, $F_p = \prod_{\langle ij \rangle \in p} \sigma^z_{\langle ij \rangle} = 1$ (при этом важно, что $[A_i, A_j] = [F_p, F_{p'}] = [A_i, F_p] = 0$). Таким образом, спиновая жидкость является суперпозицией всех конфигураций замкнутых струн с равными положительными коэффициентами.

Спинонное возбужденное состояние в данном формализме является нарушенной связью в базисе $\sigma^z_{\langle ij \rangle}$. Поскольку здесь нет сохраняющегося спинового квантового числа, спинон не несет спина, но меняет локальную обменную энергию; в $Z_2$-калибровочной теории спинон несет соответствующий электрический заряд. Состояние спинона имеет собственные значения $A_i = -1$ и $A_j = 1$ для всех $j \neq 1$. При этом $F_p = 1$ на всех плакетах, так что спинон является суперпозицией всех конфигураций струн на решетке с единственным свободным концом на узле $i$.

Визон имеет дополнительную структуру: для него $F_p = -1$ на одном плакете $p$ и $F_{p'} = 1$ для всех $p' \neq p$, однако $A_i = 1$ для всех узлов $i$. Волновая функция по-прежнему является суперпозицией всех конфигураций замкнутых струн, как и в основном состоянии.

Гамильтониан (61) по форме напоминает гамильтониан Хаббарда (1). Первый член в (61) — "электрическая энергия" порядка $U$, которая является натяжением струны, в то время как "магнитная энергия" порядка $t$ – кинетическая энергия струны. Поэтому данный микроскопический гамильтониан эффективно описывает динамику струн.

Конфайнментная фаза соответствует большой электрической энергии и, следовательно, большому натяжению струны $U \gg t$. Таким образом, основное состояние представляет собой вакуумную конфигурацию с несколькими небольшими струнами. Деконфайнментная фаза соответствует большой магнитной энергии и, следовательно, большой кинетической энергии, описывая квантовую жидкость больших струн – суперпозицию многих струнных конфигураций, струнный конденсат. Большие струнные сети имеют типичный размер порядка размера системы, заполняя все пространство (см. рис. 16). Переход между этими состояниями напоминает переход Мотта-Хаббарда.

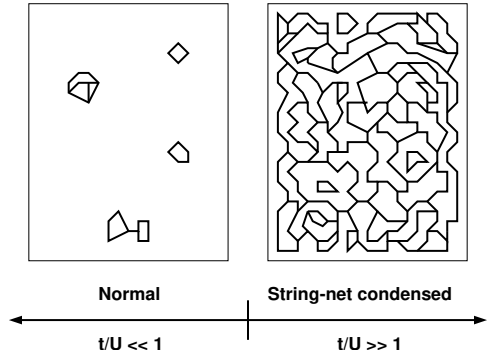

Рис. 16: Схематическая фазовая диаграмма для гамильтониана струнных сетей (61). Когда $t/U$ (отношение кинетической энергии к натяжению струны) мало, система находится в нормальной фазе, основное состояние есть вакуум с малым числом сетей. Когда $t/U$ велико, большие флуктуирующие струнные сети конденсируются. Таким образом, ожидается фазовый переход между двумя состояниями при некотором $t/U$ порядка единицы.

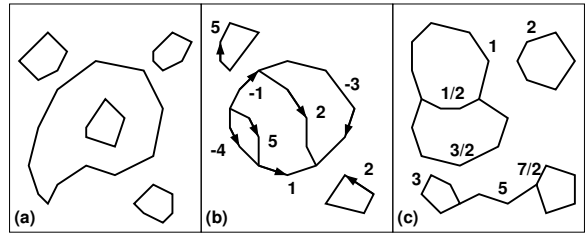

Рис. 17: Типичные конфигурации струнных сетей в дуальной формулировке (a) $Z_2$, (b) U(1) и (c) SU(2) калибровочной теории [12]. В случае (a) конфигурации состоят из замкнутых (непересекающихся) петель. В случае (b) это ориентированные графы с ребрами, помеченными целыми числами. Они подчиняются правилам ветвления $E_1 + E_2 + E_3 = 0$ для любых трех ребер, встречающихся в точке. В случае (c) сети состоят из (неориентированных) графов с ребрами, помеченными полуцелыми числами $1/2, 1, 3/2, ....$

Хотя эти выводы получены в простейшей калибровочной теории — на решетке $Z_2$, подобная, но более сложная картина существует для других деконфайнментных калибровочных теорий.

В модели U(1) рассматриваются кубиты с $N$ целыми уровнями (типами связей), в модели SU(2) возможны полуцелые значения (рис.17). На каждом узле размещается квант потока, так и возникают струнные сети. Таким образом, топологические фазы возникают при конденсации струнных сетей аналогично тому, как традиционные упорядоченные фазы возникают через конденсацию частиц.

Следует подчеркнуть, что объединение калибровочных бозонов и фермионов струнными сетями существенно отличается от теории суперструн для калибровочных бозонов и фермионов. В теории струнных сетей внутрисетевые калибровочные бозоны и фермионы происходят из кубитов, которые формируют пространство, а "струнная сеть" – это просто модель кубитов, название, которое указывает, как кубиты организованы в основном состоянии. При этом калибровочные бозоны суть волны

коллективных флуктуаций струнных сетей, а фермион соответствует концу струны. Напротив, в теории суперструн калибровочные бозоны и фермионы непосредственно происходят из струн, соответствуя их мелким кусочкам. Различные колебания этих кусочков струн приводят к разного рода частицам. Фермионы в теории суперструн получаются "вручную" путем введения грассмановых полей.

## 5.2 Тензорные сети

Новый язык в теории конденсированного состояния дают тензорные сети (Tensor Networks, TN) [29, 3]. В частности, он позволяет описать процессы запутывания и возникновения квантовых топологических состояний. Здесь за основу берется система квантовых псевдоспинов, а волновая функция системы описывается сетью связанных тензоров. Можно сказать, что такой тензор является "управляющей ДНК" для волновой функции: последняя конструируется из фундаментальных блоков-кусков квантового состояния по достаточно простым правилам [29]. В рамках этого метода удается описать широкий класс бозонных, фермионных и спиновых систем с различной размерностью, симметрией и граничными условиями, а также фазовые переходы в них.

Размерность гильбертова пространства, описывающего все связи в многочастичной системе, огромна: она экспоненциально велика по числу частиц (которое само очень велико, порядка числа Авогадро – $10^{23}$). К счастью, размерность пространства релевантных состояний в ряде случаев может быть уменьшена. Так, большинство гамильтонианов (в том числе, в теории твердого тела) являются локальными, – например, взаимодействие имеется только между ближайшими соседями. Оказывается, что для низкоэнергетических состояний таких гамильтонианов с щелевым спектром энтропия запутывания двух систем определяется не объемом, а площадью границы (здесь возникает аналогия с голографической моделью вселенной и теорией черных дыр, согласно которым вся информация закодирована на внешней поверхности). Это рассмотрение накладывает ограничение и на физически допустимый вид низкоэнергетических состояний. Более того, при эволюции квантовой многочастичной системы, описываемой локальным гамильтонианом, подавляющее большинство состояний оказывается реально недостижимыми.

Тензорная сеть (TN) – это набор тензоров, в котором некоторые (или даже все) индексы свернуты согласно некоторому правилу (эта операция называется сверткой TN). В результате получается новый тензор с несколькими открытыми индексами. При этом полное число операций, необходимых для получения окончательного результата для TN, существенно зависит от порядка свертки индексов. Формализм TN позволяет разбить волновую функцию многочастичного состояния на малые фрагменты.

Следуя обзору [29], рассмотрим квантовую систему из $N$ частиц, степени свободы каждой из которых можно описать $p$ различными состояниями (например, для системы типа квантовой модели Гейзенберга со спином $1/2$ имеем $p = 2$, так что каждая частица представляет собой двухуровневую систему или кубит). Для такой системы любую волновую функцию $|\Psi\rangle$, описывающую ее физические свойства,

можно записать в виде

$$|\Psi\rangle = \sum_{i_1 i_2 \ldots i_N} C_{i_1 i_2 \ldots i_N} |i_1\rangle \otimes |i_2\rangle \otimes \cdots \otimes |i_N\rangle \qquad (62)$$

где $|i_r\rangle$ – базис состояний каждой частицы $r = 1, ..., N$. В приведенном выше уравнении $C_{i_1 i_2 \ldots i_N}$ – наборы $p^N$ комплексных чисел $i_r = 1, ..., p$ для каждой частицы $R$, символ $\otimes$ обозначает тензорное произведение отдельных квантовых состояний для каждой из частиц в системе многих тел. Величины $C_{i_1 i_2 \ldots i_n}$ можно рассматривать как коэффициенты тензора $C$ с $N$ индексами $i_1 i_2 \ldots i_n$, т.е. ранга $N$ с коэффициентами $O(p^N)$. Отсюда видно, что число параметров, описывающих волновую функцию, экспоненциально велико по размеру системы, так что такое полное описание практически неэффективно. Введение TN-состояний позволяет уменьшить сложность в представлении состояний $|\Psi\rangle$, обеспечивая точное описание ожидаемых характеристик запутанности состояния. Это достигается путем замены "большого" тензора $C$ на сеть "меньших" тензоров. Такой подход сводится к разложению состояния $|\Psi\rangle$ на "фундаментальные блоки", т.е. сеть из тензоров меньшего ранга. Окончательное представление в терминах TN, как правило, зависит от числа модельных параметров полиномиально и, таким образом, является вычислительно эффективным описанием квантового состояния системы многих тел.

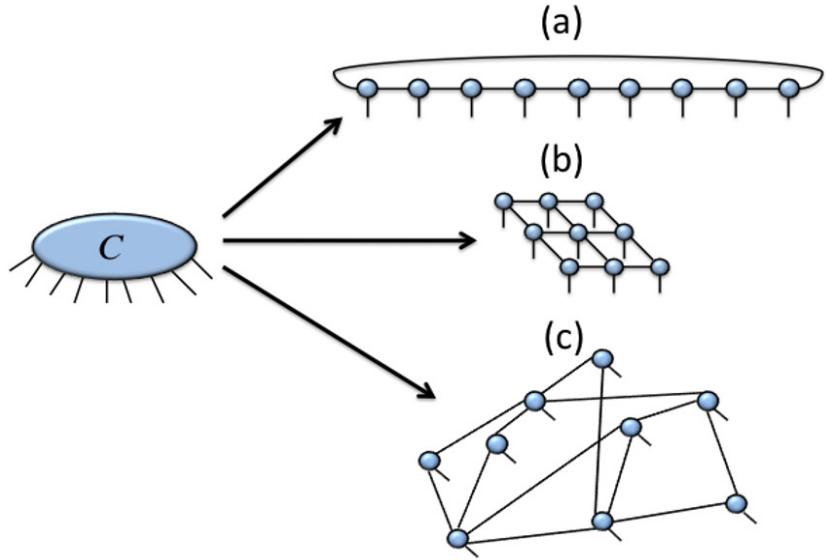

Рис. 18: Разложение в тензорные сети тензора $C$ через (a) MPS с периодическими граничными условиями, (b) PEPS с открытым граничным условием и (c) произвольную тензорная сеть.

Простой пример (рис. 18a) – состояние Матричного Произведения (Matrix Product States, MPS) с периодическими граничными условиями. Здесь количество параметров составляет $O(NpD^2)$, где открытые индексы в TN принимают $p$ значений, а остальные – $D$ значений. Тогда свертка TN дает тензор ранга $N$, и, следовательно, $p^N$ коэффициентов. Однако эти коэффициенты не являются независимыми, а получаются из свертки данной TN и поэтому имеют некую структуру.

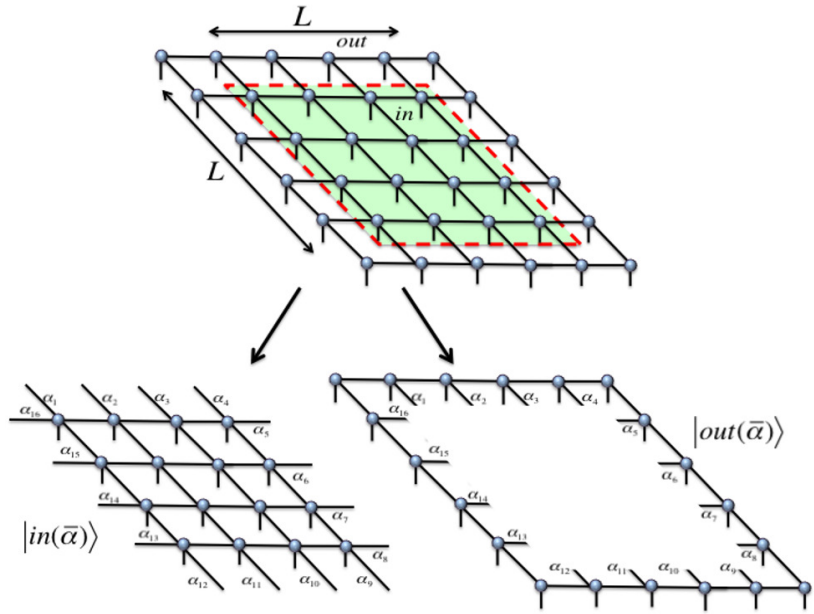

Рис. 19: Состояния $|in(\bar{\alpha})\rangle$ и $|out(\bar{\alpha})\rangle$ для блоков PEPS $4 \times 4$ и $6 \times 6$.

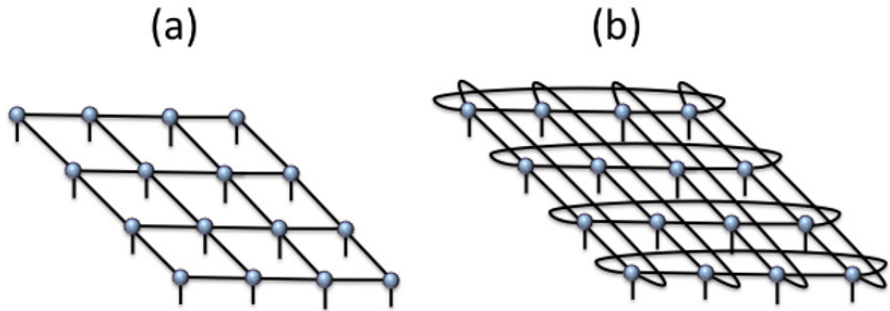

Рис. 20: $4 \times 4$ PEPS-состояния: (a) открытые граничные условия, (b) периодические граничные условия.

Состояние MPS имеет конечную корреляционную длину тогда и только тогда, когда наибольшее собственное значение ее двойного тензора невырождено. Состояние тензорного произведения (TP) является естественным обобщением состояния MPS на случай двух и большего числа измерений за счет размещения на каждом узле решетки тензоров высших ранга (вместо матриц). Оно также удовлетворяет закону площадей для запутанности. С другой стороны, оно описывает не только состояния с ближней запутанностью, но и топологически упорядоченные состояния с дальней запутанностью.

Замена тензора $C$ на TN предполагает появление дополнительных степеней свободы в системе, которые отвечают за "склеивания" различных блоков нашей ДНК. Эти новые степени свободы обозначаются индексами, соединяющими тензоры в сеть. Такие индексы называются индексами связи и имеют важный физический смысл: они представляют структуру многочастичной запутанности в квантовом состоянии $|\Psi\rangle$, а число их различных значений является количественной мерой квантовых кор-

реляций в волновой функции. Соответствующее максимальное значение $D$ – размерность связи тензорной сети (или внутренняя размерность состояния MPS).

Поясним, как запутанность соотносится с индексом связи. Пусть мы имеем TN-состояние с размерностью связи $D$ для всех индексов (рис. 19). Этот пример TN-состояния называется Проецированным Запутанным Парным Состоянием (Projected Entangled Pair State, PEPS). Оценим для этого состояния энтропию запутанности в случае линейного блока длины $L$. Введем совокупный индекс всех индексов TN $\bar{\alpha} = \{\alpha_1 \alpha_2 ... \alpha_{4L}\}$. Если индексы $\alpha_i$ принимают $D$ значений, $\bar{\alpha}$ может меняться до $D^{4L}$. Тогда в терминах состояний внутренней и внешней части блока имеем

$$|\Psi\rangle = \sum_{\bar{\alpha}=1}^{D^{4L}} |in(\bar{\alpha})\rangle \otimes |out(\bar{\alpha})\rangle . \qquad (63)$$

Редуцированная матрица плотности внутренней части запишется как

$$\rho_{in} = \sum_{\bar{\alpha},\bar{\alpha}'} X_{\bar{\alpha}\bar{\alpha}'} |in(\bar{\alpha})\rangle\langle in(\bar{\alpha}')|, \quad X_{\bar{\alpha}\bar{\alpha}'} \equiv \langle out(\bar{\alpha}')|out(\bar{\alpha})\rangle, \qquad (64)$$

причем ее ранг не более $D^{4L}$; редуцированная матрица плотности для внешней части выглядит аналогично. Энтропия запутывания $S(L) = -\text{Sp}(\rho_{in} l \rho_{in})$ блока ограничена сверху логарифмом ранга матрицы $\rho_{in}$, и мы находим

$$S(L) \leq 4L \ln D \qquad (65)$$

что является верхней границей для энтропии запутывания согласно закону площадей, который соответствует ситуации деконфайнмента (см. раздел 3.2).

MPS – это TN-состояния, которые описывают одномерные системы. В них имеется один тензор для каждого узла в системе многих тел. Соединительные индексы связей, которые склеивают эти тензоры, принимают $D$ значений, а открытые индексы, соответствующие физическим степеням свободы локальных гильбертовых пространств, – $p$ значений. Состояния PEPS (естественное обобщение MPS) позволяют описать случай более высоких пространственных измерений. На рис. 20 показано PEPS для квадратной решетки 4x4.

Для того, чтобы вычислить ожидаемую величину локальных наблюдаемых в состоянии тензорного произведения, мы должны уметь свертывать двухмерные тензорные сети, что в общем случае не является простой и эффективной процедурой. Однако с этой целью были разработаны приближенные методы, применяемые в квантовых вычислениях [30].

# 6 Заключение

Современная теория конденсированного состояния для систем с сильными корреляциями оперирует существенно многоэлектронными состояниями. В таких системах МЭ операторы и функции (раздел 2.2) оказываются первичными, а одноэлектронные операторы, в отличие от стандартной процедуры вторичного квантования, возникают при разрыве МЭ струны. Наглядный пример здесь – андерсоновское состояние

RVB, где электрон разделяется на спинон и голон (частицы, формально возникающие из представления МЭ оператора, раздел 2.4). Эти частицы возникают при разрыве связанной валентной пары, а исходно есть только полная волновая функция кристалла.

В отличие от обычной квантовой теории поля с одной стороны и классической топологии с другой стороны, квантовая топология оперирует целыми пространствами многоэлектронных операторов, причем форма и запись последних оказывается неоднозначной (примером служит представление Котляра-Рукенштейна (29) и введение хиггсова поля [116]). Каждый МЭ оператор представляет собой сложный динамический комплекс: в нем как из семени прорастают различные операторные произведения – процессы рекомбинации МЭ состояний. Таким образом описываются многочисленные квантовые фазы, которым соответствуют различные спектры возбуждений и физические модели – теории среднего поля, а флуктуации описываются калибровочными полями, определяющими устойчивость этих фаз.

Картина квантовой дальней запутанности (entanglement) меняет наши представления о структуре пространства-времени и корреляции, делая последнюю нелокальной; это существенно связано с топологическими свойствами системы (см. статью [10], где подробно обсуждается обобщенный формализм фейнмановского интеграла по траекториям, описывающий топологию пространства-времени).

Ситуация напоминает эффект Эйнштейна-Подольского-Розена, в котором имеют место непричинные квантовые корреляции на бесконечно большом расстоянии. Дополнительным фактором, способствующим квантовой когерентности, являются топологические инварианты, которые стабилизируют квантовое состояние (в обычных макроскопических системах типа кота Шредингера слабые внешние воздействия приводят к декогеренции таких состояний).

Новое описание на языке коллективных возбуждений протяженных структур – струнных сетей – позволяет единым образом ввести как калибровочные бозоны, так и фермионные частицы [3, 4]. Напротив, классическая геометрическая точка зрения (расслоенные пространства) может привести только к калибровочным взаимодействиям, так что квантовая запутанность открывает новую главу в физике.

В теории конденсированного состояния возможны языки спинов, бозонов или квантов информации – кубитов (простейшее псевдоспиновое представление последних – занятый или пустой узел решетки либо два направления стрелки). Согласно [4], на глубоком уровне материю можно рассматривать как совокупность кубитов. Такое пространство является динамической средой, оно есть океан кубитов, "кубитный эфир". Тогда материя, то есть элементарные частицы, суть возбуждения – волны, вихри, "пузырьки" и другие дефекты в этом эфире. При этом, согласно формулам квантовой физики и теории относительности, устанавливается эквивалентность массы, энергии, частоты, информации (энтропии).

В топологических подходах основное состояние рассматривается в рамках непертурбативного подхода. Основное состояние в упорядоченных фазах обычно описать не слишком трудно. Оно как правило устроено несложно и характеризуется слабой связью и голдстоуновскими возбуждениями. Наиболее интересной является как раз проблема описания "парамагнитного" (формально неупорядоченного) состояния, структура которого (в том числе топологическая, включая ряд спиновых жидкостей) может быть в реальности очень сложной, причем принципиальное значение имеют

эффекты фрустрации. В обсуждаемых проблемах мы сталкиваемся с поиском основного состояния системы, которое сформировано квантовыми флуктуациями и обладает большим топологическим вырождением (простейшим примером является синглетное основное состояние гейзенберговского антиферромагнетика, раздел 2.3). Фундаментально важен вопрос об энтропии такого состояния и содержащейся в нем информации, которая может быть колоссальной. Ведь жидкостное состояние – совокупность (суперпозиция) всех возможных положений узлов, а в жидкости струн все положения сами являются матрицами или вмещают тензоры. В такой системе естественно возникает дальнодействующее запутывание.

Топологические фазы, как и любые квантовые системы, обладают огромным пространством состояний, а добавление одного элемента существенно меняет состояние всей системы (например, андерсоновская катастрофа ортогональности [2, 48]).

Топологический порядок и запутывание приводят к ряду новых состояний квантовой материи и физических явлений, таких как дробный заряд, дробная и неабелева статистики... Если мы сможем реализовать квантовую жидкость ориентированных струн в реальных материалах, это позволит нам создать искусственные элементарные частицы, искусственный мир в искусственном вакууме [3, 4]. Эта тематика также подводит нас к философским вопросам о нелокальности и коллективной квантовой природе сознания [133], проблемам виртуальной реальности и т.д. Имеется и возможность применения концепции экзотических топологических состояний к проблеме передачи и хранения информации в живой материи (здесь возникают аналогии со структурой ДНК, см., напр., [29] и обзор [134]).

Хотя теория суперструн в применении к физике элементарных частиц и космологии встретилась с серьезными трудностями [135], аналогичные струнные подходы оказываются весьма полезными в физике конденсированного состояния. В этом контексте разными авторами предлагаются различные струнные концепции [3, 97]. Здесь возникают глубокие связи с квантовой электродинамикой [17] и хромодинамикой [136, 137], причем удобным оказывается формализм электрической и магнитной составляющих калибровочного поля. Помимо физики элементарных частиц, в настоящее время широко используются формальные методы, объединяющие теорию топологических состояний с теорией гравитации и структуры вселенной, в том числе голографические модели (AdS/CFT теория) [138, 139]. Оказывается, что теория петлевой квантовой гравитации может быть переформулирована в терминах определенного типа струнной сети, где струны помечены положительными целыми числами. Таким образом, конденсация струнных сетей в спиновой модели может привести к гравитации [6].

Современная картина спектра возбуждений в конденсированных средах дает новый взгляд и на ряд давних проблем физики твердого тела, в частности на проблему коллективизированного магнетизма и особенно описания парамагнитного состояния – как сильно коррелированных, так и обычных систем. Его природа оказывается крайне сложной: она включает сложную топологию и огромную скрытую информацию. Топология оказывается существенной и при описании корреляционного (хаббардовского) расщепления спектра в многоэлектронных системах (см. раздел 3.6).

Что касается практических применений, формализм тензорных сетей находит применение в квантовых нейронных сетях [30], а топологические фазы в линейных и плоских структурах и даже трехмерных материалах могут быть использованы в

новых поколениях электроники и потенциально в квантовых компьютерах.

# Список литературы



\end{document}